\documentclass[12pt]{article}
\usepackage[backend=biber, style=numeric, date=year, url=false, doi=false, isbn=false, eprint=true, firstinits=true, maxcitenames=5, maxbibnames=5]{biblatex}
\AtEveryBibitem{%
  \ifentrytype{online}{%
  }{%
    \clearfield{eprint}%
    \clearfield{urldate}%
  }%
}
\usepackage{graphicx}
\graphicspath{{Pictures/}}
\usepackage{transparent}
\usepackage{tikz}
\usepackage{tikz-cd}
\usepackage{tkz-euclide}
\usepackage{xparse}
\usetikzlibrary{arrows.meta}
\usetikzlibrary{quotes}
\usetikzlibrary{positioning,arrows,patterns}
\usetikzlibrary{decorations.markings}
\usetikzlibrary{calc}
\usepackage{fullpage}
\usepackage{amsfonts}
\usepackage{amsthm}
\usepackage{amsmath}
\usepackage{paralist}
\usepackage{comment}
\usepackage{subcaption}
\theoremstyle{plain} 
\newtheorem{thm}{Theorem}

\newtheorem{prop}[thm]{Proposition}

\theoremstyle{definition}
\newtheorem{defn}[thm]{Definition}
\newtheorem{expl}[thm]{Example}
\newtheorem{exc}{Exercise}

\theoremstyle{remark}

\newcommand{\R}{\mathbb{R}}
\newcommand{\C}{\mathbb{C}}
\newcommand{\g}{\mathfrak{g}}
\newcommand{\tr}{\operatorname{tr}}
\begin{document}
\title{Six lectures on Geometric Quantization}
\author{Konstantin Wernli}
\maketitle
\abstract{These are the lecture notes for a short course on geometric quantization given by the author at the XVIII Modave Summer School on Mathematical Physics, Sep 5 - Sep 9.}
\newpage
\section{Lecture 1: Introduction}
\subsection{Quantization...}
Well over a century ago, physicists realized that some of their observations could not be explained by the laws of physics as they knew them. One of many examples is the description of Black body radiation through classical statistical mechanics, the Rayleigh-Jeans law, which was wrong in the ultraviolet (high frequency) region - it predicted that for very short wavelenghts, the spectral radiance was approaching infinity, which is unphysical. Around 1900,\footnote{See for instance \cite{Planck1901}} Max Planck derived a new law for black body radiation, by assuming electromagnetic radiation can only be emitted (or absorbed) in discrete amounts, with energy $E = h\nu$ proportional to a constant $h$ now known as the Planck constant.\footnote{Around five years later, this was explained by Albert Einstein by postulating that those discrete amounts of energy correspond to physical particles called \emph{photons} \cite{Einstein1905}. } This was one of many discoveries ushering in the dawn of a new physical theory called the quantum theory. This theory was very much under active development 100 years ago: In 1922, the Stern-Gerlach experiment discoveredthe quantization of spin, Compton was doing research on what became known as Compton scattering\footnote{It was published in 1923 \cite{Compton1923}, and earned him the 1927 Physics Nobel prize.} and in 1923 de Broglie postulated the wave-particle duality. In 1930, Dirac published the seminal textbook ``The principles of Quantum Mechanics'' \cite{Dirac1930}, which quickly became one of the cornerstones of the subject. The first chapter - ``The Need for a Quantum Theory'' - is a beautiful explanation why Quantum Mechanics is needed and important. After setting up the general theory, Dirac notes that observables are represented by quantities that no longer commute: 
\begin{quote}
It now becomes
necessary for us to obtain equations to replace the commutative law
of multiplication, equations that will tell us the value of $\xi\eta - \eta\xi$ when
$\xi$ and $\eta$ are any two observables or dynamical variables. Only when
such equations are known shall we have a complete scheme of
mechanics with which to replace classical mechanics. These new
equations are called quantum conditions or comnutation relations.
\end{quote}
Now comes a crucial point. How could one possibly obtain these relations in this completely new theory? In principle, there is no need that this new theory be related to anything we have previously known. However, Dirac observes 
\begin{quote}
...classical mechanics provides a
valid description of dynamical Systems under certain conditions,
when the particles and bodies composing the Systems are sufficiently
massive for the disturbance accompanying an Observation to be
negligible. Classical mechanics must therefore be a limiting case of
quantum mechanics. [...]in particular we may hope to get the quantum conditions appearing as a simple generalization of the classical law that all dynamical
variables commute.
\end{quote}
This is the idea of \emph{quantization}:\footnote{Dirac calls it the method of the ``classical analogy''.}  To extract the quantum description of a dynamical system from its classical one. This suggests the rough relationship between classical and quantum physics sketched in Figure \ref{fig:wishfulthinking} below: 
\begin{figure}[!h]
\[
\begin{tikzcd}
\text{classical physics} \arrow[rr, "\text{quantization}"', dashed, bend right] &  & \text{quantum physics} \arrow[ll, "\text{classical limit}"', dashed, bend right]
\end{tikzcd} 
\]
\caption{A schematic and conjectural diagram on different approaches to QFT.}\label{fig:wishfulthinking}
\end{figure}
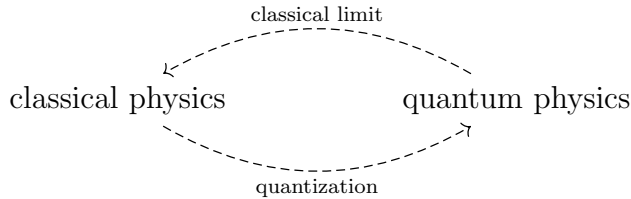
It is clear that this relationship is wishful thinking at best. Classical Physics is only an approximation to the real-world quantum physics, so it is unreasonable to expect we can extract the latter entirely from the former. But, as Matthias Blau \cite{Blau1992} observes, 
\begin{quote}
[u]nfortunately,
 however,
 it is conceptually very diffcult to describe a quantum theory from scratch
 without the help of a reference classical theory. Moreover
 there is enough to the analogy between classical and quantum mechanics to make quantization a worthwile approach. Perhaps
 ultimately
 the study of quantization will tell us enough about quantum theory itself to allow us to do away with the very concept of quantization.
\end{quote}
Let us contemplate in slightly more mathematical terms what a quantization should satisfy. A classical mechanical system with $n$ degrees of freedom can be described in the hamiltonian formalism by the position coordinates $q^i$ and their canonical momenta $p_i$, which together give us coordinates $(q^i,p_i)$ on $
\mathbb{R}^{2n}$. The observables of the dynamical system are given by $C^\infty(\mathbb{R}^{2n})$, and the dynamics are governed by a particular observable $H \in C^{\infty}(\mathbb{R}^{2n})$ called the \emph{hamiltonian}, typically given by the sum of kinetic and potential energies, for instance $H(q,p) = \frac{1}{2m}|p|^2 + V(q)$, where $V \in C^\infty(\R)$ gives the potential energy. \\ 
On the other hand, quantum systems are described by a Hilbert space $\mathcal{H}$ and observables are given by self-adjoint operators on this Hilbert space. A quantization should therefore amount to a map 
\begin{equation}
Q \colon C^{\infty}(\mathbb{R}^{2n}) \to \mathcal{L}(\mathcal{H}), \qquad f \mapsto Q_f
\end{equation}
where we denote by $\mathcal{L}(\mathcal{H})$ the space of linear (but potentially unbounded) operators on $\mathcal{H}$.
What properties, then, should we expect such a map to have? Obivously, we should expect such a map to be linear. Two more properties are natural from both a mathematical and physical viewpoint: Firstly, the map should send the constant function to the identity operator in $\mathcal{H}$: If an observable always evaluates to a certain number classically, we would expect the same from the quantum theory.  Secondly, if we allow for complex-valued functions, then the operator  corresponding to the complex conjugate function should be the adjoint of the operator corresponding to the function, in particular, real functions (which are the observables in the classical system) should be sent to self-adjoint operators. \\
These three properties themselves are not very restrictive. However, Dirac observes that on a classical system we have another piece of information available, namely the Poisson bracket of two functions, defined by 
\begin{equation}
\{f,g\} = \sum_{i \to 1}^n \frac{\partial f}{\partial q^i}\frac{\partial g}{\partial p_i} - \frac{\partial f}{\partial p_i}\frac{\partial g}{\partial q^i}
\end{equation}
which governs the dynamics of the classical system. By asking that the quantum operators have a version of ``quantum Poisson bracket'' that satisfies essentially the same properties as the classical one, Dirac derives the formula 
\begin{equation}
Q_fQ_g-Q_gQ_f := [Q_f,Q_g] = -i\hbar Q_{\{f,g\}}.
\end{equation}
Mathematically, this equation means that the quantization should be a homomorphism of the Lie algebras $(C^\infty(\R^{2n}),-i\hbar\{\cdot,\cdot\})$ and $(\mathcal{L}(\mathcal{H}),[\cdot,\cdot])$.
Finally, one should expect that a family of observables that ``know everything about'' the classical system also ``knows everything about'' the quantum system. One way to formulate this property is to define a \emph{complete} set of classical observables as a set of functions $f_1,\ldots,f_k$ such that every function $g$ which Poisson commutes with all $f_i$ (i.e. all the Poisson brackets $\{g,f_i\}$ vanish) is constant, and similarly a set $A_1,\ldots, A_k$ of quantum observables to be complete if any operator that commutes with all of them is a multiple of the identity. Anticipating that we might not be able to quantize all observables, we arrive at the following definition.
\begin{defn}\label{def: quantization}
A \emph{quantization} of a Lie subalgebra $\mathcal{A} \subset (C^\infty(\R^{2n}),\{\cdot,\cdot\})$ is a Hilbert space $\mathcal{H}$ and a map $Q \colon \mathcal{A} \to \mathcal{L}(\mathcal{H})$ satisfying 
\begin{enumerate}[Q1)]
\item $Q$ is linear,
\item $Q_1 = \mathrm{id}_{\mathcal{H}}$, 
\item $Q_{\bar{f}} = (Q_f)^*$
\item $[Q_f,Q_g] = - i\hbar Q_{\{f,g\}}$ 
\item $f_1,\ldots,f_k$ complete $\Rightarrow Qf_1,\ldots Qf_k$ complete.
\end{enumerate}
\end{defn}
Denote by $P^{\leq n} \subset C^{\infty}(\mathbb{R}^{2n})$ the Lie subalgebra of polynomials of degree less than or equal to $n$. 
As a first example, we can consider the Schr\"odinger representation, defined on $\mathcal{A} = \mathcal{P}^{\leq 1}$ by 
$\mathcal{H} = L^2(\mathbb{R}^n) \ni f(q)$ and 
\begin{align}
q^i &\mapsto \hat{q}^i\colon f(q)\mapsto q^if(q) \\
p_i &\mapsto \hat{p} = -i\hbar\frac{\partial}{\partial q^i}
\end{align}
It is an elementary exercise to check this prescription satisfies $Q1) - Q5)$, and indeed the Stone-von Neumann theorem (see for instance \cite[Section 14]{Hall2013}) tells us any quantization of $P^{\leq 1}$ must be unitarily equivalent to this one. But what about operators of higher order?
A first limitation to how many observables we can expect to consistently quantize Q1) - Q5) is given by the Groewenwald-van Hove theorem.\footnote{The original texts are \cite{Groenewold1946}, \cite{Hove1951,Hove1951a}. See \cite{Gotay1999}, \cite[Section 13.4]{Hall2013} for reviews.}
\begin{thm} 
Suppose $\mathcal{A}$ strictly contains $P^{\leq 4}$. Then there exists no quantization of $\mathcal{A}$. 
\end{thm}
Since this theorem may sound very hard to prove, we sketch here a proof for $n=1$. 

The \emph{Weyl quantization} is defined by 
$(aq + bp )^n \to (a\hat{q} + b\hat{p})^n$. 
\begin{itemize} 
\item Step 1: Any quantization satisfying Q1) - Q5) on $P^{\leq 3}$ must be equal to $Q_{Weyl}$.  For a proof of this fact we refer to \cite[Section 13.4]{Hall2013} (it is not hard but a bit long). 
\item  Step 2: We can write $p^2q^2$ as a Poisson bracket in two ways: 
$p^2q^2 = \frac19{p^3,q^3} = \frac14{p^2q,q^2p}$. 
\item Step 3: The two operators $\frac19[Qp^3,Qq^3]$ and $\frac14[q^2p,p^2q]$ don't agree. 
\end{itemize}
\begin{exc} Prove Step 2 and 3 of this proof.
\end{exc}
This theorem shows that we cannot, in general, expect to quantize all observables in the sense of Definition \ref{def: quantization} above.\footnote{However, it is possible to do so if one relaxes condition 4). Mathematically one then speaks about \emph{deformation quantization} \cite{Bayen1978,Bayen1978a}. For the standard Poisson bracket on $\R^{2n}$, a possible quantization is \emph{Moyal quantization}. With the right techniques, one extend this to general symplectic manifolds \cite{Fedosov1996}, arbitrary Poisson structures on $\R^{2n}$ \cite{Kontsevich2003} and even arbitrary Poisson manifolds\cite{Kontsevich2003} \cite{Cattaneo2001} \cite{Cattaneo2002}.} However, in many situation we may not actually need or want to quantize all possible observables, only a physically relevant subset. The bigger problems of Schr\"odinger quantization are that it is obviously very coordinate-dependent and that it is a priori unclear how to incorporate constraints and symmetries. Those are problems that geometric quantization addresses well, and on a rigorous mathematical footing. 

\subsection{... and Geometry}
Geometric quantization, as the name says, relies heavily on geometry, in particular the mathematical language of \emph{Differential Geometry}. It is a language that is well adapted to questions of coordinate independence and symmetries. Classical mechanics is naturally formulated in the context of \emph{symplectic geometry},\footnote{Or more generally Poisson geometry, but this will not be important for these notes} introduced in more detail in Section \ref{sec: sympgeom}. It is the author's explicit intention that these notes can be read by people without prior exposure to differential geometry, although it is certainly quite helpful.  The fundamental concept in differential geometry is that of a \emph{manifold}. We will get down to the details and definitions in the next lecture, but conceptually, it can be considered either as a geometric object (e.g. the 2-sphere $S^2 = \{(x,y,z)\in\R^3,x^2+y^2 +z^2 = 1\} \subset \R^3\})$ or as a collection of coordinate charts, with rules how to transition from one coordinate chart to the other. The two viewpoints are equivalent, and both are helpful at times. In other approaches to quantization the choice of coordinates sometimes poses a difficult problem, and one advantage of geometric quantization is that it is inherently coordinate independent, because it is formulated in the language of diffferential geometry from the beginning.\footnote{However, in the some sense the problems of the choice of coordinates return in the guise of the choice of a polarization, see Lecture \ref{sec: quantization}. } The differential geometric approach also gives us more mathematical tools to deal with quantization in the presence of symmetries and constraints (we will not deal with these questions systematically in these notes, but in some sense they are touched upon in the final lecture). \\
The main advantage of geometric quantization is also its main disadvantage: It is formulated in a language that, while it is powerful, may seem too complicated or abstract at first sight. This is the main reason that these lecture notes contain a ``quick and dirty'' introduction to differential and symplectic geometry. Another unsatisfying feature is that when we are developing the theory it seems like two new problems pop up for every problem that we solved. However, in the end we will be able to address all of them, and come up with certain classes of (relevant) examples where the formalism works nicely and reproduces known results in a satisfyingly conceptual way. 
\subsection{About these lecture notes} 
Those notes grew out of the ones I had prepared for the XVIII Modave Summer school, to an audience of students of high energy physics. The six lectures in these notes correspond roughly to the six lectures I gave there, however, as I was writing this text, it grew considerably beyond what I discussed there, and now contains a lot more material. However, I wanted to avoid what I dislike about many geometric quantization texts, namely presenting some of the many problems in geometric quantization without solution, or even worse, not mentioning them at all. At the same time, I tried to build up the text in a pedagogical way, discuss some recent results in the final lecture, but still keep it to a reasonable length. I leave the judgement whether I have achieved those goals to the reader. There are plenty of exercises scattered throughout the text, in an attempt to encourage the reader to work their way through it rather than simply consume it. 
\subsubsection{What's in these notes}
Any text about geometric quantization is in particular a text about quantization, and so this text starts with a discussion of what quantization is and what its problems are.  \\
The next Lecture (Lecture \ref{sec: sympgeom}) is a crash course of differentiable and symplectic geometry, in an order to make this text as self-contained as possible, in particular with an audience with a background in physics in mind. In Lecture \ref{sec: prequantization} I discuss the prequantization of a symplectic manifold - prequantum line bundles, their associated Hilbert spaces and the prequantization map. Since it is relatively simple, and one of the few general results available in Geometric Quantization, I include a classification of prequantum line bundles. In Lecture \ref{sec: quantization} I discuss quantization, i.e. the process of choosing a polarization and selecting only polarized sections - this is where the difficulties in geometric quantization start: there are various reasons why our naively defined Hilbert space could be empty: if the polarization has non-compact real directions, covariantly constant sections are not square integrable, while if the polarization has compact non-simply connected directions, we encounter the problem that there may be no smooth covariant sections at all. Also, the quantization map has to preserve polarized sections. We discuss these problems and their (sometimes partial) solutions. Lecture \ref{sec: examples} presents various examples which exhibit the different problems encountered in Section \ref{sec: quantization}: $\R^{2n}$, cotangent bundles (in particular the cylinder, where the horizontal polariztaion has circle leaves) and the 2-sphere (a model for the quantization of angular momentum). In the final section, with the goal of connecting to current research, I discuss geometric quantization in the context of Chern-Simons theory. In particular I mention the role of symmetries, and show how to construct a state from a Feynman diagram computation. 
\subsubsection{What's not in these notes}
The mathematical origins of geometric quantization, around 1970, lie in representation theory (as put down by Kostant and Souriau \cite{Kostant1970},\cite{Souriau1970}, see also the later account by Kirillov \cite{Kirillov2001}). I barely touch upon those aspects of geometric quantization here. While I tried to provide a complete account of ``classic'' geometric quantization, I fell short of including the metaplectic correction in full detail. Also, I don't mention many of the newer developments in geometric quantization,  such as geometric quantization of presymplectic \cite{Guenther1980, Vaisman1983, Silva2000} and Poisson manifolds \cite{Huebschmann1990, Vaisman1991}, relations to the Poisson Sigma Model \cite{Bonechi2006} and A-model\cite{Gukov2009}, ``higher'' geometric quantization (e.g. through symplectic groupoids, \cite{Hawkins2008} or ``shifted'' geometric quantization \cite{Safronov2020}. What is also absent is a more thorough discussio of geometric quantization in field theory, to some extent this is discussed in \cite{Woodhouse1997}. 
\subsection{Acknowledgements}
I would like to thank the organizers of the XVII Modave Summer School in Mathematical Physics for organizing a wonderful school and all the attending students for the profound interest and interesting conversations. While writing these notes I was supported by the ERC Grant ``ReNewQuantum''. 
\section{Lecture 2: Symplectic Geometry}\label{sec: sympgeom}
In this Lecture we present some elements of the theory of differentiable and symplectic manifolds. Of course, this text should not be considered a standalone introduction to either of these topics there are plenty of excellent textbooks and lecture series on the topic. Hopefully, this text can be something like a ``working introduction'' to those subjects, or, to put it simply, ``learning by doing''. Some good introductions to differential geometry are \cite{Nicolaescu2020},\cite{Cattaneo2018}, an excellent introduction to symplectic geometry is \cite{Silva2008}, but of course there are countless others. 
\subsection{Symplectic geometry and classical mechanics}
Consider again classical mechanics in the Hamiltonian formulation, i.e. we have coordinates $(q^i,p_i) \in \R^{2n}$ and a hamiltonian $H$, for instance $H(q,p) = \frac{1}{2m}|p|^2 + V(q)$, where $V \in  C^\infty(\R^n)$ is a smooth function of the positions $q^i$. Then time evolution of observables $f \in C^\infty(\R^n)$ is governed by the Poisson bracket with the hamiltonian, 
\begin{equation}
\frac{df}{dt} = \{f,H\} = \sum_{i = 1}^n \frac{\partial f}{\partial q^i}\frac{\partial H}{\partial p_i } - \frac{\partial f}{\partial p_i}\frac{\partial H}{q^i}
\end{equation}
But what happens if the problem we are trying to describe does not admit global coordinates, or we want to use different coordinates for some reason? 

\subsection{Differentiable manifolds}
Let $M$ be the set of all possible configurations and momenta, i.e. phase space. We want to use the Hamiltonian formulation of classical mechanics on $M$ without assuming that $M = \R^{2n}$. This is done through symplectic geometry, the basis of which are differentiable manifolds, which we now review. \\
\subsubsection{Manifolds}
We say that a topological space\footnote{A topological space is a set $M$ together with a notion of what subsets $U\subset M$ are open. In all examples in this text, $M \subset \R^N$ for some $N$ and $U\subset M$ is open if and only if $U = M \cap V$, where $V$ is an open ball in $\R^N$.} $M$ is a manifold of dimension $d$ if we can cover $M$ with open sets $U_{\alpha}$ and there are homeomorphism (continuous bijective maps with continuous inverse) $\phi_\alpha\colon U_\alpha \to \phi_\alpha(U_\alpha)\subset \R^d$ such that for all $\alpha,\beta$ with $U_\alpha \cap U_\beta = U_{\alpha\beta} \neq \emptyset$ the map 
\begin{equation}
\phi_{\alpha\beta} = \phi_\beta \circ \phi_\alpha^{-1} \colon \phi_\alpha(U_{\alpha\beta}) \to \phi_{\beta}(U_{\alpha\beta}),
\end{equation}
called the transition function, is a diffeomorphism (i.e. a smooth bijective map with a smooth inverse). Notice that $\phi_{\alpha\beta}$ is a map between open sets of $\R^d$, thus it makes sense to speak about differentiability.   

\begin{figure}[h!]
\centering
\includegraphics[scale=0.5]{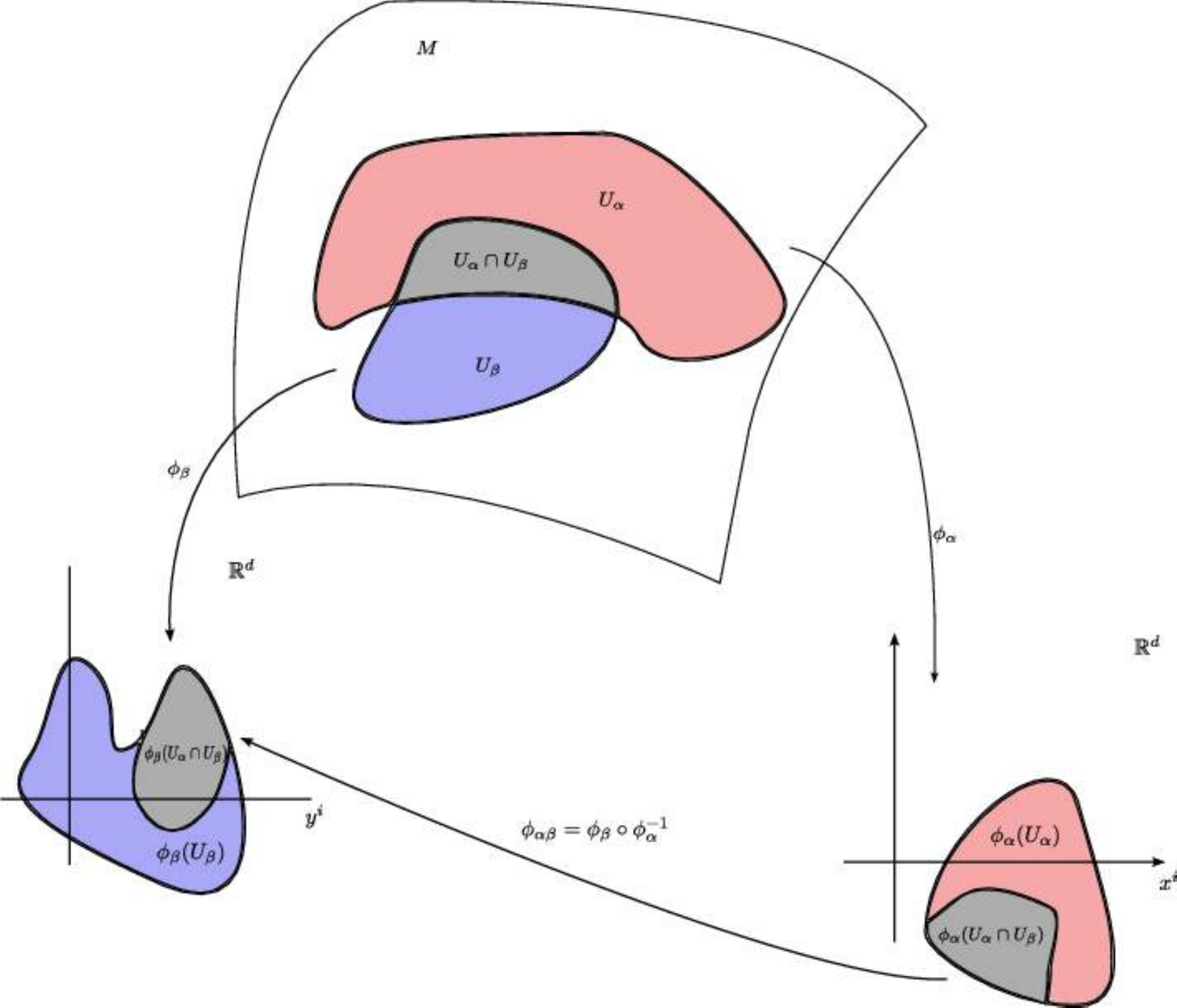} 
\caption{Transition functions on overlaps of charts.}
\end{figure} 

The pairs $(U_\alpha, \phi_\alpha)$ are called \emph{charts} on $M$. Writing $\phi_\alpha(p) = (x^1(p),\ldots,x^d(p))$ we obtain \emph{local coordinates} $(x^1,\ldots,x^n)$ on $U_\alpha$. If $(U_\beta,\phi_\beta = (y^1,\ldots,y^d))$ is another chart with $U_\alpha \cap U_\beta \neq\emptyset$, we have the \emph{coordinate change} $$\phi_{\alpha\beta}(x) = y(x) = (y^1(x^1,\ldots,x^d), \ldots, y^d(x^1,\ldots,x^d))$$ and its differential $d\phi_{\alpha\beta} = \left(\frac{\partial y^i}{\partial x^j}\right)_{i,j=1}^d$. 
\begin{expl}[Open subsets of $\R^n$] 
Any open subset $U \subset \R^n$ is a manifold: It is covered by the open set $U$ with the chart $\phi = \mathrm{id}$.
\end{expl}
\begin{expl}[The circle]
As a nontrivial example, one can consider the circle $S^1 = \{(x,y)\in \R^2\colon x^2 + y^2 = 1\}$. There are many ways to put coordinates on the circle, the easiest is probably by using an angular coordinate $\theta$, with $(x,y) =(\cos\theta,\sin\theta)$. In principle, it is possible to associate to every point in the circle an angle, for instance in the half-open interval $[0,2\pi)$. \

\begin{figure}[h!]
\centering
\def\svgwidth{\columnwidth}
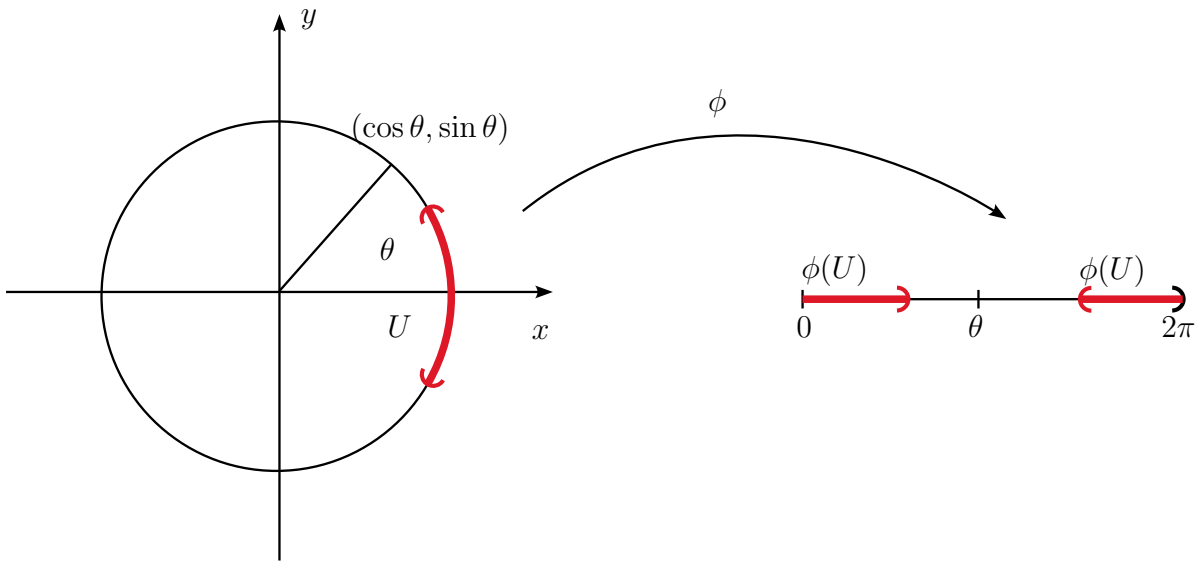
\caption{Parametrizing the circle by an angle is not continuous on the whole circle: the image of the connected set $U$ (in red) is disconnected in the interval.}
\end{figure}

 However, this does not represent the geometry of the circle correctly: On the circle we can go in both directions at the point $(1,0)$, but in $[0,2\pi)$ we can only go in one direction at $0$. Mathematically, the map $\phi \colon S^1 \to [0,2\pi)$ is not continuous at the point $(1,0)$. However, we can restrict the map $\phi$ to a map $\phi_1$ on the subset $U_1=S^1 \setminus \{(0,1)\}$, and it is a homeomorphism there. However, now we need another chart to cover every point in the circle: For instance, on $U_2 = S^1 \setminus \{(0,-1)\}$, we can define the continuous angle $\phi_2\colon U_2 \to (-\pi,\pi)$. 
 
 \begin{figure}[h!]
\centering
\def\svgwidth{\columnwidth}
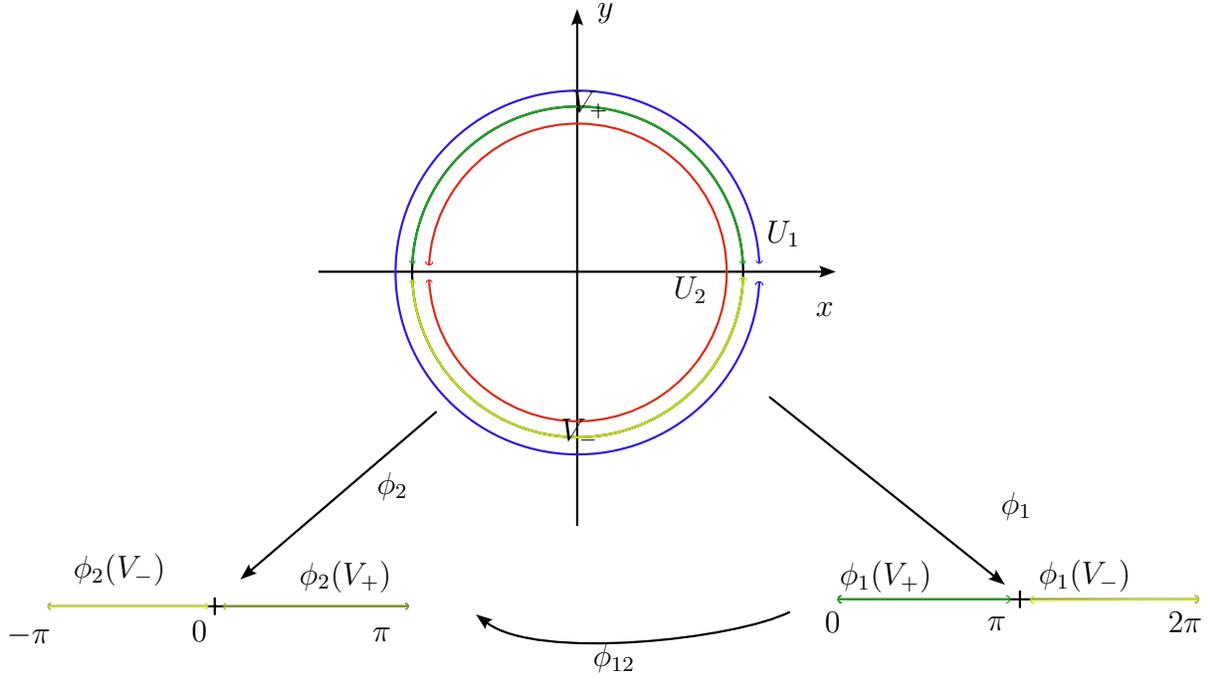
\caption{The circle can be covered by two charts $U_1$ (blue) and $U_2$ (red) whose intersection is the disjoint union of $V_+$ (green) and $V_-$ (yellow). }\label{fig: charts circle}
\end{figure} 

 Obviously, $U_1 \cup U_2 = S^1$, and $U_1 \cap U_2 = V_+ \sqcup V_-$ is a disjoint union of two connected components, $V_+$ above the $x$-axis and $V_-$ below the $x$-axis (See Figure \ref{fig: charts circle}). Then we have $\phi_1(V_+) = (0,\pi)$, $\phi_1(V_-) = (\pi,2\pi)$ and 
\begin{equation}
\phi_{12}(\theta) = \begin{cases} \theta & \theta \in (0,\pi) \\ \theta - 2\pi & \theta \in (\pi,2\pi) \end{cases}
\end{equation} 
\end{expl}

Next, we consider another atlas on the circle that generalizes easily to higher dimensions. 
\begin{exc}[The stereographic projection]\label{exc: stereo}
Let $S^n \subset \R^{n+1}$ be the unit sphere, i.e $S^n = \{x \in \R^{n+1} : \|x\| = 1\}$.
\begin{enumerate}

\item Define 
\begin{align*}
P_N \colon S^1-\{(0,1)\} &\to \R \\
(x,y) &\mapsto \frac{x}{1-y} 
\end{align*} 

and similarly 
\begin{align*}
P_S \colon S^1-\{(0,-1)\} &\to \R \\
(x,y) &\mapsto \frac{x}{1+y}. 
\end{align*} 
Show that $\{P_N,P_S\}$ is an atlas for $S^1$. 
\item Show that the intersection of the straight line through the north pole $N = (0,1)$ and the point $(x,y) \in S^1$ is given by $(P_N(x,y),0)$. This is the geometric prescription of the stereographic projection (see Figure \ref{fig: stereo}).

 \begin{figure}[h!]
\centering
\def\svgwidth{\columnwidth}
\begingroup%
  \makeatletter%
  \providecommand\color[2][]{%
    \errmessage{(Inkscape) Color is used for the text in Inkscape, but the package 'color.sty' is not loaded}%
    \renewcommand\color[2][]{}%
  }%
  \providecommand\transparent[1]{%
    \errmessage{(Inkscape) Transparency is used (non-zero) for the text in Inkscape, but the package 'transparent.sty' is not loaded}%
    \renewcommand\transparent[1]{}%
  }%
  \providecommand\rotatebox[2]{#2}%
  \newcommand*\fsize{\dimexpr\f@size pt\relax}%
  \newcommand*\lineheight[1]{\fontsize{\fsize}{#1\fsize}\selectfont}%
  \ifx\svgwidth\undefined%
    \setlength{\unitlength}{289.62679405bp}%
    \ifx\svgscale\undefined%
      \relax%
    \else%
      \setlength{\unitlength}{\unitlength * \real{\svgscale}}%
    \fi%
  \else%
    \setlength{\unitlength}{\svgwidth}%
  \fi%
  \global\let\svgwidth\undefined%
  \global\let\svgscale\undefined%
  \makeatother%
  \begin{picture}(1,0.62737127)%
    \lineheight{1}%
    \setlength\tabcolsep{0pt}%
    \put(0,0){\includegraphics[width=\unitlength,page=1]{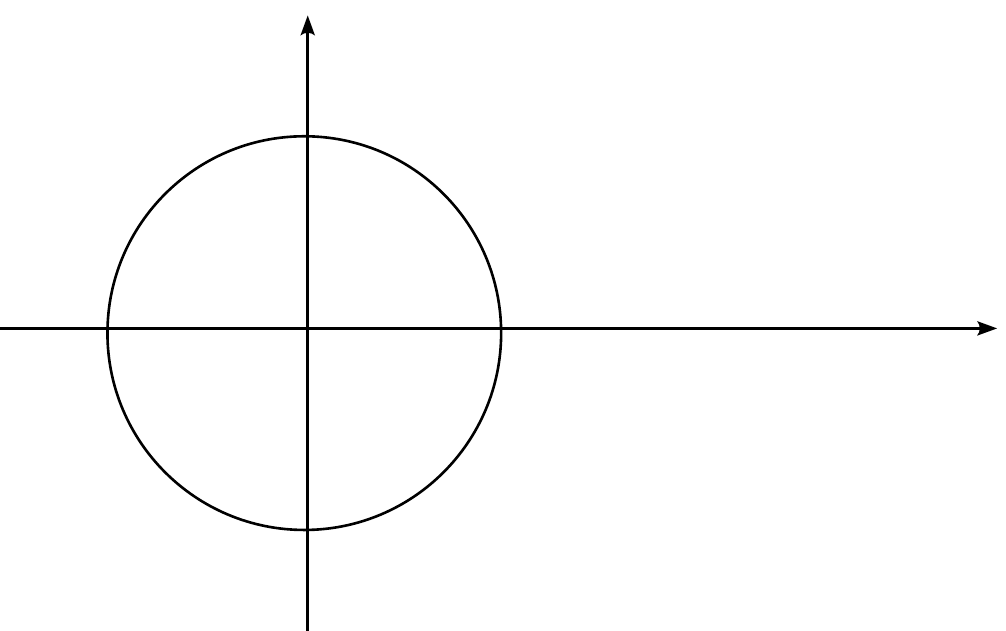}}%
    \put(0.94335178,0.26464222){\color[rgb]{0,0,0}\makebox(0,0)[lt]{\lineheight{1.25}\smash{\begin{tabular}[t]{l}$x$\end{tabular}}}}%
    \put(0.32982449,0.60246515){\color[rgb]{0,0,0}\makebox(0,0)[lt]{\lineheight{1.25}\smash{\begin{tabular}[t]{l}$y$\end{tabular}}}}%
    \put(0,0){\includegraphics[width=\unitlength,page=2]{stereo.pdf}}%
    \put(0.31493306,0.50375618){\color[rgb]{0,0,0}\makebox(0,0)[lt]{\lineheight{1.25}\smash{\begin{tabular}[t]{l}$N$\end{tabular}}}}%
    \put(0.45023234,0.4390848){\color[rgb]{0,0,0}\makebox(0,0)[lt]{\lineheight{1.25}\smash{\begin{tabular}[t]{l}$p$\end{tabular}}}}%
    \put(0.74891216,0.31059291){\color[rgb]{0,0,0}\makebox(0,0)[lt]{\lineheight{1.25}\smash{\begin{tabular}[t]{l}$P_N(p)$\end{tabular}}}}%
  \end{picture}%
\endgroup%

\caption{Stereographic projection on the circle}\label{fig: stereo}
\end{figure} 

\item Define 
\begin{align*}
P_N \colon S^2-\{(0,0,1)\} &\to \R^2 \\
(x,y,z) &\mapsto \frac{1}{1-z}(x,y) 
\end{align*} 

and similarly 
\begin{align*}
P_S \colon S^1-\{(0,-1)\} &\to \R^2  \\
(x,y) &\mapsto \frac{1}{1+z}(x,y). 
\end{align*} 
Show that $\{P_N,P_S\}$ is an atlas for $S^2$. What is the analog of the geometric description of those maps?
\item Find an analogous atlas, together with its geometric description for the $n$-sphere $S^n$. 
\end{enumerate}
\end{exc}

\subsubsection{Functions and the definition of objects on manifolds} 
\begin{figure}[h!]
\centering
\def\svgwidth{\columnwidth}
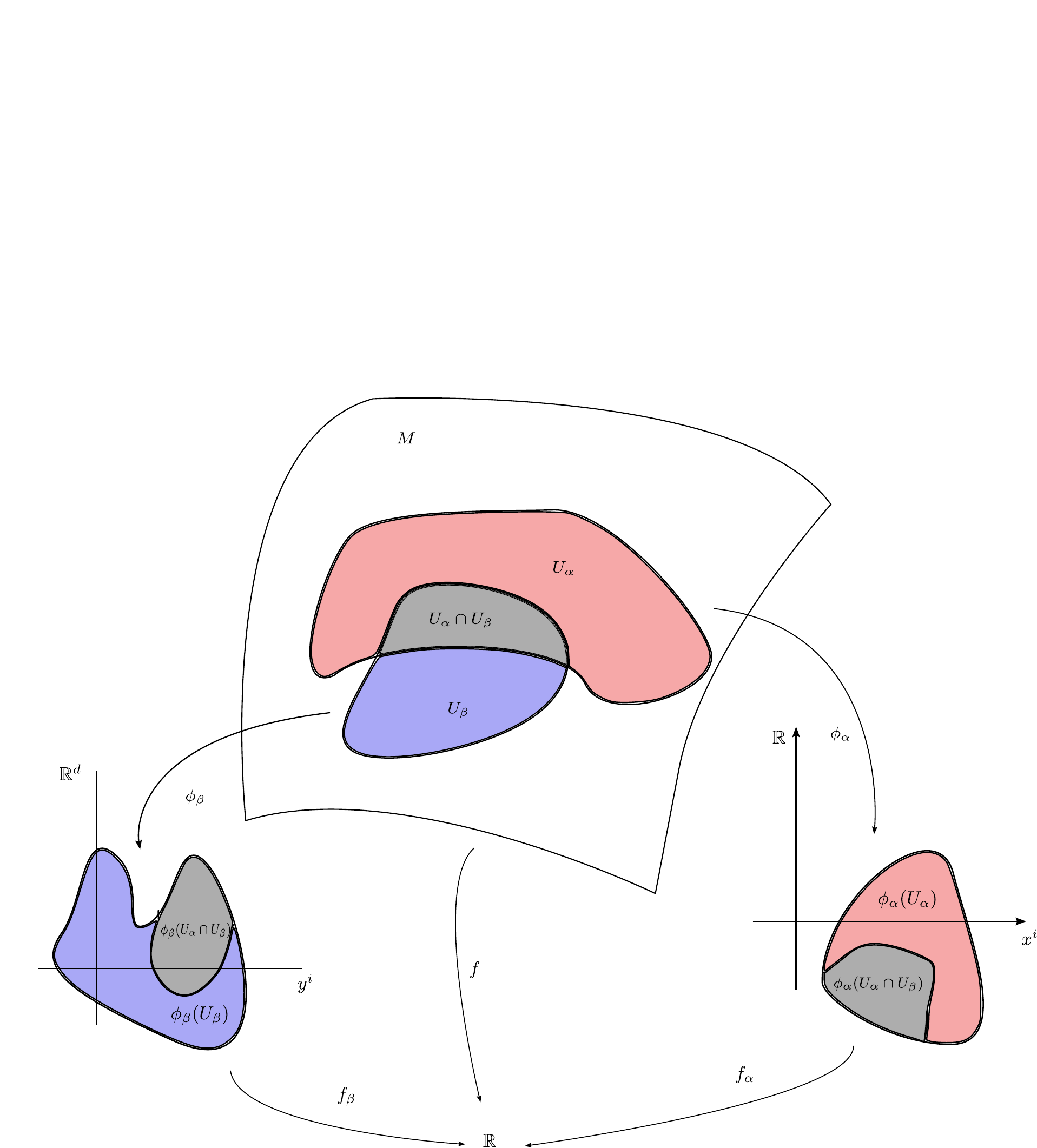
\caption{Definition of a smooth function on $M$.}\label{fig: fct}
\end{figure}

A function $f\colon M \to \R$ is called smooth if, for every $x \in M$, there is a chart $(U_\alpha, \phi_\alpha)$ with $x \in U_\alpha$ and $f_\alpha := f \circ \phi_\alpha^{-1}\colon \phi_\alpha(U_\alpha) \colon \R$ is a smooth function on the open set $\phi_\alpha(U_\alpha)$. Notice that if $U_\alpha \cap U_\beta \neq \emptyset$, we have the relation \begin{equation}
f_\beta(\phi_{\alpha\beta}(x)) = f_\alpha(x).\label{eq:fct transform}
\end{equation} 
(see Figure \ref{fig: fct}).   This suggests that one method to define an object $O$ on a manifold is to define it by specifying a family of objects $O_\alpha(x)$ on $\phi_\alpha(U_\alpha)$ and their transformation rule under coordinate change, i.e. the relationship between $O_\beta(y=\phi_{\alpha\beta}(x)) $ and $O_\alpha(x)$. For instance, if $f\colon M \to \R$ is a real-valued function, the transformation rule is \eqref{eq:fct transform}. 

 I.e., to give a function $f \colon M \to \R$ is equivalent to giving a collection of functions $f_\alpha\colon \phi_\alpha(U_\alpha) \to \R$ satisfying \eqref{eq:fct transform}. All objects on manifolds have two descriptions: Local ones (in coordinates or systems of ``trivializing neighbourhoods'', see line bundles and vector bundles that are introduced later), and global ones. It is usually handy to understand both of them and the relations between them, but in this text we will often work with the local description, since it could be considered slightly easier.  
\subsubsection{Tensors} 
Among the most important objects on a manifold are \emph{tensor fields}, usually just called tensor. A function is a tensor field of rank 0. A rank 1 covariant tensor $\omega$ is a collection of functions $$\omega_{\alpha,i} \colon \phi(U_\alpha) \to \R, \qquad i=1,\ldots,d $$ satisfying the transformation rule 
\begin{equation}
\omega_{\beta,j}(y(x)) = \left(\frac{\partial x^i}{\partial y^j}\right) \omega_{\alpha,i}(x).\label{eq: cov transform}\end{equation}
Here we use the Einstein summation convention of summing over a repeated index. 
Similarly, a rank 1 contravariant tensor is a collection of functions $v_{\alpha}^i(x) \colon \phi_\alpha(U_\alpha) \to \R$ such that 
$$ v_{\beta,j}(y(x)) = \left(\frac{\partial y^j}{\partial x^i}\right)v_{\alpha}^i(x).$$ 
In general, we can have a rank $(r,s)$ tensor $T$ on $M$, which is a collection of functions  
\begin{equation}
T_{\alpha, j_1\ldots j_s}^{i_1\ldots i_r} \colon \phi_\alpha(U_\alpha) \to \R, \qquad i_1,\ldots,i_r,j_1,\ldots,j_s = 1,\ldots d
\end{equation}
(i.e. a collection of $d^{r+s}$ functions in every coordinate chart!) 
subject to the glorious transformation rule 
\begin{equation} 
T_{\beta,i_1'\ldots i_s'}^{j_1'\ldots j_r'}(y(x)) = \left(\frac{\partial y^{j_1'}}{\partial x^{j_1}}\right)\cdots \left(\frac{\partial y^{j_r'}}{\partial x^{j_r}}\right)\left(\frac{\partial x^{j_1}}{\partial y^{j_1'}}\right) \cdots \left(\frac{\partial x^{i_r}}{\partial y^{i_r'}}\right)T_{\alpha, j_1\ldots j_s}^{i_1\ldots i_r}(x). \end{equation} 
We say that a rank $(r,s)$ tensor has $r$ contravariant and $s$ covariant indices.  There are two important operations on tensors: We can take the \emph{tensor product} $T \otimes T'$ of a rank $(r,s)$ tensor $T$ and a rank $(r',s')$ tensor $T'$ by simply multiplying the corresponding coefficients, 
\begin{equation}
(T \otimes T')_{\alpha,j_1 \ldots j_{s + s'}}^{i_1 \ldots i_{r+r'}} = T_{\alpha,j_1 \ldots j_s}^{i_1 \ldots i_r} {T'}_{\alpha, j_{s+1}\ldots j_{s+s'}}^{i_{r+1}\ldots i_{r+r'}}. 
\end{equation}
From a rank $(r,s)$ tensor $T$, we can obtain a rank $(r - k, s - k)$ tensor by ``contracting'' $k$ of the indices. This is a generalization of the concept of trace: A rank $(1,1)$ tensor $T$ is given by a collection of matrices $T_{\alpha,j}^i$ and contracting this single index is the same thing as taking the trace in every coordinate chart, 
\begin{equation}
(\tr T)_\alpha = T_{\alpha,i}^i
\end{equation}
Similarly, we can contract $k$ indices, say at positions $(l_1,\ldots, l_k)$ in a rank $(r,s)$ tensor. We denote this operation as follows 
\begin{equation}
(\tr^{l_1\ldots l_k}T)_{\alpha, j_1 \ldots j_{s-k}}^{i_1 \ldots t_{r-k}} = T_{\alpha, j_1 \ldots n_1 \ldots n_k \ldots j_{s-k}}^{i_1 \ldots n_1 \ldots n_k \ldots j_{r-k}}
\end{equation}
\subsubsection{Differential forms} A special type of tensors are the \emph{differential forms}: a differential $p$-form $\omega$ is a completely antisymmetric $(0,p)$-tensor, i.e. it is given by a collection of functions  $\omega_{\alpha, i_1\ldots i_p}$ with the property that for all pairs $(i_k,i_j)$ we have 
\begin{equation}
\omega_{\alpha,i_1\ldots i_k \ldots i_j \ldots i_p} = - \omega_{\alpha,i_1\ldots i_j \ldots i_k \ldots i_p} \label{eq: antisymmetric}
\end{equation}
i.e. whenever we exchange a pair of indices, we obtain a minus sign. The space of differential $p$-forms on $M$ is denoted $\Omega^p(M)$. 
\begin{exc}
Convince yourself there are no differential $p$-forms for $p >d$. 
\end{exc} In particular, for $p=0,1$ condition \eqref{eq: antisymmetric} is void: A 0-form is therefore just the same as a function of $M$, and a 1-form is just the same as a covariant rank 1 tensor. In particular, to any 0-form $f$ we can associate the 1-form $df$ given by $(df)_{\alpha,i} = \frac{\partial f_\alpha}{\partial x^i}$. 
\begin{exc}
Check that $df$ really defines a covariant rank 1 tensor, i.e. that the transformation property \eqref{eq: cov transform} is satisfied.
\end{exc}
Continuing in this way, we can try associate to $f$ the object tensor $\tilde{d}^2f$ given by $ (\tilde{d}^2f)_{\alpha,ij} = \frac{\partial }{\partial x^i}\frac{\partial f_\alpha}{\partial x^j} $. However, this is not a tensor - it does not transform in the right way. However, if we antisymmetrize it, we get 0, namely 
$$d^2f_{\alpha,ij} = \tilde{d}^2f_{\alpha,[ij]} =  \frac{1}{2}\frac{\partial }{\partial x^i}\frac{\partial f_\alpha}{\partial x^j} -  \frac{\partial }{\partial x^j}\frac{\partial f_\alpha}{\partial x^i} = 0 $$
where the square brackets around a set of indices denotes that we are antisymmetrizing those indices, i.e. summing over all possible permutations of the indices, multiplying the contributions with the sign of the permutation and diving by the number of total permutations. 
Similarly, to a differential $p$-form $\omega$ we can associate a $p+1$-form $d\omega$ by setting 
\begin{multline}
(d\omega)_{\alpha,i_0 i_1 \ldots i_p} = (p+1)\partial_{[i_0}\omega_{\alpha,i_1\ldots i_p]} \\ 
= \frac{p+1}{(p+1)!}\sum_{\sigma \in S_{p+1}}\partial_{i_\sigma(0)}\omega_{\alpha,i_{\sigma(1)}\ldots i_{\sigma(p)}} = \sum_{k=0}^p (-1)^k \partial_{i_k}\omega_{i_0 \ldots \hat{i}_k \ldots i_p}\label{eq: de rham diff}
\end{multline} 
where $\partial_i = \frac{\partial }{\partial x^i}$ and we are antisymmetrizing over all the $i_k$ indices (not the $\alpha$ index). \footnote{The factor of $(p+1)$ in the first equality is to obtain a sum without a factor in the final expression in the right hand side}.
As in the $p=0$ case we have that $d$ defines a linear operator $d\colon \Omega^p(M) \to \Omega^{p+1}(M)$ and that $d^2 = 0$ (it is  a good exercise to check this!). Differential $p$-forms $\omega$ satisfying $d\omega$ are called \emph{closed}, while $p$- forms $\omega$ for which there exists $\alpha$ such that $\omega = d\alpha$ are called \emph{exact}. In particular, exact forms are always closed (since $d^2 =0$), but the converse is not true! See Exercise \ref{exc: closed and exact} below. 

Another important operation on differential forms is the \emph{wedge or exterior product}: If $\omega$ is a $k$-form, and $\tau$ is an $l$-form, then $\omega \wedge \tau$ is the $k+l$-form given in coordinate charts by 
\begin{equation}
(\omega \wedge \tau)_{\alpha,i_1\ldots i_{k+l}} = \omega_{\alpha,[i_1\ldots i_k}\tau_{\alpha,i_{k+1}\ldots i_{k+l}]}.
\end{equation}
A particular instance of contraction of tensors is important for differential forms as well. Namely, if $v$ is a rank 1 contravariant tensor and $\omega$ a $p$-form, the following special contraction resulting in a $(p-1)$-form is sometimes called \emph{interior product} and denoted $\iota_v\omega$: 
\begin{equation}
(\iota_v\omega) = \tr^1 (v\otimes \omega), \qquad (\iota_v\omega)_{\alpha, i_1 \ldots i_{p-1}}= v^i\omega_{ii_1\ldots i_{p-1}}.
\end{equation}
The following is a simple exercise with antisymmetrization: 
\begin{exc}
Let $\omega$ be a $k$-form and $\tau$ be an $l$-form. Show that the wedge product satisfies 
\begin{align*}
\omega \wedge \tau &= (-1)^{kl}\tau\wedge \omega \\ 
\iota_v(\omega \wedge \tau ) &= \iota_v\omega \wedge \tau + (-1)^k \omega \wedge \iota_v\tau \\
d(\omega \wedge \tau) &= (d\omega) \wedge \tau + (-1)^k \omega \wedge (d\tau)
\end{align*}
\end{exc}
Finally, suppose we have a map between two manifolds, $f\colon M \to N$. Then, for any differential $p$-form $\omega$ on $N$, we can define the \emph{pullback} $f^*\omega$. In coordinate charts $(U_\alpha,(x^1,\ldots,x^m))$ on $M$ and $(V_\beta,(y^1, \ldots, y^n))$ on $N$, we have 
\begin{equation}
(f^*\omega)_{\alpha,i_1\ldots i_p} = \omega_{j_1\ldots j_p}\frac{\partial f^{j^1}}{\partial x^{i_1}}\cdots\frac{\partial f^{j^p}}{\partial x^{i_p}}.
\end{equation}
The following exercise summarizes the properties of the pullback: 
\begin{exc}
Let $f\colon M \to N$ be a smooth map and $\omega, \tau$ differential forms on $N$.
\begin{enumerate}
\item Prove that we have 
 $f^*(\omega \wedge \tau) = f^*\omega \wedge f^*\tau$. 
\item Prove that $df^*\omega = f^*d\omega$. 
\end{enumerate}
\end{exc}
\subsubsection{Tangent and cotangent bundles} 
\begin{figure}[h!]
\centering
\def\svgwidth{7cm}
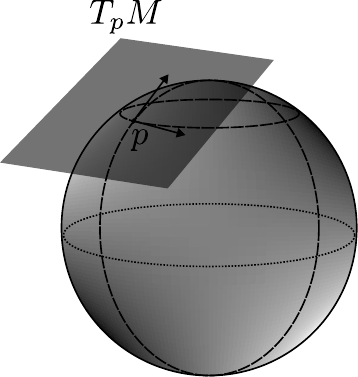
\caption{The tangent space of the 2-sphere at some point $p\in S^2$ is spanned by tangent vectors to coordinate curves.}\label{fig: 2sphere}
\end{figure}

To every manifold $M$ one can associate two other manifolds two other manifolds that will be important for us, the \emph{tangent} and \emph{cotangent} bundles. Since $M$ is not necessarily a subset of some bigger ambient space, we need some extra idea to define tangent vectors. Namely, we simply define them as directional derivatives along a curve: If $\gamma \colon (-\epsilon, \epsilon) \to M$ is a curve in $M$, we define $\dot{\gamma}(0)$ as the map $\dot{\gamma}(0)\colon C^\infty(M) \to \R$  by 
\begin{equation}
\dot{\gamma}(0)f = \frac{d}{dt}\bigg|_{t=0} f(\gamma(t))
\end{equation}
 If $p \in M$, then the \emph{tangent space} to $M$ at $p$ is the set of all tangent vectors to curves through $p$ (see also Figure \ref{fig: 2sphere}): 
\begin{equation}
T_pM = \{\dot{\gamma}(0), \gamma \colon I \to M, \gamma(0) = p\}.
\end{equation}

\begin{figure}[h!]
\centering
\def\svgwidth{\columnwidth}
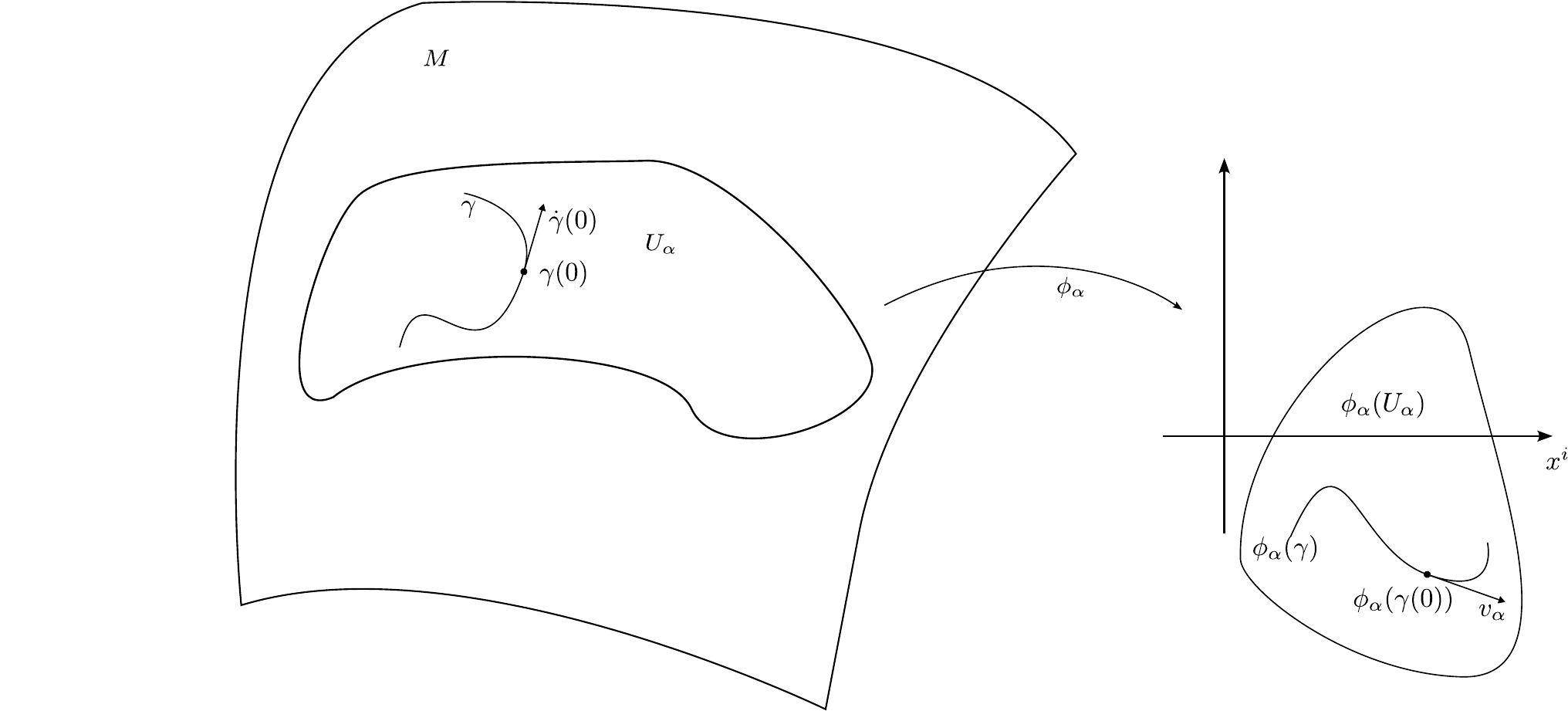
\caption{A tangent vector in a chart $M$.}\label{fig: tangent vector}
\end{figure}

The \emph{tangent bundle} of $M$ is the set $TM = \bigsqcup_{p \in M} T_pM$, i.e. the set of all tangent vectors. 
To get coordinate representations of a tangent vector $v = \dot{\gamma}(0)$, we look at a curve $\gamma$ in a chart: $f(\gamma(t)) = f_\alpha(\phi_\alpha(\gamma(t)))$. Setting $\gamma_\alpha = \phi_\alpha(\gamma)$, we obtain a curve in $\R^n$ which has a usual tangent vector $v_\alpha = (v^1_\alpha,\ldots,v^d_\alpha) = (\dot{\gamma}_\alpha^1(0),\ldots,\dot{\gamma}_\alpha^d(0))$. See Figure \ref{fig: tangent vector}.  Taking derivative at 0 we get $\dot{\gamma}(0)f = v^i_\alpha\frac{\partial f_\alpha}{\partial x^i}$. 
We therefore introduce the notation 
\begin{equation}
\dot{\gamma}(0) = v^i_\alpha\frac{\partial}{\partial x^i} \label{eq:}
\end{equation}
 We can think of $\frac{\partial}{\partial x^i}$ is the tangent vector at $p$ corresponding to the curve $\phi_\alpha^{-1}(x^1 + t,x^2,\ldots,x^d)$. In particular, $T_pM$ is a vector space and for any coordinate system $x^i$ the vectors $\frac{\partial}{\partial x^i}$ span $T_pM$ (see Figure \ref{fig: 2sphere}), so $\dim T_pM = \dim M$. Moreover in this way we obtain a chart $\hat{\phi}_\alpha$ on $TM$ by mapping $(p,v) \mapsto (\phi_\alpha(p),v^1_\alpha,\ldots v^d_\alpha)$, and we can check $v^i_\beta = \frac{\partial y^i}{\partial x^j}v^j_\alpha$ - i.e. the tangent bundle is a manifold and the transition functions are given by $\hat{\phi}_{\alpha\beta} = (\phi_{\alpha\beta},d\phi_{\alpha\beta})$. The tangent bundle has a natural map $\pi\colon TM \to M, (p,v) \mapsto p$, and a  map $v\colon M \to TM$ such that $\pi \circ v (p)  = p$ is called a \emph{vector field} on $M$. 
 
\begin{expl}
If $M \subset \mathbb{R}^n$, then the notion of tangent vector coincides with the usual notion of tangent vector of a curve in $\R^n$. For instance, if $\gamma \colon (-\varepsilon,\varepsilon) \to S^n$ is a curve in $S^n$ then $\gamma(t) = (x^1(t), \ldots, x^n(t))$ with $\sum (x^i(t))^2 \equiv 1$. Differentiating this equation at 0 we obtain 
$$\sum x^i(0)\dot{x}^i(0) = 0 = \gamma(0)\cdot \dot{\gamma}(0),$$
i.e. $T_pS^n \subset p^\perp$ consists of vectors orthogonal to $p$. On the other hand, $\dim p^\perp = n-1 = \dim S^n = \dim T_pS^n$ and therefore $T_pS^n = p^\perp$. See Figure \ref{fig: 2sphere}.
\end{expl} 

The cotangent bundle can be defined in a similar way. The vector space dual to the tangent space $T_pM$ is called the cotangent space and denoted $T^*_pM$. The cotangent bundle is then defined by $T^*M = \bigsqcup_{p \in M} T^*_pM$. We can put charts on $T^*M$ by taking a chart $\phi_\alpha$ and mapping $(p, \lambda) \mapsto (\phi_\alpha(p),\lambda\left(\frac{\partial}{\partial x^1}\right), \ldots, \lambda\left(\frac{\partial}{\partial x^d}\right)$.  The transition function are then 
$\hat{\phi}_{\alpha\beta} = (\phi_{\alpha\beta},((d\phi_{\alpha\beta})^*)^{-1})$. In terms of coordinates $(x^1,\ldots,x^d,p_1,\ldots, p_d)$ and $(y^1,\ldots,y^d,p_1',\ldots,p_d')$ we have $p_j = \frac{\partial y^i}{\partial x^j}p_i$. A 
\emph{covector field} is a map $\omega \colon M \to T^*M$ such that $\pi \circ \omega(p) = p$ for all $p \in M$.
\begin{exc}
Verify that a vector field is the same thing as a rank 1 contravariant tensor via the identification $v^i_\alpha \leftrightarrow v^i_\alpha\frac{\partial}{\partial x^i}$. Similarly, a rank 1 covariant tensor is the same as a covector field via $\omega_i \leftrightarrow \omega_idx^i$.
\end{exc} 
We can use vector fields to differentiate objects. For instance, if $f$ is a function, and $X$ is a vector field then we have the derivative introduce above: 
\begin{equation}
(Xf)_\alpha = X^i_\alpha \frac{\partial f_\alpha}{\partial x^i}.
\end{equation}
It is easy to check that $Xf$ is again a function. If we try to apply this simple rule to a general tensor, the result is not again a tensor. For instance, if $Y$ is another vector field, then 
\begin{equation}
(XY)^j_\alpha = X^i_\alpha \frac{\partial Y^j_\alpha}{x^i}
\end{equation}
is not a vector field: trying to transform to another chart we will obtain a second derivative of the transition function. 
However, it turns out that 
\begin{equation}
[X,Y]^j_\alpha = X^i_\alpha \frac{\partial Y^j_\alpha}{x^i} - Y^i_\alpha \frac{\partial X^j_\alpha}{x^i}
\end{equation}
is a vector field, called the \emph{Lie bracket} of $X$ and $Y$.\footnote{This vector field is the commutator of the derivations $X$ and $Y$ on the ring of smooth functions.} For a general rank $(r,s)$-tensor, we can form the \emph{Lie derivative} given by the following formula 
\begin{multline}
(L_XT)_{\alpha,j_1\ldots j_s}^{i_1\ldots i_r} = X^i_\alpha\frac{\partial T_{\alpha,j_1\ldots j_s}^{i_1\ldots i_r}}{\partial x^i} - \frac{\partial X^{i_1}}{x^i} T_{\alpha,j_1\ldots j_s}^{i i_2\ldots i_r} - \ldots - \frac{\partial X^{i_r}}{x^i} T_{\alpha,j_1\ldots j_s}^{i_1\ldots i_{r-1}} \\
+ \frac{\partial X^{i}}{x^{j_1}} T_{\alpha,i\ldots j_s}^{i_1\ldots i_r} + \ldots + \frac{\partial X^{i}}{x^{j_s}} T_{\alpha,j_1\ldots j_{s-1} i}^{i_1\ldots i_{r-1}}
\end{multline} 
In particular, for a function $f$ we have $L_Xf = Xf$ and for a vector field $Y$ we have $L_XY = [X,Y]$. The following nice formula is left as an exercise: 
\begin{exc}[Cartan's magic formula]
If $\omega$ is a differential form, then 
\begin{equation}
L_X\omega = d\iota_X\omega + \iota_Xd\omega.\label{eq: Cartan magic formula}
\end{equation}
\end{exc}
\subsubsection{Submanifolds, orientiation, integration, Stokes theorem}
If $M$ is a manifold, then a \emph{submanifold} $S \subset M$ is a subset such around every $x \in S$, there is a coordinate chart $(U_\alpha,\phi_\alpha)$ and $k\leq d $ such that $\phi_\alpha(U_\alpha \cap S) = \{(x^1,\ldots,x^k,0,\ldots)\}$. Such coordinate charts are called \emph{adapted to $S$}. Then $S$ is also a manifold, of dimension $k$. For instance, the circle is a submanifold of the manifold $\R^2$: around every point in the circle, we can find polar coordinates $(r,\theta)$, and then the map $\phi(x,y)= (\theta, r-1)$ is a chart with the desired properties. 
\begin{exc}
Show that the 2-sphere is a submanifold of $\R^3$. 
\end{exc}
An \emph{orientation} of a manifold $M$ is an atlas $(U_\alpha,\phi_\alpha)$ such that all transition function $\phi_{\alpha\beta}$ have positive Jacobian determinant $\det d\phi_{\alpha\beta} \geq 0$. We say that two such atlases are equivalent if their union is also an orientation.  If $M$ has an orientation, it is called orientable, and in this case it has exactly two orientations (up to equivalence). The coordinates $(x^1,\ldots,x^n)$ in any chart $U_\alpha$ in an orientation are called positive coordinates. On oriented manifolds, one can define the integral of a top differential form, i.e. a differential whose degree coincides with the dimension of the manifold. For a coordinate chart $U_\alpha$ with positive coordinates we have 
\begin{equation}
\int_{U_\alpha} \omega:= \int_{\phi_\alpha(U_\alpha)} \omega_{\alpha,1\ldots n} dx^1 \ldots dx^n.
\end{equation}
Here $n = \dim M$ and on the right hand side we have an ordinary integral in $\R^n$. In particular, if there is a coordinate chart $U$ covering all but a measure zero subset of $M$, then $\int_M\omega = \int_U\omega$. This is also the only case in which one can practically compute an integral of a differential form.  If $S \subset M$ is an orientable $k$-dimensional submanifold, then we can pull back a differential $k$-form $\omega$ on $M$ to $S$ via the inclusion $\iota\colon S \to M$ and the compute the integral $\int_S\iota^*\omega$. In particular, if $(U_\alpha,\phi_\alpha)$ is an adapted chart for $S$  giving positive coordinates for $S$, we have 
\begin{equation}
\int_{U_\alpha \cap S} = \iota^*\omega \int_{\phi_\alpha(U_\alpha \cap S)} \omega_{\alpha,1\ldots k} dx^1 \ldots dx^k. 
\end{equation}
Often we simply omit writing $\iota^*$ in this case. 
\begin{exc}\label{exc:volume form sphere}
Consider the 2-form on $\R^3$ given by $\omega_{ij} = \frac{1}{8\pi}\varepsilon_{ijk}x^k$, with $\varepsilon_{ijk}$ the Levi-Civita symbol (specified by $\varepsilon_{123} = 1$ and antisymmetry in all indices). 
Show that 
\begin{equation}
\int_{S^2}\omega =1. 
\end{equation}
Hint: One possibility is to use the adapted coordinate chart given by spherical coordinates: $ x = r \cos\theta\sin\phi, y = r\sin\theta\sin\phi,z = r \cos \phi$, $0 < \theta < 2\pi, 0 < \phi < \pi$. 
\end{exc}
A central result concerning integration on manifolds is Stokes' theorem, that generalizes many theorems in multivariable analysis. To state in one requires the concept of \emph{manifolds with boundary}. Put shortly, we now consider a set $M$ covered by charts taking values in $\overline{\mathbb{H}^n} = \{(x^1,\ldots,x^n) \in \R^n, x^n \geq 0\}$, the closed upper half space. The union of all points $x \in M$ such that there is a chart $\phi_\alpha$ such that $\phi_\alpha(x) = (x^1,\ldots,x^{n-1})$ is called the \emph{boundary} of $M$ and denoted by $\partial M$. For instance, the closed unit disk $\overline{D} = \{(x,y) \in \R^2, x^2 + y^2 \leq 1\}$ is a manifold with boundary $\partial \overline{D} = S^1$, the unit circle.

 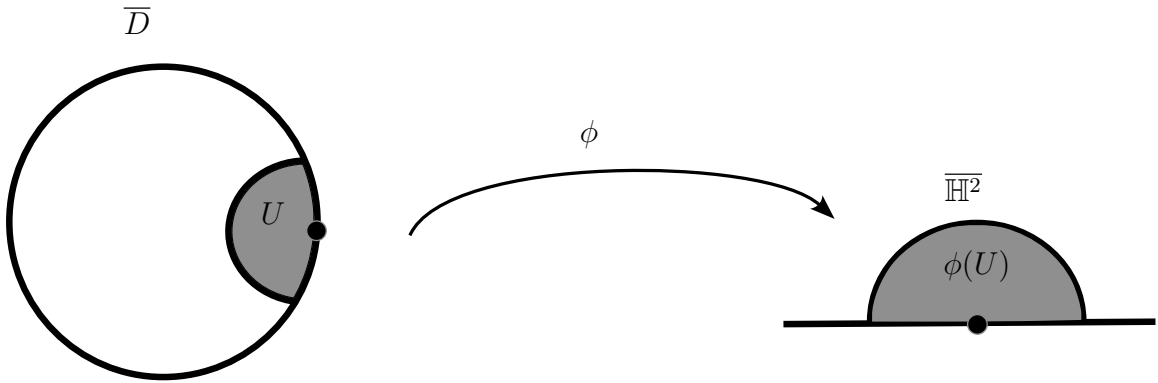
\begin{figure}[h!]
\centering
\def\svgwidth{\columnwidth}
\begingroup%
  \makeatletter%
  \providecommand\color[2][]{%
    \errmessage{(Inkscape) Color is used for the text in Inkscape, but the package 'color.sty' is not loaded}%
    \renewcommand\color[2][]{}%
  }%
  \providecommand\transparent[1]{%
    \errmessage{(Inkscape) Transparency is used (non-zero) for the text in Inkscape, but the package 'transparent.sty' is not loaded}%
    \renewcommand\transparent[1]{}%
  }%
  \providecommand\rotatebox[2]{#2}%
  \newcommand*\fsize{\dimexpr\f@size pt\relax}%
  \newcommand*\lineheight[1]{\fontsize{\fsize}{#1\fsize}\selectfont}%
  \ifx\svgwidth\undefined%
    \setlength{\unitlength}{539.39069132bp}%
    \ifx\svgscale\undefined%
      \relax%
    \else%
      \setlength{\unitlength}{\unitlength * \real{\svgscale}}%
    \fi%
  \else%
    \setlength{\unitlength}{\svgwidth}%
  \fi%
  \global\let\svgwidth\undefined%
  \global\let\svgscale\undefined%
  \makeatother%
  \begin{picture}(1,0.29291212)%
    \lineheight{1}%
    \setlength\tabcolsep{0pt}%
    \put(0,0){\includegraphics[width=\unitlength,page=1]{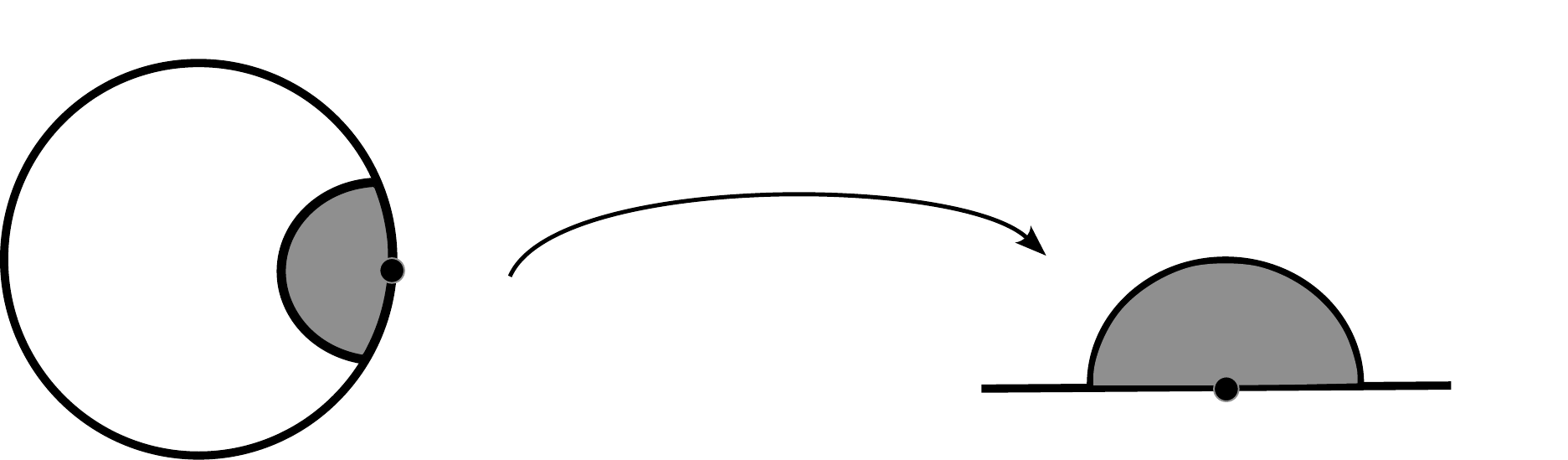}}%
    \put(0.46214743,0.19167061){\color[rgb]{0,0,0}\makebox(0,0)[lt]{\lineheight{1.25}\smash{\begin{tabular}[t]{l}$\phi$\end{tabular}}}}%
    \put(0.20524292,0.12574795){\color[rgb]{0,0,0}\makebox(0,0)[lt]{\lineheight{1.25}\smash{\begin{tabular}[t]{l}$U$\end{tabular}}}}%
    \put(0.09472549,0.27892124){\color[rgb]{0,0,0}\makebox(0,0)[lt]{\lineheight{1.25}\smash{\begin{tabular}[t]{l}$\overline{D}$\end{tabular}}}}%
    \put(0.75589109,0.14416755){\color[rgb]{0,0,0}\makebox(0,0)[lt]{\lineheight{1.25}\smash{\begin{tabular}[t]{l}$\overline{\mathbb{H}^2}$\\\end{tabular}}}}%
    \put(0.75492168,0.08600049){\color[rgb]{0,0,0}\makebox(0,0)[lt]{\lineheight{1.25}\smash{\begin{tabular}[t]{l}$\phi(U)$\end{tabular}}}}%
  \end{picture}%
\endgroup%

\caption{Any point in the boundary of the closed disk $\overline{D}$ has a neighbourhood $U$ homeomorphic to an open set $\phi(U)\subset \overline{\mathbb{H}^2}$, the closed upper half-plane.}\label{fig: disk}
\end{figure} 

 If $M$ is an $n$-dimensional manifold with boundary and $\omega$ is an $n-1$-form on $M$, then Stokes' theorem says
\begin{equation}
\int_{M} d\omega = \int_{\partial M} \omega. 
\end{equation}
In particular, if $M$ is a closed manifold, i.e. it does not have boundary, then $\int_M d\omega = 0$. We can use this as a criterion for exactness. Namely, if $\omega$ is a closed $k$-form (i.e. $d\omega = 0$) but there exists some $k$-dimensional submanifold $S$ such that $\int_S \omega \neq 0$, then $\omega$ is not exact. 
\begin{exc}[Closed and exact forms]\label{exc: closed and exact}
Show that the 1-form $d\theta$ on $S^1$ is closed but not exact. 
\end{exc}
It is a slightly nontrivial fact that \emph{all} closed forms on $\R^n$ are exact, and in fact on any contractible subset of $\R^n$, are closed. See for instance \cite{Bott1982}. This implies that any closed form on a manifold $M$ is exact when restricted to a contractible open set $U \subset M$.  
\subsection{Symplectic manifolds}
Consider now a manifold $M$ together with a closed 2-form $\omega$, i.e. in every chart $U_\alpha$ we have 
\begin{equation}
(d\omega)_{\alpha,ijk}= 3 \partial_{[i}\omega_{\alpha,jk]} = \partial_{i}\omega_{\alpha,jk} + \partial_k\omega_{\alpha,ij} + \partial_j\omega_{\alpha,ki} = 0. 
\end{equation}
Such a manifold is called \emph{presymplectic}. We say that $M$ is \emph{symplectic} if $\omega$ is additionally nondegenerate, i.e. for all charts $U_\alpha$ and all $x \in \phi_\alpha(U_\alpha)$, the matrix $\omega_{\alpha,ij}(x)$ is invertible. It follows immediately that the dimension $d$ of $M$ is even, since antisymmetric matrices can be non-degenerate in even dimensions only. 
\begin{expl}\label{exp: std symplectic}
\begin{enumerate}
\item The simplest example is $M = \R^2$ with coordinates $(p,q)$ and the symplectic form $\omega = dp \wedge dq$. 
\item Similarly, we have $M=\R^{2n}$ with coordinates $p_i,q^i$ and the symplectic form $\omega = \sum_i dp_i \wedge dq^i$. 
\end{enumerate}
\end{expl}
The examples above are called the standard symplectic space. A basic but important result, the \emph{Darboux theorem}, says that every symplectic manifold is locally standard: For every point $x$ in a symplectic manifold $M$ there exists a coordinate chart $(U, \phi=(q^1,\ldots,q^n,p_1,\ldots,p_n))$ around $x$ such that in this coordinate chart the symplectic form is locally standard, 
\begin{equation}
\omega_U = \sum_i dp_i \wedge dq^i.
\end{equation}
\begin{expl}\label{exp: cotangent bundle}
Example \ref{exp: std symplectic} is a special case of a cotangent bundle, $\R^{2n} \cong T^*\mathbb{R}^n$. Cotangent bundles carry a standard symplectic form, which can be described in local coordinates as follows. First, remember that for local coordinates $(q^1,\ldots,q^n)$ on $U$ we have associated coordinates $(p_1,\ldots,p_n,q^1,\ldots,q^n)$ on $T^*U \subset T^*M$. Define the \emph{tautological 1-form} $\theta$ by 
\begin{equation}
\theta_U = \sum p_i \wedge dq^i.
\end{equation}
It is a simple exercise to show that $\theta$ is globally defined. Letting $\omega_U = d\theta_U = \sum_i dp_i\wedge dq^i$ we obtain the standard symplectic form $\omega_{std}$ on $T^*M$. 
\end{expl}
\begin{exc} We leave an alternative definition of the tautological 1-form (that does not require choosing coordinates) as an exercise. Recall that we have the projection $\pi \colon T^*M \to M$. To define a 1-form it is enough to say what it does to a tangent vector at every point. For $(q,p)\in T^*\R^n$ and $v \in T_{(q,p)}T^*M$ define 
$$\theta_{(q,p)} = p(d\pi_{(q,p)}v).$$
Show that this definition of $\theta$ coincides with the definition using coordinates above.
\end{exc}
In all these cases the symplectic form was actually exact. An easy class of examples where the symplectic form is not exact is given by surfaces with volume forms. 
\begin{expl}[2-dimensional manifolds]\label{ex: 2dim}
In dimension 2, any 2-form is necessarily closed. It is non-degenerate if it does not vanish. This means that in two dimensions a symplectic form is the same thing as a volume form. For instance, we can take 2-sphere $S^2 = \{ x \in \R^3, |x| = 1\}$ with its standard volume form $\omega$ (cf. Exercise \ref{exc:volume form sphere}) given in spherical coordinate $0<\theta<\pi,0<\phi<2\pi$ by $\omega = \frac{1}{4\pi}\sin(\theta)d\theta \wedge d\phi$ or, using a complex coordinate $z$ and thinking of $S^2 = \C \cup \{\infty\}$, by $\omega = \frac{1}{2\pi i}\frac{d\bar{z}\wedge dz}{(1+|z|^2)^2}$. Another example is the two-torus $S^1 \times S^1$ with its standard volume form $\omega = d\theta_1 \wedge d\theta_2$. All these forms have integral equal to 1, hence they are not exact. 
\end{expl}
\subsubsection{Special submanifolds}
We introduce some standard terminology concerning submanifolds of symplectic manifolds. \\

First, for a vector space $V$ together with an antisymmetric bilinear form $\omega$ (a symplectic vector space) and a subspace $W \subset V$ we define 
\begin{equation}
W^\perp = \{v \in V, \omega(v,w) = 0, \forall w \in  W\}
\end{equation}
\begin{exc}
Show that $\dim W^\perp = \dim V - \dim W$.
\end{exc}
Then we say that a subspace $W \subset V$ is \begin{itemize}
\item \emph{isotropic} if $W \subset W^\perp$, 
\item \emph{coisotropic} if $W \supset W^\perp$, 
\item \emph{lagrangian} if $W = W^\perp$.
\end{itemize}
\begin{exc}
Show that $W \subset V$ is lagrangian if and only if $\dim W = \frac{1}{2}\dim V$ and $\omega(w_1,w_2) = 0 $ for all $w_1,w_2 \in W$.
\end{exc}
\begin{expl}
\begin{itemize}
\item Any 1-dimensional subspace of $(\R^2, \omega_{std})$ is lagrangian. 
\item The set of lagrangian subspaces $\Lambda(n)$ of $\R^{2n}$ is already quite interesting, and called the \emph{lagrangian grassmannian}. It is a manifold of dimension $\frac{1}{2}n(n+1)$. In general $\Lambda(n) = U(n)/O(n)$, in particular $\Lambda(1) = U(1)/O(1) = S^1/\mathbb{Z}_2 = \R\mathbb{P}^1 \cong S^1$.  
\end{itemize}
\end{expl}
Now we say that a submanifold $N \subset M$ of a symplectic manifold $(M,\omega)$ is \emph{lagrangian} (resp. \emph{isotropic} or \emph{coisotropic}) if for every $x \in  N$, $T_xN \subset T_xM$ is a lagrangian (resp. isotropic or coisotropic) subspace of the symplectic vector space $(M,\omega)$. 
\begin{expl}\begin{itemize}
\item Any 1-dimensional submanifold of a 2-dimensional symplectic manifold is lagrangian.
\item The zero section $M \subset T^*M$ is lagrangian. 
\end{itemize}
\end{expl}
\subsubsection{Poisson brackets}
Let $(M,\omega)$ be a symplectic manifold. The nondegeneracy of $\omega$ allows us to define the \emph{Poisson bracket} of two functions $f$ and $g$, namely we set 
\begin{equation}
\{f,g\}_\alpha = (\omega_{\alpha}^{-1})^{ij} \partial_i f_\alpha\partial_j g_\alpha = \tr (\omega^{-1}))\otimes df \otimes dg \label{eq: def PB}
\end{equation}
\begin{exc}
\begin{enumerate}
\item Show this defines a global function on $M$, i.e. the transformation property \eqref{eq:fct transform} is satisfied. 
\item Show that $\{f,g\}$ is antisymmetric , i.e. $\{f,g\} = -\{g,f\}$, and bilinear, i.e. $\{\lambda f + g,h \} = \lambda \{f,h\} + \{g,h\}$ (here $f,g,h \in C^\infty(M)$ and $\lambda \in \R$). 
\item Show the Poisson bracket satisfies the Leibniz identity, i.e. for all $f,g,h \in C^\infty(M)$ we have 
\begin{equation}
\{f,gh\} = g\{f,h\} + h\{f,g\}.
\end{equation}
\item Show the Poisson bracket satisfies the Jacobi identity, i.e. for all $f,g,h \in C^\infty(M)$ we have 
\begin{equation}
\{f,\{g,h\}\} + \{g, \{h,f\} \} + \{h,\{f,g\}\} = 0 \label{eq: Jacobi Poisson}
\end{equation}
(this uses that $\omega$ is closed!)
\end{enumerate}
\end{exc}
Of fundamental importance in symplectic geometry are the \emph{hamiltonian} vector fields. If $f$ is a function, then its hamiltonian vector field $X_f$ can be defined by its action on functions: $X_f(g) = \{f,g\}$. More explicitly, one can define it in charts by 
\begin{equation}
(X_f)_\alpha^j = (\omega^{-1}_\alpha)^{ij}\partial_i f_\alpha \label{eq: ham vect field coords}
\end{equation} 
\begin{exc}\label{exc: ham vect PB}
Show that 
\begin{equation}
X_{\{f,g\}} = [X_f,X_g] \label{eq: ham vect PB}
\end{equation}
(you can either use Equations \eqref{eq: ham vect field coords} and \eqref{eq: def PB} and work your way through the coordinates or simply use the Definition $X_f(g) = \{f,g\}$ and the Jacobi identity \eqref{eq: Jacobi Poisson}).
\end{exc}
\begin{expl}
Let $M = \R^{2n}$ with the standard symplectic structure. Then 
\begin{equation}
\{f,g\} = \sum_{i=1}^n \frac{\partial f}{\partial q^i}\frac{\partial g}{\partial p_i} - \frac{\partial f}{\partial p_i}\frac{\partial g}{\partial q_i}. 
\end{equation}
In particular 
\begin{align*}
X_{q^i} &= \frac{\partial}{\partial p_i} \\
X_{p_i} &= -\frac{\partial}{\partial q^i}
\end{align*}
\end{expl}
\subsubsection{K\"ahler manifolds}\label{sec: Kaehler}
A particularly nice class of symplectic manifolds is given by K\"ahler manifolds, those are symplectic manifolds admitting a compatible complex structure, in the following sense: An almost complex structure on a manifold $M$ is a map $J_x\colon T_xM \to T_xM$, defined for every $x \in M$, with the property that $J_x^2 = -1$, which varies smoothly with $x$.\footnote{I.e. it is a smooth section of the endomorphism bundle $\mathrm{End}(TM)$, concretely, for every coordinate chart $U_\alpha$ we obtain a map $J_\alpha \colon U_\alpha \to GL(\dim M)$ which is required to be smooth} The almost complex structure is said to be compatible with $\omega$ if the bilinear form $g_x$ on $T_xM$ defined by $g_x(v,w) = \omega(v,Jw)$ is symmetric and positive definite (i.e. a Riemannian metric on $M$). A complex structure on $M$ is a complex atlas, i.e. a collection $(U_\alpha,\phi_\alpha)$ such that $\phi_\alpha(U_\alpha) \subset \C^n$ and the transition functions $\phi_{\alpha\beta} = \phi_\beta \circ \phi_\alpha^{-1}$ are \emph{biholomorphic}. On a complex manifold we have a natural almost complex structure, given by multiplying tangent vectors by $i$.\footnote{The fact that the transition functions are biholomorphic means that their differentials are $\C$-linear rather than only $\R$-linear, which gives the tangent space the structure of a $\C$-vector space}. A K\"ahler manifold is a symplectic manifold $(M,\omega)$ with a compatible complex structure that we also denote by $J$. The symplectic form on a K\"ahler manifold is called a K\"ahler form. 
\begin{expl}\label{ex: std Kahler} 
The easiest example is $M = \R^{2n}$ with its standard symplectic form $\omega = \sum_i dp_i \wedge dq^i$. The natural complex coordinates are 
$$z^i = q^i + i p_i$$ and the complex structure $J$ is given by 
\begin{align*}
\frac{\partial}{\partial q^i} \mapsto \frac{\partial}{\partial p_i} \\
\frac{\partial}{\partial p_i} \mapsto -\frac{\partial}{\partial q^i} 
\end{align*}
It is clear that this map squares to $-1$. Using that 
$\omega\left(\frac{\partial}{\partial p_i},\frac{\partial}{\partial q^i}\right) = +1$
we get immediately that 
\begin{align*}
\omega\left(J\left(\frac{\partial}{\partial p_i}\right),\frac{\partial}{\partial p_j}\right) &= \delta_{ij} = \omega\left(J\left(\frac{\partial}{\partial q^i}\right) \frac{\partial}{\partial q^j}\right), \\
 \omega\left(J\left(\frac{\partial}{\partial q^i}\right),\frac{\partial}{\partial p_j}\right) &= 0 = \omega\left(J\left(\frac{\partial}{\partial p_i}\right),\frac{\partial}{\partial q^j}\right)
\end{align*}
 which means that $g(v,w) = \omega(Jv,w)$ is the standard Riemannian metric on $\R^{2n}$. This is called, of course, the standard K\"ahler structure on $\R^n$. 
\end{expl}
Since the map $J_x\colon T_xM \to T_xM$ squares to $-1$, it has no real eigenvectors in $T_xM$. However, if we complexify the tangent space, $(T_xM)_{\mathbb{C}} = T_xM \otimes \C$, the complexification $J\colon (T_xM)_{\mathbb{C}} \to (T_xM)_{\mathbb{C}}$ has the eigenvalues $\pm i $ and the complexified tangent space splits as the sum of the two eigenspaces, 
$$(T_xM)_{\mathbb{C}} = \underbrace{T_xM^{(1,0)}}_{{+i\text{-eigenspace}}} \oplus \underbrace{T_xM^{(0,1)}}_{-i\text{-eigenspace}}$$ 
called the \emph{holomorphic} and \emph{antiholomorphic} tangent spaces respectively.\footnote{It is an eternal source of confusion to complexify a vector space that already had a complex structure. It is therefore beneficial to denote the complex structure on $T_xM$ by $J$ (even though one can think about it as multiplication by $i$) and reserve multiplication by $i$ for the complexified vector space. That is, we think of $T_xM$ as a \emph{real} vector space - with a complex structure $J$, and only of $T_xM \otimes \C$ as a complex vector space.}  The following exercise is simple but fundamental for complex geometry. 
\begin{exc}
Consider again the example \ref{ex: std Kahler} of $\R^{2n}$ with its standard complex structure. 
\begin{itemize}\label{ex: cx geo 101}
\item Show that the $+ i$ eigenspace of $J$ is spanned by $\frac{\partial}{\partial z^i} = \frac12\left({\frac{\partial}{\partial q^i} - i\frac{\partial}{\partial p^i}}\right)$ and the $-i$ eigenspace is spanned by $\frac{\partial}{\partial \bar{z}^i} =  \frac12\left({\frac{\partial}{\partial q^i} + i\frac{\partial}{\partial p^i}}\right)$. 
\item Let $dz^i = dq^i + i dp_i$, $d\bar{z}^i = dq^i - i dp_i$. Show that $dz^i\left(\frac{\partial}{\partial z^j}\right)=\delta_{ij} = d\bar{z}^i\left(\frac{\partial}{\partial \bar{z}^j}\right)$ while $dz^i\left(\frac{\partial}{\partial \bar{z}^i}\right) = d\bar{z}^i\left(\frac{\partial}{\partial z^i}\right) = 0$.
\item Show that $dz^i \wedge d\bar{z}^i= 2i\ dp_i \wedge dq^i$.
\end{itemize}
\end{exc}

The results of this exercise also hold in a complex coordinate chart on a complex manifold $(M,J)$. In $\R^{2n}$, the last point implies in particular that we can write the symplectic form on $\R^{2n}$ as $\omega = \sum_i\frac{i}{2}d\bar{z}^i\wedge dz^i$. Here we have the special situation that the Darboux coordinates $p,q$ are also imaginary and real parts of complex coordinates. On a general K\"ahler manifold, a Darboux chart will not give rise to complex coordinates. However, we have the results of the following exercise. 
\begin{exc}
Let $(M,\omega,J)$ be a K\"ahler manifold. 
\begin{itemize}
\item Show that, for any point $x \in  M$, we have $\omega_x(J_xv,J_xw) = \omega_x(v,w)$. 
\item Show that $J^*dz^i = dz^i \circ J = i dz^i$ and $J^* d\bar{z}^i  = -i d\bar{z}^i$. 
\item In a complex coordinate chart $U$, we can write $\omega$ as  $\sum_{ij}a_{ij}dz^i \wedge dz^j + b_{ij} d\bar{z}^i \wedge dz^j+ c_{ij}d\bar{z}^i \wedge d\bar{z}^j$. Using the first two points, show that $a_{ij} = c_{ij} = 0$. Using that $\omega$ is real-valued, i.e. $\omega = \overline{\omega}$, show that $h_{ij} = \frac{i}{2}b_{ij}$ is a hermitian $n\times n$ matrix. 
\end{itemize}
\end{exc}
That is, in any complex coordinate chart $U$, the symplectic form can be written as 
\begin{equation}\label{eq: Kahler}
\omega = \sum_{i,j}\frac{i}{2}h_{ij}d\bar{z}^i \wedge dz^j
\end{equation}
where $h_{ij}$ is a matrix of complex functions such that for any point $x \in U$, $h_{ij}(x)$ is a \emph{nondegenerate} hermitian matrix. On the other hand, if there is a closed 2-form $\omega$ on a complex manifold $M$ which has the form \eqref{eq: Kahler} in every chart, then $\omega$ is a K\"ahler form.  
Another important fact about K\"ahler manifolds is that they locally admit a so-called \emph{K\"ahler potential}. For this we first need to introduce the Dolbeault operators. On a complex manifold, we can introduce the type $(k,l)$-forms, those are the $k+l$ complex-valued differential forms which in every complex cooridate chart are spanned (over $C^(U,\C)$) by monomials of the form $dz^{i_1} \wedge \cdots \wedge dz^{i_k} \wedge d\bar{z}^{i_1} \cdots \wedge d\bar{z}^{i_l}$. In particular, above we have shown that the K\"ahler form is a (1,1)-form. The space of $(k,l)$-forms is denoted $\Omega^{k,l}(M)$ and we have
\begin{equation}
\Omega^m(M,\C) = \bigoplus_{k+l=m} \Omega^{k,l}(M). 
\end{equation}
We can restrict the de Rham differential to $\Omega^{k,l}(M)$, then it will land in $\Omega^{k+1,l}(M)\oplus \Omega^{k,l+1}(M)$. We denote the composition of $d$ with the projection to the two subspaces by $\partial$ and $\bar\partial$, they are called the Dolbeault operators and satisfy $\partial^2 = \bar\partial^2 = 0$. On a complex manifold we have $d = \partial + \bar\partial$,\footnote{This is in fact the major difference setting apart almost complex manifolds from complex manifolds. We can define $(k,l)$-forms and the Dolbeault operators using just $J$, but we have $d = \partial + \bar\partial$ if and only if $J$ comes from a complex structure.} and, in a complex coordinate chart we have
\begin{equation}
\partial = \sum_i\frac{\partial}{\partial z^i}dz^i, \qquad \bar{\partial}=\sum_i\frac{\partial}{\partial \bar{z}^i}d\bar{z}^i.
\end{equation}
A (local) K\"ahler potential on some open $U \subset M$ is a function $f \colon U \to \R$ (real-valued!) such that $\omega = \frac{i}{2}\bar\partial\partial f$. Any  
\begin{exc}
Show that $f(z^1,\ldots z^n) =\sum_{i=1}^n |z_i|^2$ is a global K\"ahler potential for the standard K\"ahler form on $\R^{2n}$. 
\end{exc}
\begin{exc}\label{ex: 2 sphere Kaehler}
In this exercise we show explicitly that the 2-sphere $S^2 = \{( x {\in} \R^3, |x| =1\}$ is a K\"ahler manifold. 
\begin{enumerate}
\item Recall from Exercise \ref{exc: stereo} the atlas on the two-sphere given by stereographic projection, $\phi_{N}(x) = \frac{1}{1-x^3}(x^1,x^2)=(a,b)$ and $\phi_S(x) = \frac{1}{1+x^3}(x^1,x^2) = (c,d)$. Let $z = a+ib$ and $w = c-id$. Show that $w = \frac{1}{z}$ and conclude $(S^2 - \{N\}, z)$ and $(S^2 - \{S\}, w)$ is a complex atlas for $S^2$. 
\item Let $\omega_N = \frac{1}{2\pi i }\frac{d\bar{z}\wedge dz}{(1+|z|^2)^2}$ and $\omega_S = \frac{1}{2\pi i }\frac{d\bar{w} \wedge dw}{(1+|w|^2)}$. Check that they agree on the overlap and conclude this defines a K\"ahler form on $S^2$. This form is called the Fubini-Study form on $S^2$ (although this terminology is used more often when thinking about $S^2$ as complex projective space $\C\mathbb{P}^1$). 
\item Let $f_N(z) =\frac{1}{\pi} \log (1+|z|^2)$. Show that $f_N(z)$ is a K\"ahler potential for $\omega$ on $S^2 - \{N\}$. 
\item Show that $\int_{S^2}\omega=1$ and conclude that $\omega$ cannot admit a global K\"ahler potential.
\end{enumerate}
\end{exc}
 \section{Lecture 3: Prequantization}\label{sec: prequantization}

Having established the geometric terminology we start trying to find a quantization prescription satisfying Dirac's quantization condition. Namely, we want to find a Hilbert space $\mathcal{H}$ map $Q \colon C^\infty(M) \to \mathcal{L}(\mathcal{H}), f \mapsto Q_f$ such that 
\begin{enumerate}[Q1)]
\item $Q$ is linear,
\item $Q_1 = \mathrm{id}_{\mathcal{H}}$, 
\item $Q_{\bar{f}} = (Q_f)^*$
\item $[Q_f,Q_g] = - i\hbar Q_{\{f,g\}}$ 
\item $f_1,\ldots,f_k$ complete $\Rightarrow Q_{f_1},\ldots Q_{f_k}$ complete
\end{enumerate}
hold. Let $(M,\omega)$ be a $d = 2n$-dimensional symplectic manifold. Then $\epsilon = \frac{1}{(2\pi\hbar)^n}\omega^{n}$ is a volume form on $M$ and we therefore have an associated Hilbert space 
$\mathcal{H} = L^2(M, \epsilon)$ of complex-valued functions which are square-integrable with respect to $\omega$. Notice that this Hilbert space is ``too big'' physically: In the simplest example of $\R^2$ with standard symplectic form $dp \wedge dq$, this is the Hilbert space of square-integrable functions of both the $p$ and $q$ variables, but from Schrödinger quantization we would expect only one of them. In particular, in this situation we cannot expect the axiom Q5 to hold. Our strategy is now to disregard this problem and focus on the axioms Q1 - Q4, and call a map satisfying those a \emph{prequantization}. This will be the content of this lecture. The main result - sometimes called Weil integrality condition - is that not all symplectic manifolds admit a prequantization, but only those whose symplectic forms satisfy 
\begin{equation}
\int_N \omega = 2\pi \hbar k
\end{equation} 
where $N$ is any 2-dimensional submanifold of $M$ and $k$ is any integer. We will also classify all possible prequantizations. 
\subsection{First attempts at prequantization}\label{sec: pre quant I }
Thus, we set out on the quest to find a map $P \colon C^\infty(M) \to \mathcal{L}(\mathcal{H}), f \mapsto P_f$, satisfying Q1) - Q4). We recall that every function $f \in C^\infty(M)$ naturally acts on functions via its Hamiltonian vector field $X_f$ by $X_f(g) = L_{X_f}g =  \{f,g\}$, and that we have the  result $[X_f,X_g] = X_{\{f,g\}}$ (Exercise \ref{exc: ham vect PB}). Our first guess could therefore be to let 
\begin{equation}
P^{(1)}_f = -i\hbar X_f. 
\end{equation}
With this normalization, we see that $P^{(1)}$ satisfies Q4). Q1) is trivial to verify, and to see that Q3) holds we can do the following simple computation 
$$0 = \int_M L_{X_f}( g\bar{h}\omega^n ) = \int_M L_{X_f} (g\bar{h}) \omega^n = \int_M L_{X_f}g + g L_{X_f} \bar{h}\omega^n = \int_M L_{X_f}g + g \overline{L_{X_{\bar{f}}} h}\omega^n  $$
which implies that $X_f^* = - X_{\bar{f}}$. However, we have $P^{(1)}_1 = 0$, and therefore Q2) is violated. \\ 
Trying to ameliorate this we now define 
$$P^{(2)}_f = i\hbar X_f + \hat{f}$$ 
where $\hat{f}$ is the multiplication operator $\hat{f}g = fg $. We can now check Q1) - Q3) easily, but because $$[X_f,\hat{g}] h = X_f(gh) - g X_f(h) = \{f,gh\} - g\{f,h\} = \{f,g\}h = \widehat{\{f,g\}}h$$ now we get 
$$[P^{(2)}_f,P^{(2)}_g] = -i\hbar X_{\{f,g\}} - i\hbar [X_f,\hat{g}] - i\hbar [\hat{f},X_g] = -\hbar X_{f,g} - 2i\hbar \widehat{\{f,g\}} = -i\hbar P^{(2)}_{\{f,g\}} - i\hbar \widehat{\{f,g\}}.$$ 
For this we use the following identity: Let $X,Y$ be vector fields and $\theta$ a 1-form, then we have
\begin{equation}
X \iota_Y\theta - Y \iota_X\theta = \iota_{[X,Y]}\theta - d\theta(X,Y) \label{eq: some identity}
\end{equation}
\begin{exc}
Prove \eqref{eq: some identity} using that $[L_X,\iota_Y] = \iota_{[X,Y]}$ and Cartan's magic formula $L_X = d\iota_X + \iota_Xd$. 
\end{exc}
Suppose now that $\omega = d\theta$ for some 1-form $\theta$. Then we claim that 
\begin{equation}
P_f = -i\hbar X_f + \widehat{-\iota_{X_f}\theta} + \widehat{f} \label{eq: prequantization trivial}
\end{equation}
satisfies Q1) - Q4). 
\begin{exc}
Show that $P$ defined by \eqref{eq: prequantization trivial} satisfies Q1) - Q4). Use \eqref{eq: some identity} for Q4).
\end{exc}
\begin{expl}\label{ex:std prequant Rn}
Let us consider the case $M = T^*\R^n$. Then $\omega = \sum_{i = 1}^n dp_i \wedge dq^i$ and we can choose $\theta = \sum_{i = 1}^n p_i dq^i$. Let us look at the quantization of the coordinate functions $p_i,q^i$. We have $X_{p_i} = \frac{\partial}{\partial q^i}$ and $X_{q^i} = -\frac{\partial}{\partial p_i}$ so that \begin{align}
P_{q^i} &= -i\hbar \frac{\partial}{\partial p_i} + \widehat{q}^i \label{eq:std prequant Rn q} \\
P_{p_i} &= -i\hbar \frac{\partial}{\partial q^i} - \widehat{p}_i + \widehat{p}_i = -i\hbar\frac{\partial}{\partial q_i}. \label{eq:std prequant Rn p}
\end{align}
Here we can explicitly see that Q5) is violated - the functions $p_i$ and $q^i$ form a complete set but the operators $P_{q^i}$ and $P_{p_i}$ do not, as for instance the operators $\frac{\partial}{\partial p_i}$ commute with all of the $P_{q^i}, P_{p_i}$. But, we can also see that \eqref{eq:std prequant Rn p},\eqref{eq:std prequant Rn q}, reduce to the Schr\"odinger quantization when restricted to $L^2(\R^n)_q$, i.e. functions of the $q^i$ variables alone. We will return to this in Section \ref{sec: quantization}. 
\end{expl}
Apart from the fact that \eqref{eq: prequantization trivial} does not satisfy Q5), there are two more obvious drawbacks to \eqref{eq: prequantization trivial}. The first one is rather obvious: We had to assume that $\omega$ is exact. This is satisfied for $\R^{2n}$ or more generally contangent bundles $T^*M$, but it is \emph{never} the case for compact symplectic manifolds: Those have a finite symplectic volume 
$$ \mathrm{Vol}(M) = \int_M\epsilon = \frac{1}{(2\pi\hbar)^n}\int_M \omega^n.$$
But if $\omega = d\theta$ is exact, then $\omega^n = d (\theta \wedge \omega^{n-1})$ is also exact and 
$$\int_M \omega^n = \int d(\theta \wedge \omega^n) = 0.$$
The second drawback is that $\theta$ is not uniquely specified by the condition $\omega$. In fact, we can add any closed 1-form $\alpha$ to $\theta$ since $d(\theta + \alpha) = d\theta + d\alpha = \omega$. Have we already come to an end to our quest to quantize general symplectic manifolds? 
\subsection{Prequantization line bundles}
It turns out both problems can be partially addressed by generalizing our operators to act on sections of a line bundle instead of just functions. Let us try to derive this. Even if $\omega$ is not globally exact, suppose that we have some $U \subset M$ such that $\omega\big|_U = d\theta_U$ (any contractible $U$ will do). Then we can define $P_U(f)$ by just applying the formula \eqref{eq: prequantization trivial} in $U$, i.e. we define
$$ P_U(f) = -i\hbar X_f +  \widehat{-\iota_{X_f}\theta_U + f}$$
to be an operator acting on $C^\infty(U)$.
Now suppose that we have another subset $V$ where $\omega = d\theta_V$ and $U\cap V \neq 0$. Suppose $g$ is a function on $M$, then by restriction to $U$ (resp. $V$) we obtain functions $g_U$ and $g_V$ on which $P_{U,f}$ and $P_{V,f}$ act. Then, notice that on the intersection we have $P_{U,f}g_U - P_{V,f}g_V = \iota_{X_f}(\theta_U- \theta_V)g\big|_{U\cap V}$ and in general there is no reason for this to vanish. Here the point is that $g_U$ and $g_V$ agree on $U \cap V$. But suppose now that we instead have another object $\sigma$ such that we have instead $\sigma_V = g_{UV}\sigma_U$ on $U\cap V$, where $g_{UV}\colon U\cap V\to \mathbb{C}$ is a complex-valued function. Can we then achieve 
\begin{equation} P_V(f)\sigma_V = g_{UV}P_U(f)\sigma_U?\label{eq:question}
\end{equation}
Because $d\theta_U = d\theta_V$, we know that $\theta_U - \theta_V$ is \emph{closed} on $U\cap V$. If $U,V$ are such that $U\cap V$ is contractible, then we can conclude that $\theta_U - \theta_V = d\phi_{UV}$ is \emph{exact} and then we have 
\begin{align*}P_{V,f}\sigma_V &= (P_{U,f} - \widehat{X_f\phi_{UV}})g_{UV}\sigma_U \\
&= -i\hbar X_f(g_{UV})\sigma_U + g_{UV}P_{U,f}\sigma_U + X_f(\phi_{UV}) g_{UV}\sigma_U
\end{align*}
We therefore have \eqref{eq:question} if and only if 
$$ i\hbar X_f(g_{UV}) = X_f(\phi_{UV})g_{UV}. $$
Since $X_f$ is a first order differential operator, this holds if 
$$g_{UV} = \exp\left(\frac{i}{\hbar}\phi_{UV}\right). $$
To obtain a global version of this construction, suppose we can cover $M$ by open sets $U_\alpha$ such that $\omega\big|_{U_\alpha}$ is exact and such that all intersections $U_{\alpha} \cap  U_{\beta}$ for ($\alpha \neq \beta$) are contractible. Then we have $\theta_\alpha - \theta_\beta = d\phi_{\alpha\beta}$, for some complex-valued functions $\phi_{\alpha\beta} \colon U_{\alpha\beta} \to \C$. If $\sigma_\alpha$ is a collection of complex-valued functions on $U_\alpha$ satisfying $\sigma_\beta = g_{\alpha\beta}\sigma_\alpha = \exp(i/\hbar \phi_{\alpha\beta})\sigma_\alpha$, then 
$\sigma_\alpha = g_{\gamma\alpha}g_{\beta\gamma}g_{\alpha\beta} \sigma_\alpha$ which implies $g_{\gamma\alpha}g_{\beta\gamma}g_{\alpha\beta} = 1 $ and therefore
\begin{equation}
\phi_{\alpha\beta\gamma} := \phi_{\alpha\beta} + \phi_{\beta\gamma} + \phi_{\gamma\alpha} \in 2\pi\hbar \mathbb{Z}.\label{eq:integral cech}
\end{equation}

\begin{figure}[!ht]
\centering 
\includegraphics[scale=1]{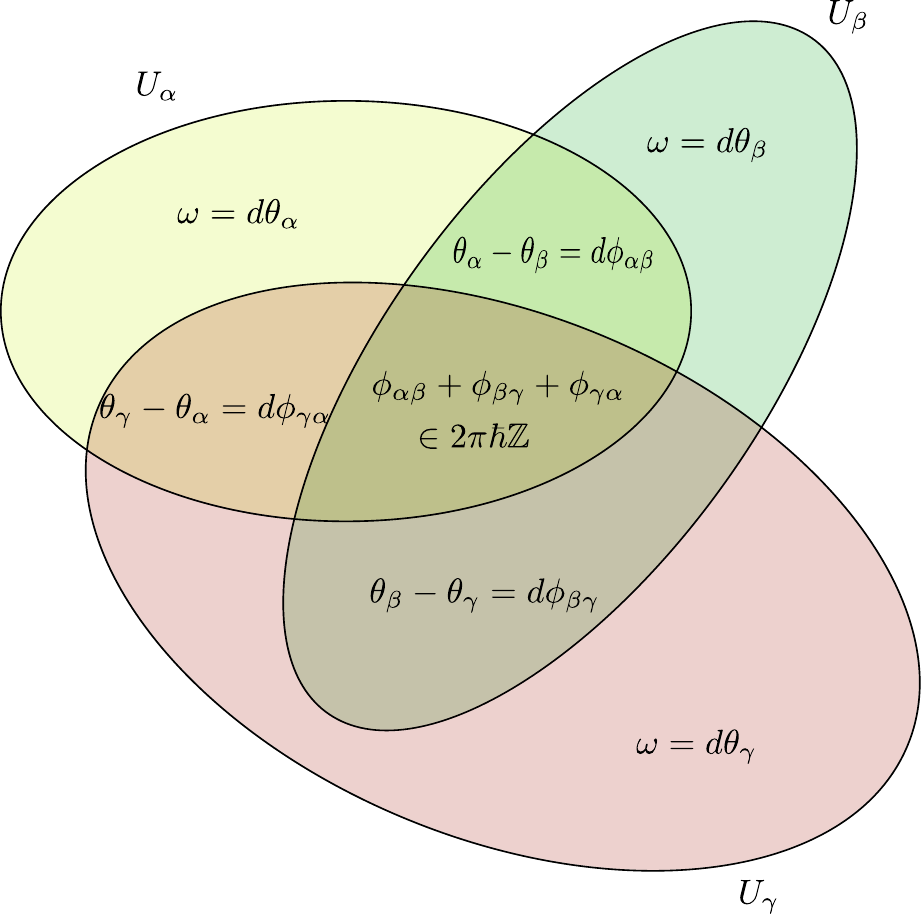}
\caption{Data involved in defining a line bundle}
\end{figure}

\begin{figure}
 \centering 
 \includegraphics[scale=0.8]{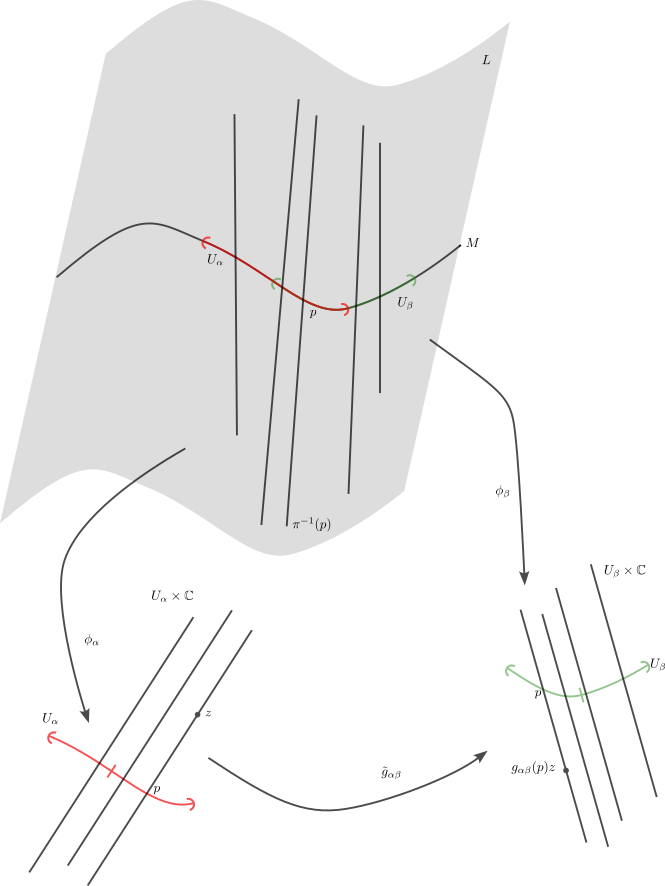}
 \caption{The geometric definition of complex line bundle: a total space $L$ with a map $\pi \colon L \to M$ such that $\pi^{-1}(p)$ is a complex line for every $p$, and maps $\phi_\alpha \colon \pi^{-1}(U_\alpha) \to U_\alpha \times \C$ that are linear isomorphism restricted to fibers. Then one defines $\tilde{g}_{\alpha\beta}(p,z) = \phi_\beta \circ \phi_{\alpha}^{-1}(p,z) = (p,g_{\alpha\beta}(p)z)$. The collection $(U_\alpha,g_{\alpha\beta})$ then defines a line bundle in the previous sense.} \label{fig: linebundle2}
 \end{figure}

Summarising, if we have 
\begin{itemize}
\item A  cover $U_\alpha$ of $M$ of open sets,
\item functions $g_{\alpha\beta} \colon U_{\alpha}\cap U_\beta \to \C$ 
\item A family of 1-forms $\theta_\alpha$ 
\end{itemize}
such that 
\begin{align}
g_{\gamma\alpha}g_{\beta\gamma}g_{\alpha\beta} &= 1 \label{eq:lb1} \\ 
g_{\beta\alpha}g_{\alpha\beta} &=1  \label{eq:lb2}\\ 
\frac{i}{\hbar}\theta_\beta - \frac{i}{\hbar}\theta_\alpha = \frac{i}{\hbar}d\phi_{\alpha\beta} &= g^{-1}_{\alpha\beta}dg_{\alpha\beta} \label{eq:lb3}
\end{align}
we can give the following definition:
\begin{defn} \label{def: line bundle}
\begin{enumerate}
\item The data $(L,\nabla) := (U_\alpha,g_{\alpha\beta},i\theta_\alpha/\hbar)$
satisfying \eqref{eq:lb1},\eqref{eq:lb2}, \eqref{eq:lb3} are called a \emph{complex line bundle with connection}. 
\item  A collection $\sigma_\alpha$ such that $\sigma_\beta = g_{\alpha\beta} \sigma_\alpha$ is called a \emph{section} of $L$. The collection of all sections of $L$ is denoted by $\Gamma(L)$. 
\item For any vector field $X$ on $M$, we define the \emph{covariant derivative} $\nabla_X \colon \Gamma(L) \to \Gamma(L)$ by 
\begin{equation}
(\nabla_X\sigma)_\alpha = X(\sigma_\alpha) - \frac{i}{\hbar}\theta_\alpha(X)\sigma_\alpha
\end{equation}
\item The \emph{curvature} of $L$ is the 2-form $\Omega$ defined  by 
\begin{equation}
F_\nabla(X,Y) = i([\nabla_X,\nabla_Y] - \nabla_{[X,Y]}) 
\end{equation}
\end{enumerate}
\end{defn}
In the following exercise we state some important properties of line bundles.
\begin{exc}
\begin{itemize}
\item Check that $\nabla_X\sigma$ is indeed a section, i.e that we have $(\nabla_X\sigma)_\beta = g_{\alpha\beta}(\nabla_X\sigma)_\alpha$. 
\item Show that $F_\nabla\big|_{U_\alpha} = d\theta_\alpha/\hbar$. In particular, we have that $F_\nabla = \frac{\omega}{\hbar}$. 
\end{itemize}
\end{exc}
 Line bundles can also be defined using an equivalent, geometric approach. Define the set $$L = \frac{\bigsqcup_{p{ \in} M}\bigsqcup_{\alpha, p \in U_\alpha}\mathbb{C}}{(p,\alpha,z)\sim (p,\beta,g_{\alpha\beta}z)}.$$
One can show that $L$ is a manifold and $\pi \colon L \to M, [(p,\alpha,z)] \mapsto p$, is a smooth surjective map with the property that $\pi^{-1}(p)$ is a line for every $p \in M$.  This set is called the \emph{total space} of the line bundle $L$ and its points are equivalence classes of triples $[(p,\alpha,z)]$. For every $\alpha$ we get a map $\phi_\alpha\colon\pi^{-1}(U_\alpha) \to U_\alpha \times \C$ by setting $\phi_\alpha([p,\alpha,z]) = (p,z)$ which is a linear isomorphism when restricted to a fiber. See  Figure \ref{fig: linebundle2}. 
 
\begin{exc}
 Show that a map $\sigma\colon M \to L$ such that $\pi(\sigma(p)) = p$ for all $p \in M$ is the same thing as a section in item 2 in the Definition above.
\end{exc}
\begin{expl}
On any manifold $M$ we have the trivial line bundle $L = M \times \C$, given by covering $M$ by the simple open set $U=M$. A connection in this case is the same thing as a 1-form $\theta$ on $M$, and its curvature is $F_\nabla = id\theta$. 
\end{expl}
We then have that for any function $f \in C^\infty(M)$, the operator 
\begin{equation}
P_f = -i\hbar\nabla_{X_f} + \widehat{f}
\end{equation}
is well-defined when acting on sections of $L$, and satisfies Q4) by construction. It is also clear that $Pf$ satifies Q1) and Q2), but for Q3) (and in fact any Hilbert-space structure on $\Gamma(L)$) we need the additional data of a compatible \emph{hermitian structure} on $L$, i.e. pairing $\langle\cdot,\cdot\rangle_L \colon \Gamma(L) \times \Gamma(L)\to C^\infty(M)$ that is hermitian and sesquilinear over $C^\infty(M)$ such that for all vector fields $X$ on $M$ we have 
\begin{equation}
X\langle \sigma_1,\sigma_2\rangle_L = \langle\nabla_X\sigma_1,\sigma_2\rangle_L + \langle \sigma_1,\nabla_X\sigma_2\rangle_L. 
\end{equation}
We then have an inner product on $\Gamma(L)$ given by $\langle\sigma_1,\sigma_2\rangle = \int_M\langle\sigma_1,\sigma_2\rangle_L\epsilon$ and we can define the prequantization Hilbert space 
\begin{equation} 
\mathcal{H}_L^{pre} = L^2(L,\langle\cdot,\cdot\rangle),
\end{equation} i.e. the (completion of) the space of square-integrable sections of $L$.
\begin{exc}
Show that $P_f$ satisfies Q3 when acting on $\mathcal{H}_L$. 
\end{exc}
We have therefore arrived at the following result. Suppose that there exists a hermitian line bundle $L$ with a compatible connection $\nabla$ with curvature $\omega/\hbar$ (such a bundle is called a \emph{prequantum line bundle} for $(M,\omega)$). Then we can define a prequantization map $P \colon C^\infty(M) \to \mathcal{L}(\mathcal{H})$ by 
\begin{equation}
Pf = -i\hbar\nabla_{X_f} + \widehat{f}
\end{equation}
satisfying the Dirac conditions Q1) - Q4). Notice that this map is defined on \emph{all} functions! 

\subsection{Existence of prequantum line bundles}
\begin{figure}
\centering
\includegraphics[scale=1]{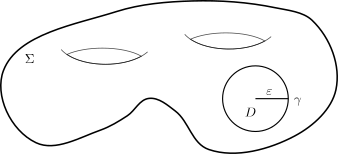}
\caption{Removing a small disk from a surface $\Sigma \subset M$. }\label{fig: surface}
\end{figure}

Our first construction of the prequantum map was possible only for exact symplectic manifolds. How much did we gain by changing our viewpoint to line bundles? At the first glance, it could seem that we can always construct such line bundles, since we can always find a cover $U_\alpha$ such that $\omega\big|_{U_\alpha} = d\theta_\alpha$ is exact. However, notice that we have the condition \eqref{eq:integral cech} which is a restriction on the local primitives $\theta_\alpha$ and transition functions $g_{\alpha\beta}$. But what does this tell us? \\ 
To answer this question we introduce some other useful terminology of line bundles. Let $\gamma\colon [0,1] \to M$ be a curve in $M$, then we say that a section $\sigma$ is parallel with respect to $\nabla$ if $\nabla_{\dot{\gamma}}\sigma = 0$. If $\gamma([0,1]) \subset U_\alpha$, then the previous equation can be written explictly as 
\begin{equation}
\frac{d}{dt}\sigma_\alpha(\gamma(t)) = \frac{i}{\hbar}(\theta_\alpha)_{\gamma(t)}(\dot{\gamma}(t))\sigma_\alpha(t). \label{eq: par trs}
\end{equation}
Given  $\sigma_\alpha(\gamma(0))$, we can solve equation \eqref{eq: par trs} by 
\begin{equation}
\sigma_\alpha(\gamma(t)) = \exp\left(\frac{i}{\hbar}\int_0^t(\theta_\alpha)_{\gamma(t)}(\dot{\gamma}(s))ds\right)\sigma_\alpha(\gamma(0)).
\end{equation}
In particular if $\sigma_\alpha(\gamma(0)) \neq 0 $ we have $\sigma_\alpha(\gamma(1))\sigma_\alpha(\gamma(0))^{-1} = \exp(i/\hbar\int_\gamma\theta_\alpha) =: P_\alpha $. 
\begin{exc}
\begin{enumerate}
 \item Denote by $L_p := \pi^{-1}(p)$ the fiber of the line bundle $L$ over $p \in M$. Check that we can define a map $P_\gamma \colon L_{\gamma(0)} \to L_{\gamma(1)}$ by $[(\gamma(0),\alpha,z)] \mapsto [(\gamma(1),\alpha,P_\alpha z]$ .
\item If $\gamma$ is not contained in a single $U_\alpha$, define a similar map $P_\gamma\colon L_{\gamma(0)} \to L_{\gamma(1)}$ by dividing $\gamma$ in several curve which are contained in some $U_\alpha$ each, and generalizing the above procedure to this case. $P_\gamma$ is called the \emph{parallel transport} along $\gamma$.
\end{enumerate}
\end{exc}  
In particular, if $\gamma$ is a circle, then $P_\gamma$ is simply a nonzero complex number (equal to $\exp \int_\gamma \frac{i}{\hbar}\theta_\alpha$ if $\gamma \subset U_\alpha$). If $\gamma$ is the boundary of a compact 2-dimensional submanifold $\Sigma \subset M$, then we have the result 
\begin{equation}
P_\gamma = \exp\left(\int_\Sigma \frac{i}{\hbar}\omega\right)
\end{equation}
that follows from Stokes' theorem if $\Sigma \subset U_\alpha$.\footnote{If the general case we can divide $\Sigma$ into pieces that lie in a $U_\alpha$, and notice that when applying Stokes theorem contributions along the inner edges cancel out.} Now, let $\Sigma$ be any compact 2-dimensional submanifold of $M$. Removing a disk $D$ of radius $\varepsilon$ with boundary $\gamma$ from $\Sigma$ (see Figure \ref{fig: surface}) we obtain 
\begin{equation}
\exp\left(\int_{\Sigma - D}\frac{i}{\hbar}\omega\right) = P_\gamma = \exp\left(\int_D \frac{i}{\hbar}\omega\right).
\end{equation}
Letting the radius of the disk tend to 0, we get 
\begin{equation}
\exp\left(\int_{\Sigma}\frac{i}{\hbar}\omega\right) = \exp 0 =  1\notag
\end{equation}
and therefore conclude that 
\begin{equation}
\int_\Sigma \omega \in 2\pi\hbar\mathbb{Z}\label{eq: integrality}
\end{equation}
for all closed 2-dimensional submanifolds of $M$. 

For completeness, we also give a short answer using the theory of Cech cohomology (see for instance \cite{Bott1982}). The quantity $\phi_{\alpha\beta\gamma}$ defines what is known as a Cech 2-cocycle with coefficients in the constant sheaf $2\pi\hbar \underline{\mathbb{Z}}$, and therefore defines an element $[\phi_{\alpha\beta\gamma}] \in H^2(M,2\pi \hbar\mathbb{Z})$. But from the Cech-de Rham isomorphism we know that $[\phi_{\alpha\beta\gamma}]$ equals the first Chern class  of the line bundle $[\omega]$, and therefore $[\omega] \in H^2(M,\R)$ has to define an \emph{integral} cohomology class, i.e. lie in the image of the map $i_*\colon H^2(M,2\pi \hbar \mathbb{Z} \to H^2(M,\R)$. - which is equivalent to \eqref{eq: integrality}. \marginpar{fix}
Overall, we have arrived at the following result. 
\begin{thm}[Weil integrality condition]
Let $(M,\omega)$ be a symplectic manifold. Then there exists a prequantum line bundle if and only if for every closed 2-dimensional submanifold of $M$ we have 
\begin{equation}
\int_\Sigma \omega \in 2\pi\hbar\mathbb{Z}, \label{eq: weil integrality}
\end{equation}
i.e. $[\omega] \in H^2(M,2\pi\hbar\mathbb{Z})$. 
\end{thm}
In passing we note that it is easy to construct examples of symplectic manifolds that do not admit a prequantization, even if we allow ourselves to rescale the symplectic form. For instance, we can consider the product $S^2 \times S^2$ with symplectic form $\hbar(\pi_1^*\omega_{FS} \times \pi^*_2\omega_{FS})$, where $\omega_{FS}$ denotes the Fubini-study form from example \ref{ex: 2 sphere Kaehler}, and $\lambda \in \R \setminus \mathbb{Q}$. Then no rescaling of $\omega$ will define an integral cohomology class. 
\subsection{Different choices of prequantum line bundle}\label{sec: prequantum choices}
There were two problems with our original prequantization map $P$ defined in Subsection \ref{sec: pre quant I }. The first one was that we were only able to define it for exact symplectic forms. We have solved this problem partially by allowing ourselves to let our operators act on sections of nontrivial line bundles, and thereby finding the maximal class of symplectic forms for which such a prequantization exists. The other problem was that our construction depended on the choice of a primitive of the symplectic form, which we now recognize as the choice of a line bundle $L$ with a connection $\nabla$ with a fixed curvature $F_\nabla = \omega/\hbar$. So it is now a natural question, given a symplectic manifold $(M,\omega)$, how many different prequantum line bundles are there, and what are they? \\
One possibility is to have \emph{isomorphic line bundles with connection}. Namely, suppose we are given two line bundles with connection\footnote{We can assume that both line bundles are using the same trivializing cover $U_\alpha$ of $M$. Otherwise, we simply pass to the intersection of the two trivializing covers. } $(L,\nabla) = (U_\alpha,g_{\alpha\beta},\theta_\alpha)$ and $(L',\nabla') = (U_\alpha, g'_{\alpha\beta},\theta'_\alpha)$. We say they are isomorphic if there are maps $\tau_\alpha \colon U_\alpha \to U(1)$  such that $g_{\alpha\beta} = \tau_\beta^{-1}g'_{\alpha\beta}\tau_\alpha$ and $\theta'_\alpha = \theta_\alpha + \tau_\alpha^{-1}d\tau_\alpha$. We obviously have $F_{\nabla} = F_{\nabla'}$. Since we here are using isomorphic choices, it would be good if the result was also isomorphic. This is indeed the case: Namely, the map 
\begin{align}
\Phi\colon \mathcal{H}^{pre}_L &\to \mathcal{H}^{pre}_{L'} \notag\\
\sigma_\alpha &\mapsto \tau_\alpha\sigma_\alpha   \label{eq:uny iso}
\end{align}
is unitary and, for all $f{\in} C^\infty(M)$, intertwines the actions of $P_f$ and $P'_f$, i.e. $\Phi((P_f)\sigma) = (P'_f)(\Phi\sigma)$. 
\begin{exc}
Verify the claims in the sentence above, i.e. 
\begin{enumerate}
\item $\Phi$ maps sections to sections (i.e. $\tau_\alpha\sigma_\alpha$ has the correct transformation property)
\item $\Phi$ is unitary, i.e. $\langle \sigma_1,\sigma_2\rangle_L = \langle \Phi\sigma_1,\Phi\sigma_2\rangle$ 
\item $\Phi$ intertwines $P_f$ and $P'_f$, i.e. $\Phi((P_f)\sigma) = (P'_f)(\Phi\sigma)$.
\end{enumerate}
\end{exc} 
To summarise, if we replace our prequantum line bundle by an isomorphic one, then the corresponding prequantizations are unitarily equivalent. The question is therefore, how many \emph{isomorphism classes} of line bundles with connection are there with curvature $\omega/\hbar$? To describe the answer, one can introduce the \emph{tensor product} of line bundles with connection: Given $(L,\nabla)=  (U_\alpha,g_{\alpha\beta},\theta_\alpha)$ and $(L',\nabla') = (U_\alpha, g'_{\alpha\beta},\theta'_\alpha)$, their tensor product is 
\begin{equation}
 (L\otimes L', \nabla + \nabla') = (U_\alpha, g_{\alpha\beta} g'_{\alpha\beta}, \theta_\alpha + \theta'_\alpha).  \label{eq: tensorproduct}
\end{equation}
 \begin{exc}
 Show that $L \otimes L'$ is a line bundle and that its curvature is $F_{\nabla + \nabla'} = F_\nabla + F_{\nabla'}$. 
 \end{exc}
 In particular, any two line bundles with curvature $\omega/\hbar$ are related by tensoring with a \emph{flat line bundle}, i.e. a line bundle with connection with zero curvature. So we can classify all possible prequantum line bundles on $(M,\omega)$ by classifying flat line bundles on $M$ - note that this is independent of $\omega$: I.e. if $(M,\omega)$ is prequantizable then isomorphism classes of prequantum line bundles are in one-to-one correspondence with flat line bundles on $M$.\footnote{In slightly fancier language, we can note that flat line bundles form a group $Pic_{flat}$ under the tensor product (exercise, the inverse of $L = (U_\alpha,g_{\alpha\beta},\theta_\alpha)$ is given by the dual bundle $L = (U_\alpha,g_{\alpha\beta}^{-1},-\theta_\alpha)$) and isomorphism classes of prequantum line bundles form a \emph{torsor} over this group, i.e. $Pic_{flat}$ acts on it freely and transitively.} \\
 
On the \emph{trivial} line bundle defined by $U_\alpha = M$ and $g_{\alpha\beta} =1$, a flat connection is the same as a closed 1-from $\theta \in \Omega^1(M,\R)$. However, some of those 1-forms define isomorphic bundles with connection: Whenever $\theta - \theta' = T^{-1} dT $ for some globally defined function $T \colon M \to S^1$, the line bundles with connection $(M,1,\theta)$ and $(M,1,\theta')$ are isomorphic. It turns out that 1-forms of the form $\alpha = T^{-1}dT$ satisfy $\int_\gamma \alpha = 2\pi i k$, for every circle $\gamma \in M$. In particular, isomorphism classes of flat connections on the trivial line bundle are given by $H^1(M,\R)/H^1(M,2\pi i \mathbb{Z}) \cong (S^1)^{\dim H^1(M,\mathbb{R}})$. This means that we will have a space of inequivalent quantizations parametrized by a product of circles. In the physics literature those are sometimes known as ``vaccuum angles''. See section \ref{sec: cotangent} for an example.\\
\subsubsection{Digression on classification of all flat $U(1)$-bundles and prequantizations}
However, there can be also \emph{nontrivial} flat line bundles. For the interested reader versed in algebraic topology, let me again use a little more tools to explain what is going on (this part can be safely skipped for the remainder of the text). let us denote the set of isomorphism classes of flat $U(1)$-bundles by $MFC(M,U(1))$. For \emph{flat} $U(1)$-bundles the holonomy map is invariant under homotopies of paths, that is, a flat $U(1)$-bundle $(L,\nabla)$ defines a map $hol_{(L,\nabla)}\colon \pi_1(M) \to U(1)$, and it turns out that we get in this way an isomorphism \begin{equation}
hol\colon MFC \to \operatorname{Hom}(\pi_1(M),U(1)) \cong \operatorname{Hom}(H_1(M),U(1)) \cong H^1(M,U(1))
\end{equation}
where in the first isomorphism we have used $H_1(M) \cong \pi_1(M)/[\pi_1(M),\pi_1(M)]$ and the fact that a map to an abelian group vanishes on the commutator, and the second isomorphism we have used the universal coefficient theorem (together with the fact that $U(1)$ is divisible). To compute the latter cohomology group, wee have a short exact\footnote{I.e. a sequence of maps where the kernel of every map is the image of the previous one.} sequence of abelian groups 
\begin{equation}
0 \xrightarrow{\iota} \mathbb{Z} \xrightarrow{\exp} \R \to U(1) \to 0
\end{equation}
translating to a short exact sequence of constant sheaves on $M$, and therefore to a long exact sequence in cohomology 
\begin{equation}
\ldots \to H^1(M,\mathbb{Z}) \to H^1(M,\mathbb{R}) \to H^1(M,U(1)) \xrightarrow{\delta} H^2(M,\mathbb{Z}) \xrightarrow{\iota_*} H^2(M,\mathbb{R}) \to \ldots
\end{equation} 
from which we can extract the short exact sequence 
\begin{equation}
0 \to \frac{H^1(M,\mathbb{R})}{H^1(M,\mathbb{Z})} \to H^1(M,U(1)) \xrightarrow{\delta} \ker\iota_*  \to 0. 
\end{equation}
The connecting homomorphism $\delta$ gives exactly the (integral) Chern class of a line bundle which classifies its topological type. We can split the short exact sequence by choosing a section of  $\delta$ - i.e. a flat line bundle $(L_\phi,\nabla_\phi)$ for every class $\phi \in \ker \iota_* H^2(M,\mathbb{Z})$ (those are known as torsion classes). Every flat $U(1)$-bundle is then isomorphic to a bundle of the form 
$$(L,\nabla) = (L_\phi,\nabla_\phi + \theta)$$ 
with $\theta \in \frac{H^1(M,\mathbb{R})}{H^1(M,\mathbb{Z})}$. 
\section{Lecture 4: Quantization}\label{sec: quantization}
Having at length discussed prequantization -- i.e. the construction, given a prequantum line bundle $L$ on $M$, of a Hilbert space $\mathcal{H}^{pre}_L$ and a map $P \colon C^\infty(M) \to \mathcal{L}(\mathcal{H}^{pre}_L)$ -- we can see that from a mathematical perspective it was nice and rather simple. To summarize, given an arbitrary symplectic manifold $(M,\omega)$, we have seen 
\begin{itemize}
\item that a prequantization exists if and only if $\omega/\hbar$ satisfies the integrality condition \eqref{eq: weil integrality}, 
\item that if $\omega$ satisfies this condition, what are the choices involved in the prequantization (a hermitian line bundle with connection) and classified the choices up to isomorphism in terms of purely topological information of $M$, 
\item and, last but not least, given our choices, an explicit formula for the prequantization map.
\end{itemize}
If all we had been interested in was solving this mathematical problem, then we could end these lectures here and be very satisfied with ourselves. Alas, we are trying to solve a problem from physics, and reality and first physical principles are telling us that our work is not done here, because the Hilbert space that we constructed in prequantization is simply too big. As we saw in Example \ref{ex:std prequant Rn}, it is a sense exactly twice too big - so we expect that, loosely speaking, we will have to reduce the number of variables by a factor by a half. This is easy enough in $\R^{2n}$, where we have global coordinates at hand, but getting rid of coordinates was precisely one of the reason to embark on the journey of geometric quantization in the first place. So how are we to proceed? \\
\subsection{Polarizations}
To understand the notion of polarization we require the concept of a \emph{vector bundle}. We briefly introduce this notion.  For more detail, see \cite[Section 8]{Cattaneo2018} or \cite[Section 2.1.4]{Nicolaescu2020}.
\subsubsection{A primer on vector bundles}
By $\mathbb{K}$, we mean either $\R$ or $\C$. For a manifold $M$, a  \emph{$\mathbb{K}$-vector bundle over $M$} is a manifold $E$ together with a surjective submersion $\pi \colon E \to M$ and a $\mathbb{K}$-vector space structure on $\pi^{-1}(\{p\})$ for every $p \in M$, such that for every point $p \in M$ there is a neighbourhood $U \subset M$ and a diffeomorphism $\phi_U \colon \pi^{-1}(U) \to U \times \mathbb{K}^n$ such that 
\begin{equation}
\begin{tikzcd}
\pi^{-1}(U) \arrow[rd, "\pi"] \arrow[rr,"\phi_U"] &   & U\times \mathbb{K}^n \arrow[ld, "\mathrm{pr}_1"'] \\
                                  & U &                                
\end{tikzcd}\label{eq: def vect bdl}
\end{equation} 
commutes and the restriction to $\phi_U$ to $\pi^{-1}(\{x\})$ is a linear isomorphism from $\pi^{-1}(\{x\})$ to $\R^n$. If $U$, $V$ are two such neighbourhoods, then we have $\phi_V \circ \phi_U^{-1} (x, v) = (x, g_{UV}(x))$, for a smooth map $g_{UV}\colon U \cap V \to GL(n,\mathbb{K})$ called \emph{transition functions}. It is then obvious that the $g_{UV}$ satisfy, for all $x \in U \cap V$, 
\begin{align}
g_{UV}(x)g_{VU}(x) &= \mathrm{id} \\
g_{WU}(x)g_{VW}(x)g_{UV}(x) &= \mathrm{id}.
\end{align}
Conversely, one can construct a vector bundle over $M$ by choosing an open cover $U_\alpha$ of $M$ and specifying the transition functions $g_{\alpha\beta}$ on overlaps $U_\alpha \cap U_\beta$, c.f. the definition of line bundles (cf. Definition \ref{def: line bundle} and Figure \ref{fig: linebundle2}.\footnote{In fact if I would try to draw a picture explaining this definition it would look precisely the same as in the line bundle case (Fig. \ref{fig: linebundle2}) due to the annoying fact that screens only have two dimensions.}). The integer $n$ is called the \emph{rank} of the vector bundle, $E$ is called the \emph{total space} and $M$ the \emph{base}. A pair $(U,\phi_U)$ as above is called a \emph{local trivialization}, and a cover $U_\alpha$ of $M$ by local trivializations is called a trivializing cover. 
\begin{expl}
\begin{itemize}
\item The tangent bundle $TM$ of any manifold $M$ is a real vector bundle of rank $\dim M$. Any coordinate chart is a local trivialization and the transition functions are given by $g_{\alpha\beta}(x) = d\phi_{\alpha\beta}(\phi_\alpha(x))$ with $\phi_{\alpha\beta}$ the corresponding coordinate change. 
\item Line bundles are the same as rank 1 vector bundles. 
\item Over any manifold $M$ we have the trivial rank $n$ bundles $E = M \times \mathbb{K}^n$, with $\pi$ the projection to the first factor. 
\end{itemize}
\end{expl} Similarly to the case of line bundles, a section $\sigma$ of a vector bundle is a map $\sigma \colon M \to E$ such that $\pi \circ \sigma = \mathrm{id}_M$. With respect to a trivializing cover $U_\alpha$, a section is given by a collection of maps $\sigma_\alpha(x) = \phi_\alpha \circ \sigma $ such that $\sigma_\beta(x) = g_{\alpha\beta}\sigma_\alpha(x)$.   If We can apply any functorial construction on vector spaces to vector bundles, by applying the functor to all fibers and the maps $\phi_U$ in \eqref{eq: def vect bdl}. For instance, if $E,F$ are vector bundles, then we have the vector bundles $E^*, \mathrm{Sym}^k E, \wedge^k E, \mathrm{Hom}(E,F), E\otimes F$ (here $\mathrm{Sym}$ denotes the symmetric product and $\wedge^k$ the $k$-th exterior power of a vector space). The following is an exercise in unraveling the definitions. 
\begin{exc}
Convince yourself of the following facts. 
\begin{enumerate}
\item The dual of the tangent bundle is the cotangent bundle: $(TM)^* = T^*M$. 
\item Sections of the tangent bundle are vector fields and sections of the cotangent bundles are 1-forms. 
\item Differential $k$-forms are sections of $\wedge^k T^*M$.
\item Type $(r,s)$ tensors are sections of $(TM)^{\otimes r} \otimes (T^*M)^{\otimes s}$.
\end{enumerate}
\end{exc}
\subsubsection{The idea}
The solution to the problem of having reducible Hilbert space proposed in geometric quantization is that of introducing a \emph{polarization} of our manifold $M$ - roughly speaking, we are trying to select one half of the states in our Hilbert space. While this choice is absolutely necessary to make contact with results from physics, it introduces a number of additional problems -- both of technical and conceptual nature -- which will force us to work with (classes of) examples instead of the general case. Let us discuss this procedure a bit more before entering into the technicalities. The idea to half the number of states is to select, at every point $p$ of our manifold $M$, half of the directions in the tangent space $T_pM$ and asking the sections of the line bundle that is comprising our Hilbert space to be constant along these directions. Let us denote the space of those directions by $\mathcal{P}_p$, it forms a half-dimensional subspace of $T_pM$. The challenge is to do this consistently for all points $p \in M$. The first requirement is to ensure that the subspace $\mathcal{P}_p$ varies smoothly with $p$ -- this is ensured by asking that $\mathcal{P} = \sqcup_{p \in M} \mathcal{P}_p$ is a smooth subbundle of $TM$.\footnote{This means that for every $p \in M$ there exists a neighbourhood $U\subset M$ restricted to which $\mathcal{P}$ is spanned by smooth vector fields $X_1,\ldots X_n$ on $U$.} The vector fields tangent to this bundle are denoted $\Gamma(\mathcal{P})$. We now want to define the Hilbert space as those sections $\sigma$ of $L$ that satisfy, for all $X \in \Gamma(P)$,
\begin{equation}
\nabla_X\sigma = 0 .\label{eq:covar const}
\end{equation}
This equation is called the \emph{polarization condition}.
But then we also must have 
\begin{equation}
0 =[\nabla_X,\nabla_Y]\sigma = \nabla_{[X,Y]}\sigma - \frac{i}{\hbar}\omega(X,Y)\sigma
\end{equation}
where we have used that the curvature of $\nabla$ is $\omega/\hbar$. 
It is safest to require that those two terms vanish individually. By the polarization condition, the first term vanishes if $[X,Y] \in \Gamma(P)$. Vanishing of the second term is asking that for each $p \in M$, the subspace $\mathcal{P}_p$ (which contains $X_p$ and $Y_p$) is \emph{isotropic}, i.e. the symplectic form $\omega$ vanishes on it. Since we were also asking $\mathcal{P}_p$ to be half-dimensional, this requirement means that $\mathcal{P}$ is in fact \emph{lagrangian}. We have therefore arrived at two conditions on our bundle $\mathcal{P}$: 
\begin{enumerate}
\item For all $X,Y \in \Gamma(P)$, the Lie bracket $[X,Y]\in \Gamma(P)$ -- we say that $\mathcal{P}$ is \emph{involutive}, 
\item For all $p \in M$, $\mathcal{P}_p \subset T_pM$ is \emph{lagrangian} -- we say $\mathcal{P}$ is \emph{lagrangian}.
\end{enumerate}
It turns out that in many cases, one actually has to generalize the idea mentioned above to include ``complex directions'' in $T_pM$. To this end we introduce the complexified tangent bundle $T_\mathbb{C}M$ which is the complex vector bundle over $M$ whose fiber over $p \in M$ is $(T_\mathbb{C}M)_p = (T_pM)_\mathbb{C} = T_pM \otimes \mathbb{C}$.\footnote{The local trivializations are simply the complex linear extensions of the local trivializations of $TM$.} An element of $v  \in (T_\mathbb{C}M)_p$ is a linear combination $v = v_x + i v_y$ with $v_x,v_y \in T_pM$ and we extend the symplectic form $\omega$ bilinearly as 
\begin{equation}
\omega(v,w) = \omega(v_x,w_x) - \omega(w_x,w_y) + i\omega(v_x,w_y) + i \omega(v_y,w_x).
\end{equation}
We then define a \emph{polarization} $\mathcal{P}$ on $M$ to be an involutive Lagrangian subbundle of $TM_\mathbb{C}$. There is one extra condition usually placed on such a bundle. Namely, for every $p \in M$ we denote 
$$D_p := \mathcal{P} \cap \overline{\mathcal{P}} \cap T_pM \subset T_pM.$$ 
We call this the space of real directions of $\mathcal{P}$ at $p$. It is easy to come up with examples of involutive Lagrangian distributions where the dimension of $D_p$ varies with $p$. Consider for instance the on $M = \mathbb{R}^2$ the bundle spanned by the vector field $v(x,y) = \frac{\partial}{\partial x} + iy \frac{\partial}{\partial y}$. Then $D_{(x,y)} = \{0\}$ unless $y=0$, where $D_{(x,0)} = span(\frac{\partial}{\partial x}).$ To avoid such cases we require that $D_p$ is constant on $M$. i.e. that \begin{equation} D = \sqcup_{p \in  M} D_p = \mathcal{P} \cap \overline{\mathcal{P}}\cap TM
\end{equation} is a subbundle of $TM$. This implies that also
\begin{equation}
E := (\mathcal{P} + \overline{\mathcal{P}})\cap TM = D^\perp
\end{equation} is a subbundle of $TM$.\footnote{This is because pointwise we have $E_p = D_p^\perp = \{v \in T_pM\colon \omega(v,w) = 0 \text{ for all} w \in D_p\}$, which in turn implies that the dimension of $\dim E_p = \dim M - \dim D_p$ is constant if $\dim D_p$ is constant.} The bundles $D$ and $E$ are important information about the polarization $\mathcal{P}$. Two particularly important classes of polarizations are 
\begin{itemize}
\item those for which $\mathcal{P} = \overline{\mathcal{P}}$, in particular $D = E = \mathcal{P}\cap TM$ and $\mathcal{P} = D_\mathbb{C}$ is the complexification of an involutive lagrangian \emph{real} subbundle of $TM$, such polarization are called \emph{real polarizations},
\item those for which $\mathcal{P} \cap \overline{\mathcal{P}} = \emptyset$, i.e. $D = \{0\}$ and $E = TM$, such polarizations are called \emph{K\"ahler polarizations}.
\end{itemize} 
\subsubsection{Strongly admissible polarizations} 
Even if $D$ and $E$ are both (real) subbundles of the (real) tangent bundle, they can still be rather wild. To describe the problems that can possibly arise with them and the assumptions used to avoid those porblems we introduce a little more terminology. \\
An \emph{integral manifold} of a rank $k$ subbundle $\Delta$ of the tangent bundle is a $k$-dimensional submanifold $S$ such that $T_pS = \Delta_p$ for all $p \in S$. A subbundle of the tangent bundle is called integrable if through every point $p \in M$ there is an integral manifold. It is a classical theorem that involutive distributions are integrable. A \emph{leaf} of an integrable distribution is a maximal connected integral manifold. The \emph{space of leaves} of $\Delta$, denoted $M / \Delta$ is the quotient of $M$ by the equivalence relation $x \sim y \Leftrightarrow x,y $ are in the same leaf of $\Delta$. 
\begin{exc}
Consider $\mathbb{R}^2$ together with the vector field $v(x,y) = y\frac{\partial}{\partial x} - x\frac{\partial}{\partial y}$.
\begin{itemize}
\item Show that the span of $v$ defines a distribution $\Delta$ on $\R^2 \setminus \{(0,0)\}$, but not on $\R^2$ - why? 
\item Show that the leaves of $\Delta$ are circles $C_r$ centered at 0. 
\item Show that the space of leaves of $\Delta$ is a smooth manifold diffeomorphic to $\R_{>0}$. 
\end{itemize}
\end{exc}
For instance, we can consider the torus $S^1 \times S^1$ with coordinates $(x,y)$, its tangent bundle is trivial: $T(S^1 \times S^1) \cong S^1 \times S^1 \times \mathbb{R}^2.$ Any 1-dimensional subspace $V$ of $\R^2$ (i.e. a line through 0) defines a subbundle $\mathcal{V}$ of $T(S^1 \times S^1)$ by letting $\mathcal{V}_p \cong V$, and is trivially lagrangian and involutive.  However, if the slope of the line defining $V$ is not rational, then all leaves are dense and the quotient space $S^1 \times S^1/\mathcal{V}$ is not Hausdorff (cf. Exercise \ref{exc: torus}). To avoid these and similar problems, we define what is called \emph{strongly admissible} polarizations. Namely, those are those for which 
\begin{itemize}
\item the subbundle $E$ is also integrable, 
\item the leaf spaces $M/D$ and $M/E$ are smooth hausdorff manifolds, 
\item the quotient map $M/D \to M/E$ is a smooth submersion. 
\end{itemize}
\begin{exc}
Show K\"ahler polarizations are always strongly admissible. 
\end{exc}
In this text we will restrict ourselves to strongly admissible real or K\"ahler polarizations - those are easier to understand and more relevant to the examples we want to study here, and all the interesting ideas are present in this case. On a more fundamental note, it is a general problem of geometric quantization that once one starts dealing with polarizations one has to more or less give up developing a general theory and rely on constructions that work in examples. Certainly more examples that the ones we treat here can (and should be) discussed - we will point out references in the right places. 

\subsubsection{Real polarizations}
The most obvious source of real polarizations are the cotangent bundles $\pi \colon T^*Q \to Q$, with $\mathcal{P}$ the complexification of the subbundle of vertical vector fields $\mathfrak{X}_{ver} = \ker d \pi \subset TM$. This is a strongly admissible polarization with leaf through a point $(q,p) \in T_qQ$ is the cotangent space $T_qQ$. In this case $E = D = \ker d\pi$ and the leaf space can be identified with $Q$. We will call this the \emph{vertical polarization} of a cotangent bundle. 

\begin{figure}[!ht]
\centering
\def\svgwidth{0.3\columnwidth}
\begingroup%
  \makeatletter%
  \providecommand\color[2][]{%
    \errmessage{(Inkscape) Color is used for the text in Inkscape, but the package 'color.sty' is not loaded}%
    \renewcommand\color[2][]{}%
  }%
  \providecommand\transparent[1]{%
    \errmessage{(Inkscape) Transparency is used (non-zero) for the text in Inkscape, but the package 'transparent.sty' is not loaded}%
    \renewcommand\transparent[1]{}%
  }%
  \providecommand\rotatebox[2]{#2}%
  \newcommand*\fsize{\dimexpr\f@size pt\relax}%
  \newcommand*\lineheight[1]{\fontsize{\fsize}{#1\fsize}\selectfont}%
  \ifx\svgwidth\undefined%
    \setlength{\unitlength}{417.46896338bp}%
    \ifx\svgscale\undefined%
      \relax%
    \else%
      \setlength{\unitlength}{\unitlength * \real{\svgscale}}%
    \fi%
  \else%
    \setlength{\unitlength}{\svgwidth}%
  \fi%
  \global\let\svgwidth\undefined%
  \global\let\svgscale\undefined%
  \makeatother%
  \begin{picture}(1,0.73841545)%
    \lineheight{1}%
    \setlength\tabcolsep{0pt}%
    \put(0,0){\includegraphics[width=\unitlength,page=1]{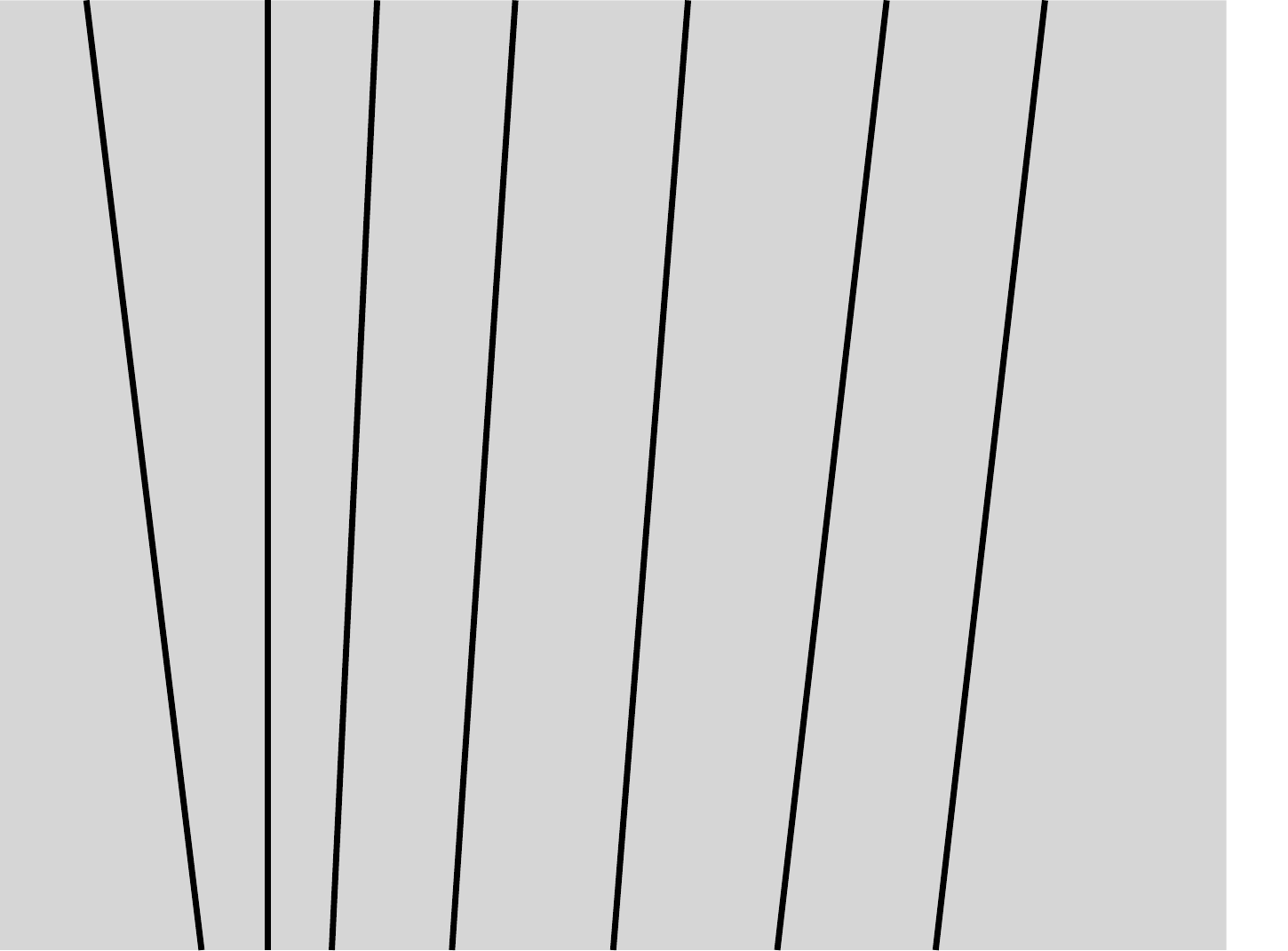}}%
    \put(0.86090042,0.70967297){\color[rgb]{0,0,0}\makebox(0,0)[lt]{\lineheight{1.25}\smash{\begin{tabular}[t]{l}$T^*M$\end{tabular}}}}%
    \put(0.95338018,0.15010367){\color[rgb]{0,0,0}\makebox(0,0)[lt]{\lineheight{1.25}\smash{\begin{tabular}[t]{l}$M$\end{tabular}}}}%
    \put(0,0){\includegraphics[width=\unitlength,page=2]{cotangent.pdf}}%
    \put(0.77430313,0.32599952){\color[rgb]{0,0,0}\makebox(0,0)[lt]{\lineheight{1.25}\smash{\begin{tabular}[t]{l}$p$\end{tabular}}}}%
    \put(0.79902553,0.56816699){\color[rgb]{0,0,0}\makebox(0,0)[lt]{\lineheight{1.25}\smash{\begin{tabular}[t]{l}$\pi^{-1}(p)$\end{tabular}}}}%
  \end{picture}%
\endgroup%

\caption{The vertical polarization of any cotangent has leaves $\pi^{-1}(p)$, $p\in M$, and therefore its leaf space is canonically identified with $M$.}
\end{figure}
However, there are other sources of real polarizations. For instance, on a cotangent bundle one can try to work instead with horizontal vector fields. This would mean choosing at every point $(q,p)\in T^*Q$ a \emph{horizontal subspace} $H_{(q,p)}$ of $T_{(q,p)}T^*M$, i.e. $H_{(q,p)} \oplus \ker d\pi_{(q,p)} = T_{(q,p)}T^*Q$, such that they fit together in a smooth subbundle $H \subset TT^*Q$, which is integrable. This is equivalent to the choice of a flat connection on $T^*Q$, and thus relies on the choice of extra data. It is not a priori clear that such polarizations are strongly admissible, but it is in some examples: the easiest being the ones where the cotangent bundle is trivial $T^*Q = Q \times \R^n$. In this case there is a trivially a horizontal subbundle $TQ \subset TT^*Q = TQ \oplus T\R^n$. The leaf through a point $(q,p)$ is $Q \times \{p\}$ and the space of leaves is identified with $\R^n$.

\begin{figure}[ht]
\begin{subfigure}{0.4\linewidth}
\def\svgwidth{\columnwidth}
\begingroup%
  \makeatletter%
  \providecommand\color[2][]{%
    \errmessage{(Inkscape) Color is used for the text in Inkscape, but the package 'color.sty' is not loaded}%
    \renewcommand\color[2][]{}%
  }%
  \providecommand\transparent[1]{%
    \errmessage{(Inkscape) Transparency is used (non-zero) for the text in Inkscape, but the package 'transparent.sty' is not loaded}%
    \renewcommand\transparent[1]{}%
  }%
  \providecommand\rotatebox[2]{#2}%
  \newcommand*\fsize{\dimexpr\f@size pt\relax}%
  \newcommand*\lineheight[1]{\fontsize{\fsize}{#1\fsize}\selectfont}%
  \ifx\svgwidth\undefined%
    \setlength{\unitlength}{380.75979158bp}%
    \ifx\svgscale\undefined%
      \relax%
    \else%
      \setlength{\unitlength}{\unitlength * \real{\svgscale}}%
    \fi%
  \else%
    \setlength{\unitlength}{\svgwidth}%
  \fi%
  \global\let\svgwidth\undefined%
  \global\let\svgscale\undefined%
  \makeatother%
  \begin{picture}(1,0.86381355)%
    \lineheight{1}%
    \setlength\tabcolsep{0pt}%
    \put(0,0){\includegraphics[width=\unitlength,page=1]{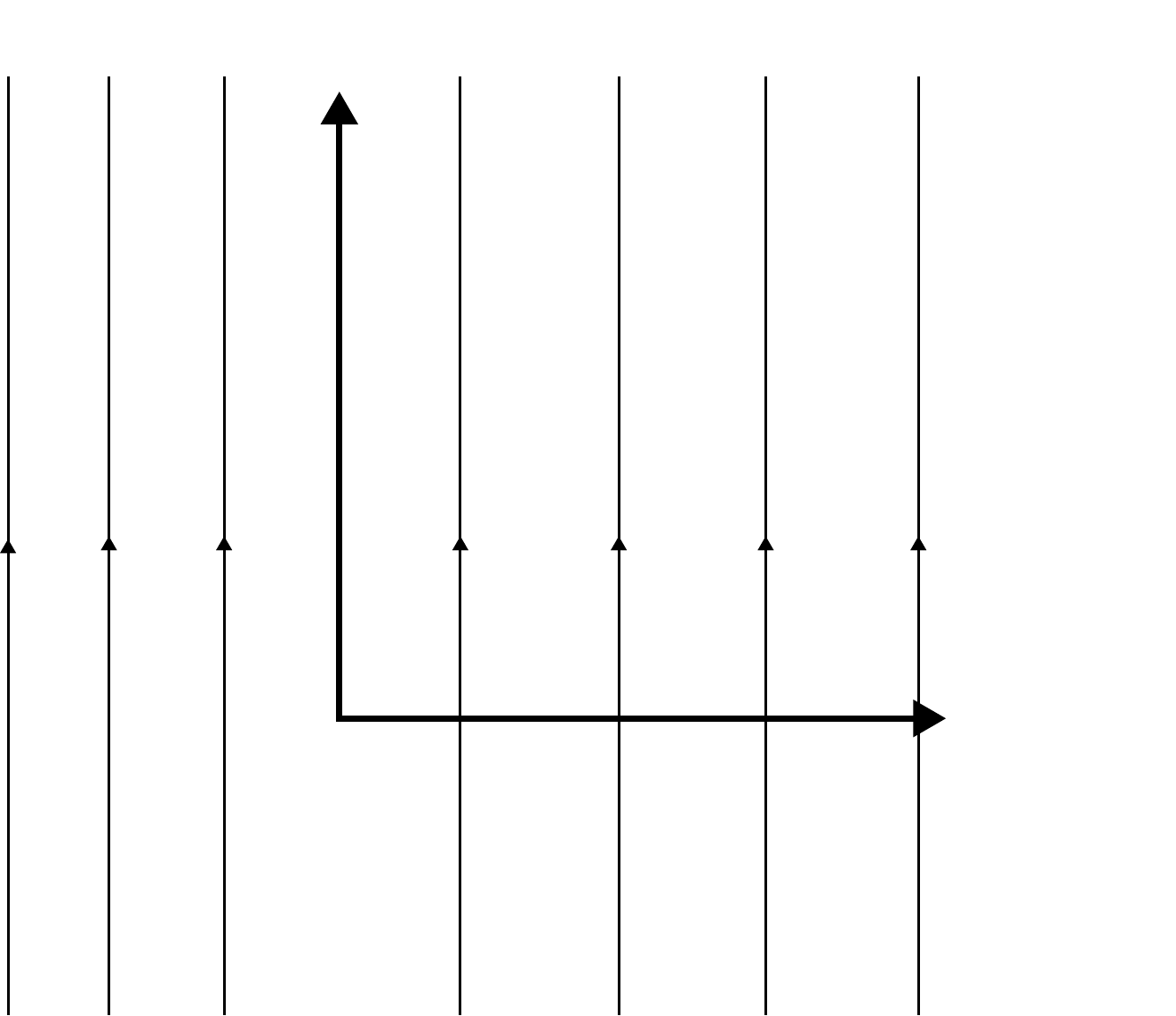}}%
    \put(0.85747712,0.81383369){\color[rgb]{0,0,0}\makebox(0,0)[lt]{\lineheight{1.25}\smash{\begin{tabular}[t]{l}$\R^ {2n}$\end{tabular}}}}%
    \put(0.83296409,0.2230689){\color[rgb]{0,0,0}\makebox(0,0)[lt]{\lineheight{1.25}\smash{\begin{tabular}[t]{l}$q^i$\end{tabular}}}}%
    \put(0.27406614,0.84324938){\color[rgb]{0,0,0}\makebox(0,0)[lt]{\lineheight{1.25}\smash{\begin{tabular}[t]{l}$p_i$\end{tabular}}}}%
  \end{picture}%
\endgroup%

\caption{Vertical polarization of $\R^{2n}$.}\label{fig: vertical}
\end{subfigure}
\qquad
\begin{subfigure}{0.4\linewidth}
\def\svgwidth{\columnwidth}
\begingroup%
  \makeatletter%
  \providecommand\color[2][]{%
    \errmessage{(Inkscape) Color is used for the text in Inkscape, but the package 'color.sty' is not loaded}%
    \renewcommand\color[2][]{}%
  }%
  \providecommand\transparent[1]{%
    \errmessage{(Inkscape) Transparency is used (non-zero) for the text in Inkscape, but the package 'transparent.sty' is not loaded}%
    \renewcommand\transparent[1]{}%
  }%
  \providecommand\rotatebox[2]{#2}%
  \newcommand*\fsize{\dimexpr\f@size pt\relax}%
  \newcommand*\lineheight[1]{\fontsize{\fsize}{#1\fsize}\selectfont}%
  \ifx\svgwidth\undefined%
    \setlength{\unitlength}{368.48261315bp}%
    \ifx\svgscale\undefined%
      \relax%
    \else%
      \setlength{\unitlength}{\unitlength * \real{\svgscale}}%
    \fi%
  \else%
    \setlength{\unitlength}{\svgwidth}%
  \fi%
  \global\let\svgwidth\undefined%
  \global\let\svgscale\undefined%
  \makeatother%
  \begin{picture}(1,0.68783483)%
    \lineheight{1}%
    \setlength\tabcolsep{0pt}%
    \put(0,0){\includegraphics[width=\unitlength,page=1]{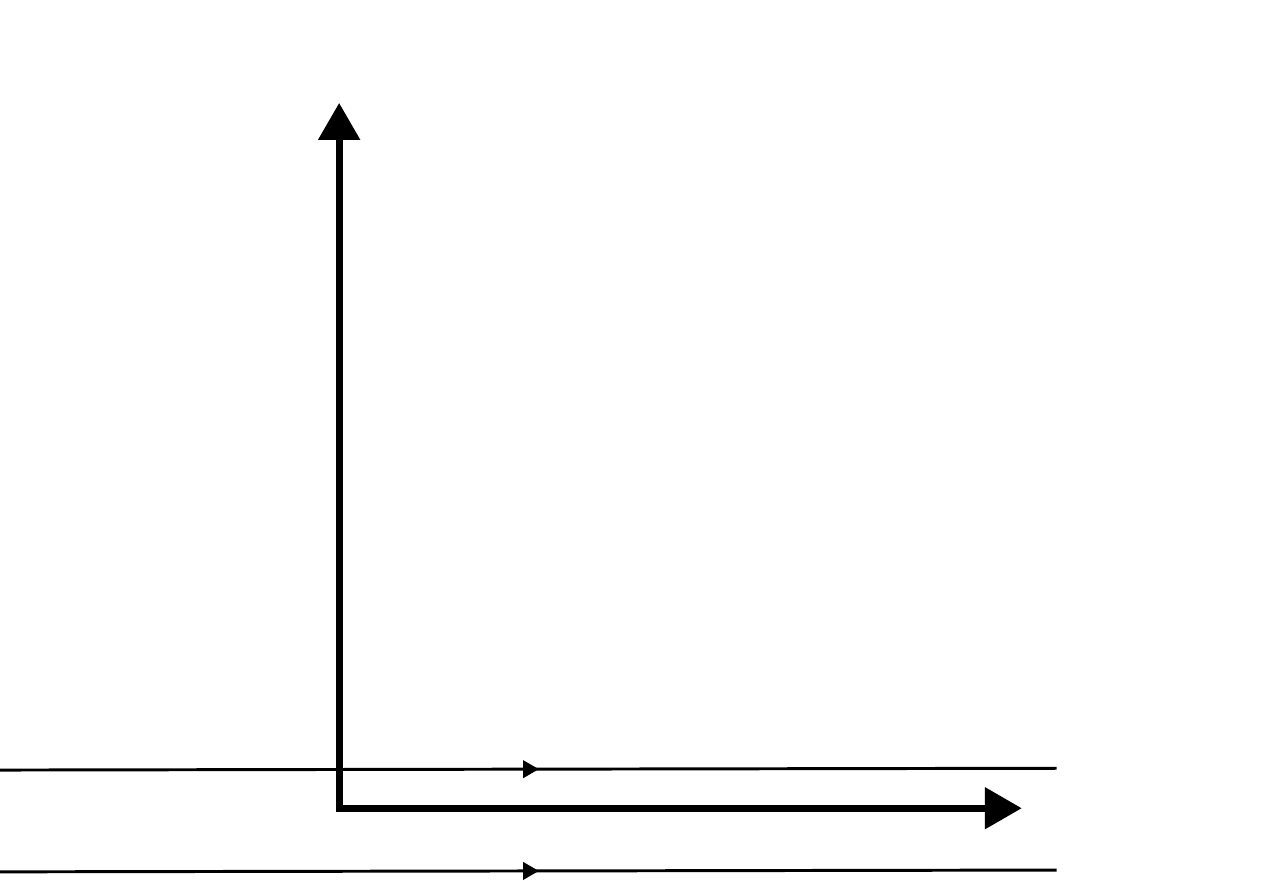}}%
    \put(0.85272849,0.63618972){\color[rgb]{0,0,0}\makebox(0,0)[lt]{\lineheight{1.25}\smash{\begin{tabular}[t]{l}$\R^ {2n}$\end{tabular}}}}%
    \put(0.82739873,0.02574171){\color[rgb]{0,0,0}\makebox(0,0)[lt]{\lineheight{1.25}\smash{\begin{tabular}[t]{l}$q^i$\end{tabular}}}}%
    \put(0.2498793,0.66658549){\color[rgb]{0,0,0}\makebox(0,0)[lt]{\lineheight{1.25}\smash{\begin{tabular}[t]{l}$p_i$\end{tabular}}}}%
    \put(0,0){\includegraphics[width=\unitlength,page=2]{horizontal.pdf}}%
  \end{picture}%
\endgroup%

\caption{Horizontal polarization of $\R^{2n}$.}\label{fig: horizontal}
\end{subfigure}
\caption{$\R^{2n}$ has both a horizontal and a vertical polarization.}
\end{figure}

Another example of a real polarization can be found by considering the torus $S^1 \times S^1$ - any 1-dimensional subbundle of $TM$ is lagrangian and integrable, but not all of them are strongly admissible. 
\begin{exc}\label{exc: torus}
Denote $(t,\theta)$ the coordinates on the torus. Consider the subbundle $\mathcal{P}\subset T(S^1 \times S^1)$ spanned by the vector $\frac{\partial}{\partial t} + \lambda \frac{\partial}{\partial \theta}$, where $\lambda \in \mathbb{R}$ is constant. Show $\mathcal{P}$ is strongly admissible if and only $\lambda \in \mathbb{Q}$.
\end{exc}
\subsubsection{K\"ahler polarizations}
In this text, we restrict ourselves to \emph{positive K\"ahler polarizations},
\footnote{General K\"ahler polarizations induce on $M$ a complex structure which is compatible with the symplectic form only in the weaker sense that $g(\cdot,\cdot) = \omega(\cdot,J\cdot)$ is a non-degenerate bilinear form, but not necessarily positive definite \cite[Section 5]{Woodhouse1997}. We call \emph{positive} K\"ahler polarizations the ones for which the induced complex structure is compatible with the symplectic form  in the sense of Subsection \ref{sec: Kaehler}, i.e. a positive definite symmetric bilinear form.}which are also called \emph{holomorphic polarizations}, and form the nicest class of polarizations. They are constructed in the following way. Let $M$ be a K\"ahler manifold, cf. Section \ref{sec: Kaehler}: A manifold with a symplectic form $\omega$ and a complex structure $J$ such that $g(v,w):= \omega(v,Jw)$ defines a Riemannian metric on $M$ ($J$ is called a compatible complex structure). Since, $J^2 = -1$, we can split the \emph{complexified} tangent spaces as 
\begin{equation}
(T_xM)_\mathbb{C} = \underbrace{T^{(0,1)}_xM}_{=v \in  T_xM,\ Jv = -iv} \oplus \underbrace{T^{(1,0)}_xM}_{=v \in T_xM,\ Jv = iv}.  
\end{equation} 
This is a global splitting into subbundles. From \ref{sec: Kaehler} we know that locally $M$ admits complex coordinates $z^1,\ldots,z^n$ such that 
\begin{equation}\label{eq: Kahler properties}
\begin{split}
\omega &= \sum_{i,j=1}^n \omega_{ij}dz^i \wedge d\bar{z}^j \\
T^{(1,0)}_xM &= span\left(\frac{\partial}{\partial z^1},\ldots, \frac{\partial}{\partial z^n}\right) \\
T^{(0,1)}_xM &= span\left(\frac{\partial}{\partial \bar{z}^1},\ldots, \frac{\partial}{\partial \bar{z}^n}\right) 
\end{split}
\end{equation}
\begin{exc}
Show that equations \eqref{eq: Kahler properties} imply that both $\mathcal{P} = T^{(0,1)}M$ and $\overline{\mathcal{P}}=T^{(1,0)}M$ are integrable lagrangian subbundles of $T_\mathbb{C}M$, and define K\"ahler polarizations of $M$. We call $\mathcal{P}$ the \emph{holomorphic polarization} of $M$.
\end{exc}

\subsection{The Hilbert space}
Once we have a symplectic manifold $(M,\omega)$ together with a prequantum line bundle $(L,\nabla)$ and a polarization $\mathcal{P} \subset T_\C M$ of $M$, we can, in following the idea sketched in the beginning of the draft, try to define our Hilbert space to be given by \emph{polarized sections}, i.e. the sections $\sigma \in \Gamma(L)$ for which $\nabla_X\sigma = 0$ for all $X \in \Gamma(P)$. Unfortunately, it turns out that this definition has a variety of new problems. We start with the case of holomorphic polarizations, where this is not the case.

\subsubsection{The Hilbert space for holomorphic polarizations}
Remember that we defined the prequantum Hilbert space as the space of square-integrable sections of the line bundle $L$. Now, let us look at the \emph{polarized sections}, i.e. those $\sigma \in \Gamma(L)$ satisfying $\nabla_X\sigma = 0$, for all $X \in \Gamma(T^{(0,1)}M)$. Suppose we have two non-vanishing polarized sections $\sigma$ and $\sigma'= f\sigma$, defined over a complex chart $(z^1,\ldots,z^n)$. Then the polarization condition reads
\begin{equation}
0 = \nabla_{\frac{\partial}{\partial \bar{z}^i}}\sigma' = \nabla_{\frac{\partial}{\partial \bar{z}^i}}f\sigma =
\underbrace{\nabla_{\frac{\partial}{\partial \bar{z}^i}}\sigma}_{=0} + \frac{\partial f}{\partial z^i} 0,
\end{equation}
i.e. $f$ is a holomorphic function! Using only polarized sections to define the transition functions of $L$, we obtain a holomorphic structure on $L$, i.e. a cover of $M$ by open sets over which $L$ is trivial such that the transition functions are holomorphic.\footnote{This argument is adapted from \cite{Woodhouse1997}.} Moreover, the polarized sections are precisely the holomorphic sections of this line bundle, i.e. in the trivialization above they correspond to holomorphic functions. It is a classical fact that holomorphic sections on a line bundle form a finite-dimensional Hilbert space if $M$ is compact. This is the Hilbert space of \emph{holomorphic quantization}:
\begin{equation}
\mathcal{H}_L := H^0(L) = \{\sigma \in \Gamma(L), \sigma \text{ is holomorphic }\}.
\end{equation} 
The pairing on this Hilbert space is simply the restriction of the hermitian metric $\langle\cdot,\cdot\rangle$ to the space of holomorphic sections. 
\subsubsection{Problmes with the Hilbert space for real polarizations - cylinder example}
Real polarizations, or more generally any polarizations containing real directions (i.e. $D \neq \{0\}$) are more complicated because they involve the actual geometry of integral submanifolds of $M$. There are two different problems that can arise here, which mean that the subspace of the prequantum Hilbert space consisting of polarized sections will be empty (zero-dimensional) for wide classes of examples.  \\
We will exhibit both problems in the (seemingly) simple example of $T^*S^1 \cong S^1 \times \R$. The fact that these problems arise even in this simple and fundamental example (the phase space of a particle constrained to a circle) tells us that we cannot ignore them and should have a detailed understanding of what is going on. \\
The standard symplectic form on $M= T^*S^1$ is $\omega = dp\wedge d\phi$ which is exact with potential $\theta = p\wedge d\phi$, so that the trivial line bundle $L = M \times \C$ with connection 1-form given by $\theta$ is a prequantum line bundle. The prequantum Hilbert space is simply the space of square-integrable functions on the cylinder. We have to natural polarizations, the vertical one (spanned by $\frac{\partial}{\partial p}$) and the horizontal one (spanned by $\frac{\partial}{\partial \varphi}$). 
\begin{exc}
Show that for $M = T^*S^1$, both the horizontal and the vertical polarization are strongly admissible. 
\end{exc}
Fist, let us look at the vertical polarization. The polarization condition is 
$$\nabla_{\frac{\partial}{\partial p}}f(\phi,p) = \frac{\partial f(\phi,p)}{\partial p} = 0,$$
which means that polarized sections are functions of $\phi$ only. This is what we want, but unfortunately, those functions do not live in $\mathcal{H}^{pre}$, since for any two such functions 
$$\langle f(q),g(q) \rangle = \int_{T^*S^1}f(q)\bar{g}(q) dp \wedge dq = \infty$$
(the integral over $p$ diverges). This means that the subset of square-integrable functions which is polarized only the zero function. In this case, it is clear that we should use instead the measure $dq$ on the space of polarized sections. In general however we need to have a measure on the quotient $M/D$, and it is not quite clear how to obtain it. This problem is present whenever the fibers of the map $\pi \colon M \to M/D$ are non-compact.\\
Next, we consider the horizontal polarization. Now the polarization condition is 
$$ \nabla_{\frac{\partial}{\partial \phi}}f(\phi,p) = \frac{\partial f(\phi,p)}{\partial p} - \frac{i}{\hbar}p\cdot f(\phi,p) = 0.$$
I.e. polarized functions are of the form 
$$f(\phi,p) = e^{\frac{i}{\hbar}p\cdot\phi}g(p).$$
However, as those are sections of the trivial line bundle, they are globally defined functions, and thus we must have 
$f(\phi+2\pi,p) = f(\phi,p)$ which implies 
$$e^{\frac{i}{\hbar}2\pi \cdot p}\cdot g(p) = g(p). $$ 
That is, we have to require $g(p) = 0$ unless $p \in \hbar \mathbb{Z}$ - i.e. the support of $g$ is \emph{discrete}. In the world of square integrable functions, such $g$s are identically zero, because they vanish outside a measure zero subset.  Also here, the answer is essentially clear from physical intuition: in the momentum representation (our states are functions of the momenta) states should be delta functions at integer multiples of $\hbar$. \\
Mathematically, the problem is due to the fact that the polarization has leaves which are not simply connected, i.e. there are leaves containing non-contractible loops. In this case, the holonomy of the connection along $\gamma$, 
$$hol_\gamma(A) = \exp\left(\frac{i}{\hbar}\int_\gamma\theta\right)$$ could be nontrivial (notice that the restriction of the connection to a leaf of the polarization is flat because the leaves are lagrangian, and therefore holonomy along contractible loops is always 1). In particular, the parallel transport along a loop with non-trivial holonomy is also nontrivial. Since polarized section are in particular parallel along any loop contained in a leaf of the polarization, they satisfy 
\begin{equation}
\sigma(p) = hol_\gamma(A)\sigma(p)
\end{equation} for all loops $\gamma$ based at $p$ and contained in the leaf of the polarization through $p$. In particular, $\sigma(p)$ must vanish whenever $hol_\gamma \neq 1$ for such a loop. This recovers the Bohr-Sommerfeld conditions 
\begin{equation}
\int_\gamma \sum p_k dq^k \in 2\pi\hbar  \mathbb{Z}.
\end{equation} which is a condition on points of $M$. Only sections which are supported on the Bohr-Sommerfeld variety are allowed, but in general this variety is discrete and such sections should be understood as distributions (in particular they are not square integrable).  A more modern take on this is that our states should be top cohomology classes of a certain complex. In any way, it is clear from those two problems that we have to ditch our nice prequantum Hilbert space with its nice Liouville measure and try something new. Interestingly, it turns out that we require this redefinition of the Hilbert space even for K\"ahler polarizations if we want to make contact with physics, see the Example in Subsection \ref{sec: quant Rn complex} below. 
\subsubsection{Noncompact leaves and half-forms}
By now it is clear that we have to leave our nice prequantum Hilbert space with its pairing induced form the symplectic volume form behind and look for something new. Instead, we want to construct from our prequantum line bundle $L$ a new line bundle $L_P = L \otimes \delta_P$, \emph{such that the product of two sections is naturally a volume form on the quotient}. There are essentially two ways to go about this: one can either use half-densities or half-forms. The former approach has the advantage that it works independently of the topology of $M$, however it fails to reproduce the correct shift in the energy spectrum of the harmonic oscillator, and therefore we present here the second version, called the bundle of half-forms. We follow here the elegant and concise presentation in \cite{Blau1992}, but adapt what is written there to the general case. 
First we consider, for any polarization $P$ (not necessarily real) its \emph{annihilator} 
\begin{equation}
P^0 = \{\alpha \in T^*M_\mathbb{C}, \alpha|_P = 0\}.
\end{equation}
Sections of this bundle are 1-forms which vanish identically when evaluated on vector fields belonging to $P$, like $P$, its rank is $n$. Its top exterior power is called the \emph{canonical bundle of $P$} and denoted 
\begin{equation}
K_P:=\bigwedge^nP^0
\end{equation}
A \emph{half-form} bundle is by definition a \emph{square root} of $K_P$, i.e. a line bundle $\delta_P$ with the property that 
\begin{equation}
\delta_P \otimes \delta_P = K_P.
\end{equation}
Such square roots may not exist, and, when they exist, they might not be unique, depending on the topological properties of $M$. The choice of such a bundle should be considered an additional piece of data we are choosing for geometric quantization. 
From now we will assume that we have fixed such a square root bundle $\delta_P$. \\
We consider the projection $\pi \colon M \to M/D =: Q$, then we can pull back the complex line bundle $\det Q = \bigwedge^{\dim Q}_\mathbb{C}$ to $M$ via $\pi$. There is a pairing 
\begin{equation}
(\cdot,\cdot)_{\delta_P}\colon\delta_{\overline{P}}\otimes \delta_{P} \to \pi^*\det Q
\end{equation} which we will describe for the real ($P=\overline{P} =D$) and K\"ahler $(P \cap \overline{P} = \{0\} =D)$ cases (for the general case we refer to \cite[Section 10.3]{Woodhouse1997}). In the real case, this is simply the observation that we can identify $K_P \cong \pi^*\det Q$ (they are the subbundles of $\bigwedge^nT^*M_\mathbb{C}$ such that $\iota_X\omega =0 $ for all $X \in \Gamma(P)$) and therefore the pairing is simply the map $\delta_P \otimes \delta_P \to K_P$. In the complex case, notice that $D =\{0\}$ and $Q = M$. Therefore sections of $\det Q$ are $2n$-forms on $M$, but the natural map $\delta_{\overline{P}} \otimes \delta_P$ only gives us a $n$-form. In this case the pairing is given by multiplying with the square root of the symplectic volume form: 
\begin{equation}
(\psi,\psi')_{\delta_P} = \sqrt{\varepsilon}\otimes\psi\otimes\psi' \in \Gamma(\wedge^{2n}T^*M_\mathbb{C})
\end{equation} 
We now want to extend the connection $\nabla$ on our prequantum line bundle $L$ to the new line bundle $L \otimes \delta_P$ to obtain a new version of the polarization condition. We do this as follows. Given a section $\mu \in \Gamma(K_P)$ and a vector field $X \in \Gamma(TM)$, we have the Lie derivative $L_X\mu$, which in general is just an $n$-form on $M$. When is this $n$-form again a section of $K_P$? Sections of $K_P$ are characterized by the fact that $\iota_{X_P}\mu = 0$ for all vector fields $X_P \in \Gamma(P)$ tangent to the polarization $P$. From the standard identity $[L_X,\iota_Y] = \iota_{[X,Y]}$ we get 
\begin{equation}
\iota_{X_P}L_X\mu = L_X\underbrace{\iota_{X_P}\mu}_{=0} + \iota_{[X,X_P]}\mu = \iota_{[X,X_P]}\mu
\end{equation}
which vanishes if and only if $[X,X_P] \in \Gamma(P)$. We call such vector fields \emph{polarization-preserving}.\footnote{They are equivalently characterized by the fact that their flow leaves the polarization invariant, and will play an important role in quantization later}. Thus, for polarization-preserving vector fields the Lie derivative acts on sections of the canonical bundle $K_P$, and therefore also sections on $\delta_P$ via $L_X\nu^2 = 2\nu L_X\nu$. When we restrict this action to vector fields $X$ tangent to $P$, we get 
\begin{equation}
L_X\mu = \underbrace{d\iota_X\mu}_{=0} + \iota_Xd\mu = \iota_Xd\mu =: \nabla^{\delta_P}_X\mu
\end{equation}
and $\nabla^{\delta_P}$ behaves like a connection, but only when evaluated on vector fields tangent to $P$.\footnote{For this reason it is often called a \emph{partial connection}.} The next bit of terminology is important so we emphasize it:
\begin{defn}\label{def: P wave fct}
Suppose we have a symplectic manifold $(M,\omega)$ with prequantum line bundle $(L,\nabla,\langle\cdot,\cdot\rangle)$, a polarization $P$ and a half-form bundle $\delta_P$. Let $L_P = L \otimes \delta_P$. A section $\tilde{\sigma} = \sigma \otimes \psi \in \Gamma(L_P)$ is called a \emph{$P$ wave function} if for all $X \in \Gamma(P)$
\begin{equation}
(\nabla_X + \nabla_X^{\delta_P})(\tilde{s}) = \nabla _X\sigma \otimes \psi + \sigma \otimes \nabla_X^{\delta_P}\psi =0.
\end{equation}
\end{defn}
In other words, the $P$ wave functions are precisely the polarized sections of $L_P$. Suppose that $\tilde{\sigma}_1 = \sigma_1 \otimes \psi_1, \tilde{\sigma}_2 = \sigma_2 \otimes \psi_2$ are $P$, wave functions, then $\nabla_X\sigma_i = \nabla^{\delta_P}\psi_i = 0$, and because of the identities 
\begin{align*}
L_X\langle \sigma_1,\sigma_2\rangle_L &= \langle \nabla_{\bar{X}}\sigma_1,\sigma_2\rangle_L + \langle \sigma_1,\nabla_X\sigma_2\rangle =0 \\ 
L_X(\psi_1,\psi_2) &= (\nabla_{\bar{X}}^{\delta_P}\psi_1,\psi_2) + (\psi_1,\nabla_X^{\delta_P}\psi_2)=0
\end{align*}
(which are the compatibility equations between the pairings and the connections), the quantity 
\begin{equation}
\langle\tilde{\sigma}_1,\tilde{\sigma}_2\rangle_{L_P}:= \langle\sigma_1,\sigma_2\rangle(\psi_1,\psi_2)\label{eq: pairing P wave}
\end{equation}
is \emph{invariant} under the flow of any \emph{real} vector field $X \in \Gamma(D)$, and therefore defines a volume form on $Q = M/D$. We are finally ready to give the general definition of the Hilbert space for a polarization with simply connected leaves. 
\begin{defn}\label{def: Hilb space half forms}
With the notation of Definition \ref{def: P wave fct}, we define the Hilbert space $\mathcal{H}:=\mathcal{H}(M,L,P)$ to be the completion of the space of $P$-wave functions $\tilde{\sigma}$ with respect to the inner product
\begin{equation}
\langle\tilde{\sigma}_1,\tilde{\sigma}_2\rangle_{\mathcal{H}}:=\int_{Q}\langle\tilde{\sigma}_1,\tilde{\sigma}_2\rangle_{L_P}.
\end{equation} 
\end{defn}
We will discuss plenty of examples in Section \ref{sec: examples}.
\subsubsection{Non-simply connected leaves and cohomological wave functions}
The introduction of half-forms takes care of the fact that there might exist no \emph{square-integrable} polarized sections. However, as we have seen above, there might not be any polarized sections because of the non-trivial holonomy of the connection $\nabla$ along the leaves of $P$. In fact, the support of any polarized section has to be contained in the Bohr-Sommerfeld variety $S \subset M$ which is the set of points $s \in M$ such that $hol_\gamma \nabla = 1$ for any loop $\gamma$ lying in a leaf of the polarization through $s$. If the leaves of $P$ are not simply connected, then this is generically a subset that would be irrelevant once we pass to equivalence classes of square integrable sections. One can instead work with sections which are distributions, but I think a slightly more modern viewpoint is to work with \emph{cohomological wave functions}. \\ 
It is a standard notion to obtain from a flat connection $\nabla$ on a vector bundle $E$ a complex of forms 
\begin{equation}
\Omega^\bullet(M,E) = \Gamma(\wedge^\bullet T^*M \otimes E).
\end{equation}
The differential is defined in the following way: over a trivializing neighbourhood $U$ of $V$, sections of $\wedge^\bullet T^*M \otimes E$ are sums of sections of the form $\omega \otimes \sigma$, with $\omega$ a differential form on $U$ and $\sigma \in \Gamma(U)$. We then define the differential $d^\nabla$ on such sections by 
\begin{equation}
d^\nabla(\omega \otimes \sigma) = d\omega \otimes \sigma + \omega \wedge \nabla\sigma
\end{equation}
where $\nabla\sigma \in  \Gamma(T^*U\otimes E)$ is the 1-form with values in $E$ whose value on a vector field $X$ is $\nabla_X\sigma$. The properties of connetions ensure this is well-defined and extends to a global differential. The flatness condition ensures that $(d^\nabla)^2 = 0$.
We would like to apply this idea to the line bundle $L_P$ with the connection $\nabla_{L_P} = \nabla^L + \nabla^{\delta_P}$, but we have two problems: 
\begin{itemize}
\item $\nabla^L$ is not flat,
\item $\nabla^{\delta_P}$ is only partially defined (on vector fields tangent to $P$).
\end{itemize}
Curiously, we can solve these problems both at the same time by restricting the differential forms to $P$. That is we say that $\omega_P \in \Omega^k(M,L_P)$ is $P$-closed if 
\begin{equation}
(d^{\nabla^{L_P}}\omega_P)\bigg|_P = 0.
\end{equation}
Similarly, we say that $\omega_P$ is $P$-exact if there is $\alpha_P$ such that $(\omega_P - d^{\nabla^{L_P}}\alpha)\bigg|_P = 0$.
Notice that $(d^{\nabla^{L_P}})^2\omega_P\bigg|_P = 0$, in particular $P$-exact forms are $P$-closed and we can form the usual cohomology groups 
\begin{equation}
H^k(M,P,L_P):= \frac{\Omega_{P-closed}^k(M,P,L_P)}{\Omega_{P-exact}^k(M,P,L_P)}
\end{equation}
whose elements are called \emph{degree $k$ cohomological wave functions}.
We leave some remarks as an exercise. 
\begin{exc}
\begin{itemize}
\item Show that any form vanishing on $P$ is automatically $P$-exact. 
\item Show that $H^0(M,P,L_P)$ is precisely the space of $P$-wave functions. 
\end{itemize}
\end{exc} 
The last point shows us that $H^\bullet(M,P,L_P)$ is a natural generalization of $P$ wave functions. For $k > 0$ it is in general not possible to define a inner product to get an honest Hilbert space, but we will see in examples how to give it a Hilbert space structure. For completeness, we mention here the following interesting results for real polarizations obtained by Sniatycki \cite{Sniatycki1977}. Namely, under some slight additional assumptions on $P$ one has that the leaves are all of the form $(S^1)^k \times\R^{n-k}$, with fundamental group $\mathbb{Z}^k$. Then $H^{m}(M,P,L_P) = \{0\}$ for $m \neq k$. Under an additional orientability assumption, $H^k(M,P,L_P) \cong S_F(S)$, where 
$S_F(S)$ denotes the polarized sections of $L_P$ restricted to the Bohr-Sommerfeld variety $S$. Again, we defer examples to Section \ref{sec: examples}.
 \subsection{Quantization}
Recall that our original goal was that, given a symplectic manifold $(M,\omega)$, to construct a Hilbert space $\mathcal{H}$ and a map $Q\colon C^\infty(M) \to \mathcal{L}(\mathcal{H})$ satisfying the Dirac axioms Q1) - Q5) explained in Definition \ref{def: quantization}. This is a good point to summarize our findings. 
\begin{itemize}
\item In Section \ref{sec: prequantization} we showed that we can define a Hilbert space $\mathcal{H}^{pre}_L$ and a prequantization map $P\colon C^\infty(M) \to \mathcal{H}^{pre}_L$ satisfying Q1) - Q4) if and only if $\omega$ satisfies the Weil integrality condition \ref{eq: weil integrality}. In this case, $\mathcal{H}^{pre}_L = L^2(L)$ was the space of square-integrable sections of a line bundle $L$ with curvature $\omega$, and 
\begin{equation}
Pf = -i\hbar\nabla_{X_f} + \widehat{f}.
\end{equation}
\item In a first attempt to satisfy Q5), we picked a polarization $P \subset TM_\C$ and tried to define the Hilbert space as the space of \emph{polarized} square-integrable sections of $L$, i.e. those sections that satisfy $\nabla_X\sigma = 0$ for all $X \in \Gamma(P)$. However, it turned out that in many cases (actually for all polarizations containing real directions) this space was often empty. To rectify this, we introduced the bundle of half-forms $\delta_P$ of $P$ and defined the Hilbert space to be, in the case where all the leaves are simply connected,
\begin{equation}
\mathcal{H} = L^2(\Gamma_P(L_P)),
\end{equation}
the space of polarized sections of $L_P$ (called $P$ wave functions), with square-integrable with respect to the pairing \ref{eq: pairing P wave}. 
\item Finally, it turns out that in the case of non-simply connected leaves we have to instead use the cohomolgical wave functions 
\begin{equation}
\widehat{\mathcal{H}}:=H^\bullet(M,P,L_P).
\end{equation}
\end{itemize}
There is now another obvious problem, we had a nice and satisfying formula for the prequantization map $P$, but then we changed the space on which the operators $Pf$ were supposed to act - so do we have to start over and find a completely new formula? \\ 
Luckily, the answer turns out to be no, we just have to use the new ingredients we have at our disposal. However, we have to accept that the space of functions we can quantize becomes quite a bit smaller. 
\subsubsection{Quantizable functions and the quantization map}
Setting aside integrability issues for the moment, let us look first at polarized sections of $L$. For a function $f \in C^\infty(M)$, does $P_f$ still define an operator on polarized sections? To obtain an operator on polarized sections, it is necessary that if $\nabla_X\sigma = 0$, for all $X \in \Gamma(P)$, then also 
$\nabla_X(P_f(\sigma)) = 0$. Let us try to check this in a computation: 
\begin{align*}
\nabla_X(P_f(\sigma)) &= \nabla_X(-i\hbar\nabla_{X_f}\sigma + f\sigma) \\ 
&= -i\hbar\left(\nabla_{X_f}\underbrace{\nabla_X\sigma}_{=0} - \nabla_{[X,X_f]}\sigma - \frac{i}{\hbar}\underbrace{\omega(X,X_f)}_{=X(f)}\sigma\right) + f\underbrace{\nabla_X\sigma}_{=0} + X(f)\sigma \\ 
&=-i\hbar\nabla_{[X,X_f]}\sigma.
\end{align*}
where we used that the curvature of $\nabla$ is $\omega/\hbar$, the Leibniz rule for connections and the definition of hamiltonian vector fields. That is, $f$ preserves polarized sections if and only if $[X,X_f] \in \Gamma(P)$, i.e. the hamiltonian vector fields of $f$ is polarization-preserving! 
\begin{exc}
Show that the functions whose vector fields are polarization-preserving form a Lie subalgebra $C^\infty_P(M) \subset C^\infty(M)$ with respect to the Poisson bracket. 
\end{exc}
Curiously, the condition that $X_f$ be polarization-preserving means we the Lie derivative with respect to $X_f$ maps the canonical bundle $K_P$ of $P$, and thereform also the half-form bundle $\delta_P$, to itself. This means that if 
$\tilde{\sigma} = \sigma \otimes \psi$ is a section of $L_P = L \otimes \delta_P$, then we can define the operator \begin{equation}
Q_f(\tilde{\sigma}) = P_f(\sigma) \otimes \psi + \sigma \otimes (-i\hbar L_{X_f}\psi).
\end{equation} Again, we need to check that polarized sections are mapped to themselves under this operator. The polarized sections are those which satisfy, for all $X \in \Gamma(P)$,
\begin{equation}
\nabla_X\sigma = 0 = \nabla^{\delta_P}_X\psi. 
\end{equation}
We only need to check that $\nabla^{\delta_P}_XL_{X_f}\psi = 0$, as we already know that $\nabla_X Pf(\sigma) = 0$. 
\begin{exc} Recall that $\nabla^{\delta_P}\psi = 0$ implies $\nabla^{\delta_P}\psi^2 = \iota_X d\psi^2 = 0$ ($\psi^2$ is just a regular differential form on $M$). Now, show that this implies $\iota_X d L_{X_f}\psi^2 = 0$, if $[X,X_f] = 0$, and therefore also $\nabla^{\delta_P}_XL_{X_f}\psi = 0$. 
\end{exc}
This means that for $f \in C^\infty(P)$, the operator $Qf$ maps the space of polarized sections of $L_P$ to itself. With a little more work, one can show this operator is unitary for the pairing \eqref{eq: pairing P wave} and therefore extends to the completion of the space of square-integrable sections. Finally, we can extend the quantization map to act on differential forms with values in $L_P$. Locally those are of the form $\tau_P = \tau \otimes \tilde{\sigma}$, and then we define
\begin{equation}
\widehat{Q}f (\tau\otimes \tilde{\sigma}) := -i\hbar L_{X_f}\tau \otimes \tilde{\sigma} + \tau \otimes Qf(\tilde{\sigma}). 
\end{equation}
More explicitly, if $\tau_P = \tau \otimes \sigma \otimes \psi$, with $\tau$ a differential form, $\sigma \in \Gamma(L), \psi \in \Gamma(\delta_P)$, the quantization map is 
\begin{align}
\widehat{Q}f (\tau \otimes \sigma \otimes \psi) &= -i\hbar(L_{X_f}\tau \otimes \sigma \otimes \psi + \tau\otimes\nabla_{X_f}\sigma\otimes \psi + \tau\otimes \sigma \otimes L_{X_f}\psi) + f(\tau \otimes \sigma \otimes \psi) \\
&:= -i\hbar \mathsf{L}_{X_f}\tau_P + f\cdot \tau_P \label{eq: coh quant easy}
\end{align}
where we have introduced the twisted Lie derivative 
\begin{equation}
L_X\tau_P = L_{X}\tau \otimes \sigma \otimes \psi + \tau\otimes\nabla_{X}\sigma\otimes \psi + \tau\otimes \sigma \otimes L_{X}\psi
\end{equation}
Then we have that $$\left[d^{\nabla_{L_P}},\mathsf{L}_X\right](\tau\otimes\sigma\otimes\psi) = \tau \wedge \iota_X\omega \otimes \sigma \otimes \psi, $$ which implies that $d^{\nabla_{L_P}}$ commutes with $\widehat{Q}f$ if $X_f$ preserves the polarization and therefore $\widehat{Q}f$ acts on $H^\bullet(M,P,L_P)$. 
Finally, another computation that we skip here 
$$[\mathsf{L}_X,\mathsf{L}_Y] = \mathsf{L}_{[X,Y]} -i\hbar\widehat{\omega(X,Y)}$$
which implies the quantization property (Q3)
\begin{equation}
\left[\widehat{Q}f,\widehat{Q}g\right] = -i\hbar\widehat{Q}_{{f,g}}.
\end{equation}
Furthermore, from equation \eqref{eq: coh quant easy} it is obvious that $\widehat{Q}_1 = 1$. 
We summarize our findings in the following theorem. 
\begin{thm}
Suppose we have a symplectic manifold $(M,\omega)$, together with a prequantum line bundle $(L,\nabla)$, a polarization $P$ and a bundle of half-forms $\delta_P$. Denote $\widehat{\mathcal{H}} = H^\bullet(M,P,L_P)$ and $C^\infty(M)_P$ the functions whose hamiltonian vector preserve the polarization.
Then, the map $\widehat{Q}\colon C^\infty(M)_P \to \mathcal{L}(\widehat{H})$ defined by \eqref{eq: coh quant easy} satisfies the quantization conditions Q1), Q2), and Q4). Moreover, $\mathcal{H} = H^0(M,P,L_P)$, which coincides with the space of smooth polarized sections of $L_P$, has a pairing given by \eqref{eq: pairing P wave}, and the map 
\begin{equation}
Q \colon C^\infty(M)_P \to \mathcal{H}, \qquad f \mapsto Q_f
\end{equation} 
satsifies Q1) - Q4) with respect to this pairing. 
\end{thm}
Now, it is of course a very good question to ask: What about Q5)? After all, violation of Q5) was the original reason we set out on the daunting task of trying to make sense of the ``Hilbert space'' of polarized sections, only to encounter a variety of problems which forced us to seriously complicate matters. So, one would hope that we end with a quantization map that satisfies Q5)? \\ 
Unfortunately, there seems to be no general result available to this end. As Woodhouse remarks \cite[Section 9]{Woodhouse1997}: 
\begin{quote}
``It should be stressed [\ldots ] that the physical justification is not based on general mathematical results (such as the Borel-Weil theorem), but on the way in which the construction works in examples.'' 
\end{quote}
With this in mind, we turn towards examples in the next Section, but before that we just very briefly discuss the idea of BKS kernels to enlarge the space of quantizable functions again. 
\subsubsection{BKS kernels, idea} \label{sec: BKS idea}
One of the many problems we encountered when we started using polarizations was that we had to restrict to functions which preserve the polarization $P$ of choice. In turns out that this is quite limiting. For instance, we can consider $\R^2 = {(q,p)}$ with the vertical polarization spanned by $X=\frac{\partial}{\partial p}$, which the hamiltonian vector field of the function $q$. Then $\{f,q\} = \frac{\partial f}{\partial p}$, and we have 
$$[X_f,X] = X_{\{f,q\}} =  \frac{\partial^2 f}{\partial q\partial p}\frac{\partial}{\partial p} - \frac{\partial^2 f}{\partial p^2} \frac{\partial}{\partial q}$$
which is tangent to $P$ again if and only if $\frac{\partial^2 f}{\partial p^2} \equiv 0$, that is, the only quantizable functions are those which are at most linear in momenta! That is, not even the hamiltonian of the harmonic oscillator, $p^2 + q^2$, is quantizable in this sense. The idea of Blattner-Kostant-Sternberg (BKS) kernels to save this is as follows. \\ 
First, one has to analyze what happens to polarized sections when acting with the quantum operator of a function that is not polarization preserving. It turns out this is quite complicated and we will only sketch some results here. Suppose $f$ is a function whose hamiltonian vector field $X_f$ is complete (i.e. the associated flow $\phi^t_f$ exists for all $t \in \R$). In the general case, $\phi^t_f$ maps the polarization $P$ to a polarization $P_t$ (if $P_m$ is spanned by $v_1,\ldots,v_n$, then $(P_t)_m)$ is spanned by $d\phi^t_fv_1,\ldots, d\phi^t_fv_n$). Recall that the sections defining our Hilbert space are of the form $\tilde{\sigma} = \sigma \otimes \psi$, where $\psi \in \Gamma(L)$ is a polarized section of our prequantum line bundle and $\psi \in \Gamma(\delta_P)$ is a section of the half-form bundle associated to $P$. Using the connection on the prequantum line bundle, one can lift the hamiltonian flow to act on sections of $L$ (see e.g. \cite[Section 3]{Sniatycki1980}), and also to a map $\phi^t_f\colon \Gamma(\delta_P)\to \Gamma(\delta_{P_t})$.\footnote{See \cite[Section 6.1]{Sniatycki1980}. This requires the choice of metaplectic structure on $M$: The metaplectic group $Mp(2n)$ is a double cover of the symplectic group $Sp(2n)$, i.e. there is a (smooth) group homomorphism $\rho\colon Mp(2n) \to Sp(2n)$. A metaplectic structure on $M$ is a principal $Mp(2n)$-bundle such that its associated vector bundle (via the map $\rho$ and the inclusion $Sp(2n) \subset GL(2n)$ is the tangent bundle $TM$.  } One can thus define a map $\phi^t_f \colon \Gamma(L_P) \to \Gamma(L_{P_t})$ which sends $P$ wave to $P_t$ wave functions. Finally, one can rewrite the action of the quantum operator $Q$ associated to a \emph{quantizable} function $f$ as 
\begin{equation}
Qf \tilde{\sigma} = i\hbar \frac{d}{dt}\bigg|_{t=0}\phi^t_f\tilde{\sigma},
\end{equation}
where $\phi^t_f$ denotes the local flow of the Hamiltonian vector field of $X_f$.\footnote{In this section we are following the presentation in \cite{Sniatycki1980} (e.q. Eq. (3.35), (6.29)). In other sources (e.g. \cite{Blau1992}, eq. (4.16) ) this equation comes with a minus sign, but the quantum operators are the same} If $f$ is quantizable, then $X_f$ is polarization preserving, and this gives a well-defined map $Qf \colon \mathcal{H}_P \to \mathcal{H}_{P}$. \\

Now, suppose that we have two different polarizations $P,P'$ and we assume that we have Hilbert spaces $\mathcal{H}_P,\mathcal{H}_{P'}$ given by polarized sections of $L_P$. Then, a BKS kernel is a sesquilinear map $K_{PP'} \colon \mathcal{H}_{P} \times \mathcal{H}_{P'} \to \C$. It induces a linear map $U_{PP'_t}\colon \mathcal{H}_{P'} \to \mathcal{H}_{P}$ with the property $K(\sigma,\sigma') = \langle \sigma,U\sigma'\rangle_{\mathcal{H}}$. \footnote{One can construct the BKS kernel in a fairly general setting if one has a metaplectic structure on $M$. Given a metaplectic structure on $M$, one obtains a half-form bundle for every polarization $P$, varying smoothly with the polarization, which allows one to construct the corresponding kernel. See \cite{Sniatycki1980}.} In good cases, this map is unitary (but in the general case this is far from guaranteed).  Also, assume that we have a family of BKS kernels $K_{PP_t}$ and that the corresponding operators $U_{P_t}\colon \mathcal{H}_{P_t} \to \mathcal{H}_P$ are unitary. Then, we can  define the quantum operator of $f$ by 
\begin{equation}
Qf \tilde{\sigma} = i\hbar \frac{d}{dt}\bigg|_{t=0} U_{P_t} (\phi^t_f\tilde{\sigma}).
\end{equation} 
We will continue the example of $f(q,p)=p^2$ below in subsection \ref{sec: BKS ex}.
\section{Lecture 5: Examples} \label{sec: examples}
\subsection{$\R^{2n}$} 
A very important example is the case of $M = \R^{2n}$ together with the standard symplectic structure $\omega = \sum_{i=1}^n dp_i\wedge dq^i$. Of course quantization here was understood in different ways by physicists much earlier, but it provides an important conceptual check for the methods we have developed in the past two chapters. 
\subsubsection{Prequantization}
The symplectic form is exact and the standard primitive is given by 
\begin{equation}
\theta_{std} = \sum_{i=1}^n p_idq^i
\end{equation}
Any line bundle on $\R^n$ is trivial, so we will take as our prequantum line bundle $L = M \times \C$ with (global) connection 1-form, so that sections of $L$ are-simply complex-valued functions on $\R^{2n}$. The hermitian structure is simply the (inner) product on fibers on $L$, so the prequantum Hilbert space is $\mathcal{H} ^{pre}_L =L^2(M,\C)$ with inner product 
\begin{equation}
\langle f ,g\rangle =  \frac{1}{(2\pi\hbar)^n}\int_{\R^n} \overline{f}\cdot g \ dp_1\ldots dp_n dq^1 \ldots dq^n.
\end{equation}
The prequantization map assigns to a function $f$ the operator 
\begin{equation}
Pf = -i\hbar\nabla_{X_f} + \widehat{f} = -i\hbar X_f + \widehat{p\{f,q\}} + \widehat{f}
\end{equation}
Some examples are 
\begin{align}
Pq &= -i\hbar\frac{\partial}{\partial p} + \hat{q} \notag\\
Pp &= -i\hbar\frac{\partial}{\partial q} \notag  \\
Pp^2 &= -i\hbar2p \frac{\partial}{\partial q} -  \widehat{p^2}. \label{eq: prequant p2}
\end{align}
\begin{exc}
\begin{itemize}
\item Repeat those steps for the symplectic potential $\theta'  = \sum_i -i q^i dp_i$. 
\item Notice that $\theta' = \theta - df$ with $f = (\sum_i q_ip^i)$. Then check explicitly that multiplication with $e^{\frac{i}{\hbar}f}$ is a unitary isomorphism from $\mathcal{H} ^{pre}_L$ to itself, intertwining the different prequantizations, i.e. $$P_{std}g = \widehat{e^{\frac{i}{\hbar}f}}\circ P'g \circ \widehat{e^{\frac{i}{\hbar}f}}, $$ for all $g \in C^\infty(M)$. 
\end{itemize}
\end{exc}

\subsubsection{Quantization in real polarizations}
We consider first the vertical polarization $P_{vert}$, spanned by the vector fields $P_i = \frac{\partial}{\partial p_i}$ inside $TM_\mathbb{C}$. In this case, the polarized sections of $L$ are constant in all $p$ directions, and are in particular never square-integrable. However, notice that the leaves of this polarization are simply connected.  The cotangent bundle $T^*\R^{2n}$ of $\R^{2n}$ is trivial, and decomposes as $$T^*{\R^{2n}} = \R^{2n} \times (\R^n)^*_q \times (\R^n)^*_p,$$ here subscripts $q,p$ indicate the coordinates dual space corresponds to.\footnote{This means that $(\R^n)^*_q$ is spanned by $dq^1,\ldots, dq^n$, and similarly for $(\R^n)^*_p$.} 
The canonical bundle of $P$ is $$K_P = \R^{2n} \times \wedge^\bullet (\R^n)_q^*$$ and its sections are of the form $ f(q,p) \cdot dq^1\wedge \ldots \wedge dq^n =: f(q,p)(dq)^n.$ A section of the half-form bundle $\delta_P$ is then of the form $f(q,p)\sqrt{(dq)^n}$. Since $L$ is the trivial bundle we have $L_P \cong \delta_P$. Polarized sections are of the form $f(q)\sqrt{(dq)^n}$, and the Hilbert space is given by square-integrable complex-valued functions on $\R^n$ with the pairing 
\begin{equation}
\left\langle f(q)\sqrt{(dq)^n}\,g(q)\sqrt{(dq)^n}\right\rangle = 
\int_{\R^n}\overline{f(q)}\ g(q)\ (dq)^n.
\end{equation}
That is, the Hilbert space is exactly as we expected previously. We leave the small generalization of the example discussed in subsection \ref{sec: BKS idea} to the reader:
\begin{exc}
Show that the quantizable functions are of the form $f(q,p) = f_0(q) + f_1^i(q)p_i$. 
\end{exc}  
From the half-form factor we will get a new contribution to the quantum operators. Let us analyse the effect of acting on $\sqrt{(dq)^n}$ by a quantum operator. Namely, we have 
\begin{align*} Qf (\sqrt{(dq)^n} &= -i\hbar  L_{X_f}\sqrt{(dq)^n} = i\hbar\frac12 \sqrt{(dq)^n}^{-1}L_{X_f}(dq)^n  \\
&=  i\hbar\frac12 \sqrt{(dq)^n}^{-1}d\iota_{X_f}(dq)^n  
\end{align*}
If $f$ is quantizable, then the coefficient of the $\frac{\partial}{\partial q^i}$ in $X_f$ is $-f_1^i(q)$ and thus 
$$ d \iota_{X_f} (dq)^n = -\sum_i\frac{\partial f^i_1}{\partial q^i}(dq)^n $$ 
with the result that $$Qf \sqrt{(dq)^n} = -i\hbar\frac12 \sum_i \frac{\partial f_i}{\partial q^i}(\sqrt{(dq)^n} = -i\hbar \frac 12 \mathrm{div} f_1 $$
where we are thinking of the collection $(f_1^1,\ldots,f_n^1)$ as defining a map $f_1\colon \R^n \to \R^n$.
Also notice that if $f$ is not quantizable the result will not be proportional to $\sqrt{(dq)^n}$.  Let us also explicitly compute the action of $Pf$ on polarized sections of $L$ (which are just functions of $q$: Namely, we have 
\begin{align} 
Pf \sigma &= -i\hbar \nabla_{X_f} \sigma + f \sigma  \notag \\
&= -i\hbar \sum f^i_1(q)\frac{\partial \sigma}{\partial q^i} - p_i f^i_1(q)\sigma + (f_0(q) + f^i_1p_i)\sigma \notag\\
&=  -i\hbar \sum f^i_1(q)\frac{\partial \sigma}{\partial q^i} + f_0\sigma. 
\end{align} Overall, the quantum operator acting on an arbitrary section $\tilde{\sigma} = \sigma \otimes \sqrt{(dq)^n} $ is 
\begin{align}
Qf \tilde{\sigma} &= -i\hbar f^i_1\frac{\partial\sigma}{\partial q^i}\otimes \sqrt{(dq)^n}- \frac{i\hbar}{2}\mathrm div f \tilde\sigma + f_0\tilde{\sigma}.
\end{align} 
Next, we turn to the horizontal polarization, spanned by $\frac{\partial}{\partial q^i}$. Here, polarized sections of $L$ are given by sections $\sigma(p,q)$ satisfying 
\begin{equation}
\frac{\partial \sigma}{\partial q^i} = \frac{i}{\hbar}p_i\sigma
\end{equation}
which are of the form 
\begin{equation}
\sigma(q,p) = f(p)e^{\frac{i}{\hbar}\sum_i p_iq^i}
\end{equation}
The corresponding half-forms are of the form $\sqrt{(dp)^n}$. The appearance of the phase factor is because our symplectic potential $\theta = \sum p_idq^i$ was not adapted to the horizontal polarization, i.e. it does not vanish on the vectors spanning it. However, we can change our symplectic potential to $\theta' = -\sum_i q^idp_i = \theta - d(\sum_{i}p_iq^i)$ and by doing so, the phase factor disappears and we see that the Hilbert space is isomorphic to square-integrable functions of the momenta $p_i$. 
\subsubsection{The 1d harmonic oscillator: An example of quantization using BKS kernels}\label{sec: BKS ex}
We now sketch a computation of a quantization of an operator which is quadratic in the momenta, using the method of BKS kernels. 

We now consider the case $n=1$ and the function $f = p^2$.  Then the hamiltonian vector field of $f$ is $X_f = p\frac{\partial}{\partial q}$ and its flow is given by 
$$\phi^t_f(q,p) = (q + tp, p)$$ 
with differential 
$$(d\phi^t_f)_{(q,p)} = \begin{pmatrix}
1 & t \\ 
0 & 1
\end{pmatrix}
$$
The polarization $P_t$ is spanned by $(d\phi^t_f) \frac{\partial}{\partial p} = t\frac{\partial}{\partial q} + \frac{\partial}{\partial p}$.
\begin{figure}
\def\svgwidth{\columnwidth}
\begingroup%
  \makeatletter%
  \providecommand\color[2][]{%
    \errmessage{(Inkscape) Color is used for the text in Inkscape, but the package 'color.sty' is not loaded}%
    \renewcommand\color[2][]{}%
  }%
  \providecommand\transparent[1]{%
    \errmessage{(Inkscape) Transparency is used (non-zero) for the text in Inkscape, but the package 'transparent.sty' is not loaded}%
    \renewcommand\transparent[1]{}%
  }%
  \providecommand\rotatebox[2]{#2}%
  \newcommand*\fsize{\dimexpr\f@size pt\relax}%
  \newcommand*\lineheight[1]{\fontsize{\fsize}{#1\fsize}\selectfont}%
  \ifx\svgwidth\undefined%
    \setlength{\unitlength}{439.57263376bp}%
    \ifx\svgscale\undefined%
      \relax%
    \else%
      \setlength{\unitlength}{\unitlength * \real{\svgscale}}%
    \fi%
  \else%
    \setlength{\unitlength}{\svgwidth}%
  \fi%
  \global\let\svgwidth\undefined%
  \global\let\svgscale\undefined%
  \makeatother%
  \begin{picture}(1,0.93632814)%
    \lineheight{1}%
    \setlength\tabcolsep{0pt}%
    \put(0,0){\includegraphics[width=\unitlength,page=1]{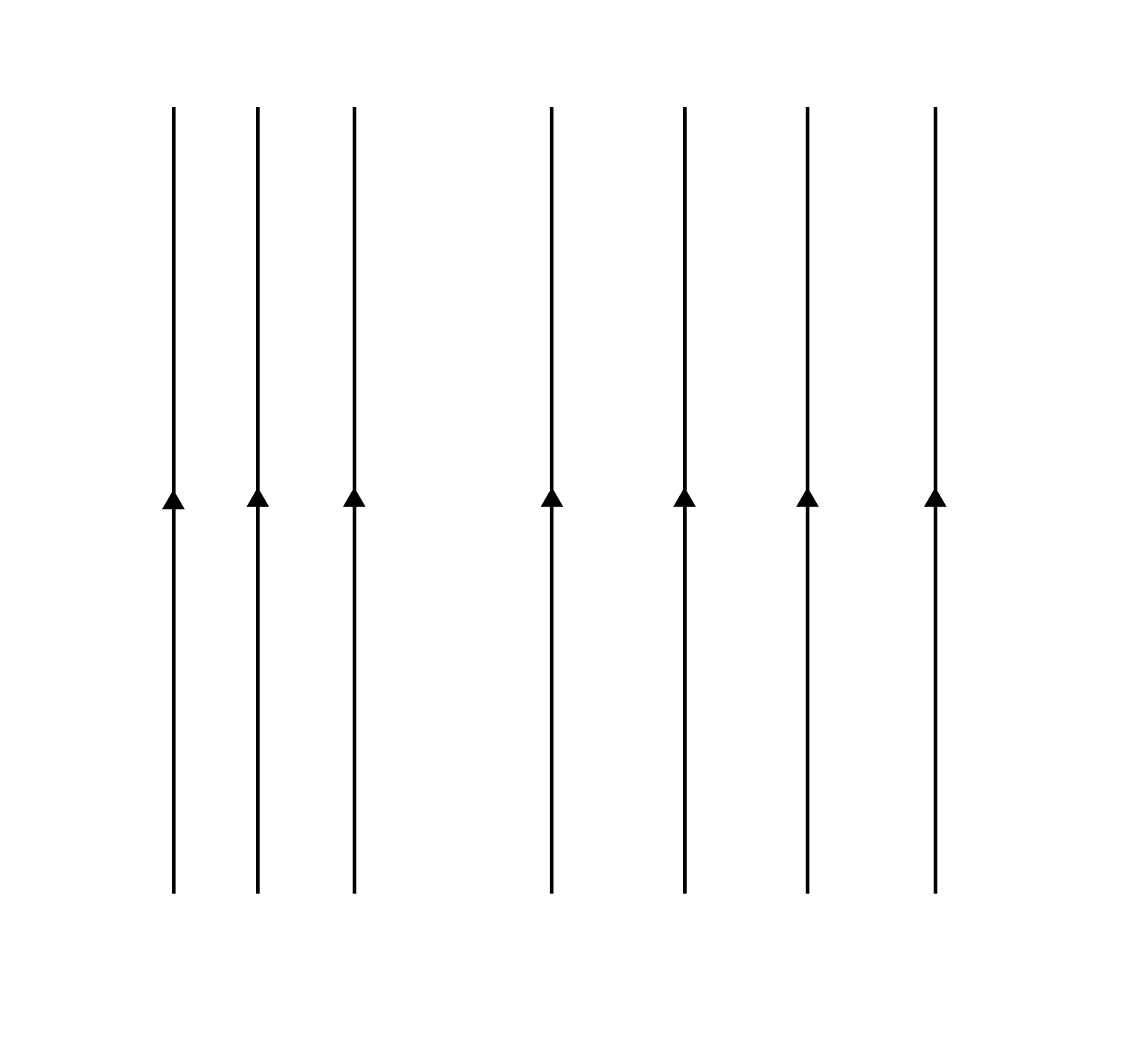}}%
    \put(0.88952245,0.85461055){\color[rgb]{0,0,0}\makebox(0,0)[lt]{\lineheight{1.25}\smash{\begin{tabular}[t]{l}$\R^ {2}$\end{tabular}}}}%
    \put(0,0){\includegraphics[width=\unitlength,page=2]{evolvingPol.pdf}}%
    \put(0.72177919,0.82063723){\color[rgb]{0,0,0}\makebox(0,0)[lt]{\lineheight{1.25}\smash{\begin{tabular}[t]{l}$P_0$\end{tabular}}}}%
    \put(0.91712567,0.56159065){\color[rgb]{0,0,0}\makebox(0,0)[lt]{\lineheight{1.25}\smash{\begin{tabular}[t]{l}$P_t$\end{tabular}}}}%
    \put(0,0){\includegraphics[width=\unitlength,page=3]{evolvingPol.pdf}}%
  \end{picture}%
\endgroup%

\caption{Changing the vertical polarization of $\R^{2n}$ to a polarization $P_t$.}\label{fig: vertical}
\end{figure}

 Let us describe the associated Hilbert space. Sections of our prequantum line bundle are just functions $\sigma(q,p)$, and the polarization condition is equivalent to 
$$\nabla_{t\frac{\partial}{\partial q} + \frac{\partial}{\partial p}}\sigma(q,p) = t\frac{\partial \sigma}{\partial q} + \frac{\partial \sigma}{\partial p} - \frac{i}{\hbar}tp f = 0.$$
Solving this differential equation, we obtain that polarized sections are of the form 
$$\sigma(q,p) = g(q-tp)e^{\frac{itp^2}{2\hbar}}.$$ 
It turns out that if $\sigma = g(q)$ is a $P$-polarized section of the trivial line bundle, then the section $\phi^t_f\sigma$ is precisely \footnote{The fact that the exponent is a multiple of our original function $f$ is a coincidence for $f=p^2$, in general, the exponential factor is the integral of the \emph{Lagrangian} $\mathcal{L}_f = \theta(X_f) - f$ along the curve $\phi^{-t}_f(q,p)$.}
$$\phi^t_f\sigma = g(q-tp)e^{\frac{itp^2}{2\hbar}}.$$
The canonical bundle is spanned by the 1-form $\alpha = dq - t dp$, and the corresponding half-form bundle by sections of the form $\sqrt{dq - t dp}$. Again, it turns out that this is precisely $\phi^t_f\sqrt{dq}$. On $\R^{2n}$ , we can also easily define the BKS pairing. Since the two polarizations $P,P_t$ are transversal, we can multiply elements of the canonical bundles $K_P$ and $K_{P_t}$ to get a pairing  $K_P \times K_{P_t} \to C^\infty(M)$: 
\begin{equation}
(\mu,\mu')\varepsilon := \mu \wedge \bar{\mu}
\end{equation} 
and then the pairing on half-forms is defined as 
\begin{equation}
(\psi,\psi') = \sqrt{(\psi^2,(\psi')^2)}.
\end{equation}
The BKS pairing is then given by 
\begin{equation} \langle \sigma \otimes \psi,\sigma'\otimes \psi'\rangle_{BKS} = \int_{\R^{2n}} \langle\sigma,\sigma'\rangle (\psi,\psi') \varepsilon, 
\end{equation}
in particular, in our case, we have $dq \wedge (dq - t dp) = t dp dq = t \omega $ and therefore $(\sqrt{dq},\sqrt{d(q-tp)}) = \sqrt{2\pi\hbar t}$, which leads to 
\begin{equation}
\left\langle g(q)\sqrt{dq}, h(q-tp)e^{\frac{itp^2}{2\hbar}}\sqrt{d(q-tp)}\right\rangle_{BKS} = \sqrt{\frac{t}{2\pi \hbar}}\int_{\R^2}g(q)\bar{h}(q-tp)e^{\frac{itp^2}{2\hbar} }dqdp
\end{equation}
which means that the map $U_t \colon \mathcal{H}_{P_t} \to \mathcal{H}_P$ is given by 
\begin{equation}
U_t (h(q-tp)e^{\frac{itp^2}{2\hbar}}\sqrt{d(q-tp)}) = \sqrt{\frac{t}{2\pi \hbar}}\int_{\R}h(q-tp)e^{\frac{itp^2}{2\hbar} }\sqrt{dq}.
\end{equation}
The quantization of $f=p^2/2$ is then given by 
\begin{equation}
Qf (g(q)\sqrt{dq}) = \frac{d}{dt}\bigg|_{t=0}\sqrt{\frac{t}{2\pi\hbar}}\int_{\R}g(q-tp)e^{\frac{itp^2}{2\hbar}}dp \cdot\sqrt{dq}
\end{equation}
Setting $u = tp$ in the integral, we have to compute 
$$i\hbar\frac{d}{dt}\bigg|_{t=0}\frac{1}{\sqrt{2\pi\hbar t}}\int_\R g(q-u)e^{\frac{iu^2}{2\hbar t}}du.$$
The asymptotic behaviour of this integral as $t\to 0$ can be computed by the method of stationary phase\footnote{This stationary phase formula requires the phase function (in this case $p^2/2\hbar$) to have non-degenerate critical points, i.e. nonvanishing second derivative. Therefore, this approach fails when trying to quantize monomials of higher degree.} (see e.g. \cite[Section 7]{Hoermander2003}): If $I(t) = \int_\R g(q-u)e^{\frac{iu^2}{2\hbar t}}$ then 
\begin{equation}
I(t) \sim_{t\to 0} (2\pi\hbar t)^{1/2}e^{\frac{i\pi}{4}}\left(1 - t \frac{i\hbar}{2}g''(q) + O(t^2)\right) 
\end{equation}
In particular, we get that 
\begin{equation}
Qf (g(q)\sqrt{dq} =e^{\frac{i\pi}{4}} \frac{\hbar^2}{2}g''(q) \sqrt{dq},
\end{equation}
that is, 
\begin{equation} Qp^2/2 = -e^{\frac{i\pi}{4}}\frac{\hbar^2}{2}\frac{d^2}{dq^2}. 
\end{equation}
It should be noted that this operator is quite different from the prequantization of $p^2$ given by \eqref{eq: prequant p2}. In particular, if we quantize $h = \frac{p^2 +q^2}{2}$, the hamiltonian of the harmonic oscillator, we obtain 
\begin{equation}
Q_h = e^{\frac{i\pi}{4}}\frac{\hbar^2}{2}\frac{d^2}{dq^2}.  + \hat{q}^2
\end{equation}
which is \emph{almost} the correct answer, apart from the unwanted phase factor $e^{i\pi/4}$, which - when taking care of the metaplectic structure - can be absorbed in the pairing of half-forms. 
\begin{exc}
In this exercise we sketch an alternative method to quantize the function $p^2$ in the vertical polarization. The idea is that it is simple to quantize $p^2$ in the horizontal polarization, and we can transform states from the vertical to the horizontal polarization and back using the BKS kernel.  
\begin{itemize}
\item Denote the Hilbert space of the vertical polarization by $\mathcal{H}_{vert}$ and the Hilbert space of the horizontal polarization by $\mathcal{H}_{hor}$.  Following the steps above, show that the BKS pairing between $\mathcal{H}_{vert}$ and $\mathcal{H}_{hor}$  is given by 
\begin{equation}
\left\langle f(q)\sqrt{dq}, g(p)\sqrt{dp}\right\rangle_{BKS} = \int_{\R^2}f(q)\overline{g}(p)e^{\frac{i}{\hbar}p\cdot q}dpdq.
\end{equation}
Conclude that the induced map $U\colon \mathcal{H}_{vert} \to \mathcal{H}_{hor}$ is given by 
\begin{equation}
U\left(f(q)\sqrt{dq}\right) = \int_{R}f(q)e^{\frac{i}{\hbar}p\cdot q}dq \sqrt{dp}, \label{eq: fourier transform}
\end{equation}
i.e. it coincides with the Fourier transform (up to the power of $\hbar$). 
\item  Compute the action of $p^2$ on a state $f(q)\sqrt{dq}$ in the vertical polarization by \begin{itemize}
\item transforming the state into the horizontal polarization \eqref{eq: fourier transform}, 
\item applying the quantum operator $Qp^2$ in the horizontal polarization 
\item transforming the result back to the vertical polarization using the inverse of \eqref{eq: fourier transform}
\end{itemize}
\end{itemize} 

\end{exc}
\subsubsection{Quantization in complex polarization}\label{sec: quant Rn complex}
Next, we analyze what happens when we instead the complex polarization given by the natural K\"ahler structure of $\R^{2n}$, cf. Example \ref{ex: std Kahler}.  The holomorphic polarization is given by $\mathcal{P} = T^{0,1}\R^{2n}$, the span of $\frac{\partial}{\partial \bar{z}^1},\ldots, \frac{\partial}{\partial \bar{z}^n}$.
To determine the polarized sections, we want to change the connection 1-form from $\theta = \sum_i p_idq^i$ to $\theta'=\sum_i \frac{i}{2}\bar{z}^i dz^i$ (the latter being \emph{adapted} to the polarization, i.e. vanishing on all vector fields tanget to $\mathcal{P}$). To do so, we compute 
\begin{align*}
\theta' - \theta &= \frac{i}{2}\sum_i\bar{z}^idz^i - p_idq^i \\
&= \frac{i}{2}\sum_i(q^i dq^i + p^idp^i + ip^i dq^i  + i q^idp^i) &= \frac{i}{4}\sum_i d((p_i)^2 + (q^i)^2 + 2 i p q) .
\end{align*}
That means that if $\sigma_0(x) = (x,1)$ is the reference section of $L$ satisfying $\nabla\sigma_0 = -\frac{i}{\hbar}\theta \otimes \sigma_0$, then 
\begin{equation}
\sigma_1 = \exp\left(\frac{-1}{4\hbar}\left(\sum_i (p_i)^2 + (q^i)^2 + 2 i p q\right)\right) s_0 = \psi(z,\bar{z})\sigma_0
\end{equation}
is the section such that $\nabla \sigma_1 = -\frac{i}{\hbar}\theta'\otimes \sigma_1$. 

Any other polarized section is then of the form $\sigma = f(z)\sigma_1(z)$, where $f(z)$ is a holormorphic function of $z$. In particular, the hermitian structure on $L$, evaluated on sections of this form, is given by 
\begin{align*}
\langle f(z) \psi(z,\bar{z}) \sigma_0, g(z)\psi(z,\bar{z})\sigma_0\rangle &= \int_{\R^{2n}}f(z)\overline{g}(z)|\psi(z,\bar{z})|^2\langle \sigma_0,\sigma_0\rangle  \epsilon \\
&= \frac{1}{(2\pi \hbar)^n}\int_{\C^n}f(z)\overline{g(z)}e^{-\frac{1}{2\hbar}\sum_i |z^i|^2}d\bar{z}^n dz^n
\end{align*}
The line bundle of half-forms is spanned by $\sqrt{dz}$ and in this case doesn't influence the Hilbert space structure, but it does affect the quantization of functions, and as we will see presently, in a fundamental way. Namely, consider again the hamiltonian of the harmonic oscillator, say for $n=1$: 
\begin{equation}
H  = \frac{1}{2}(p^2 + q^2) = \frac{z\bar{z}}{2}
\end{equation}
in particular, it preserves the polarization directly! 
 The hamiltonian vector field is 
\begin{equation}
X_H = i\left( \bar{z}\frac{\partial}{\partial \bar{z}} - z\frac{\partial}{\partial z}\right)
\end{equation}
Then, if $\tilde{\sigma}$ is a $P$ wave function, i.e. $\tilde{\sigma} = f(z)  s_1 \otimes \sqrt{dz}$, we have 
\begin{align*}
Q_H\tilde{\sigma} &= -i\hbar\nabla_{X_H}(f(z)s_1) \otimes \sqrt{dz} - f(z)s_1 \otimes i\hbar L_{X_H}\sqrt{dz} \\
&= \hbar z\frac{\partial f}{\partial z} s_1 \otimes \sqrt{dz} + \frac{\hbar}{2} f(z) s_1\otimes \sqrt{dz} \\ 
&= \left[\hbar\left(z\frac{\partial}{\partial z} + \frac{1}{2}\right)f\right] s_1 \otimes \sqrt{dz}.
\end{align*}
In particular, this operator has the correct spectrum $\hbar(n+\frac12)$, whereas if we had not taken the half-form correction into account we would be missing the crucial $\frac12$ here. This is another indication that the half-form quantization scheme is indeed necessary to obtain correct answers. 
\subsection{Cotangent bundles, cylinder}\label{sec: cotangent}
\subsubsection{Cotangent bundles in the vertical polarization}
Next, consider any manifold $Q$ with cotangent bundle $M = T^*Q$. Here, we have a canonical symplectic form $\omega = d\theta$, with $\theta$ the tautological 1-form. Since $\omega$ is exact, we can use the trivial line bundle $M\times L$ as a prequantum line bundle.  We still have a globally defined vertical polarization $P_{vert} = \ker d\pi$, where $\pi \colon T^*M \to M$ is the projection. All the computations of the previous section still go through in local coordinates  $(q^1,\ldots,q^n)$ on $Q$ and the corresponding coordinates $(q^1,\ldots,q^n,p_1,\ldots,p_n)$ on $T^*M$, since coordinates are always Darboux coordinates for the canonical symplectic form on $T^*Q$, i.e. the symplectic form has the standard expression $\omega = \sum_i dp_i \wedge dq^i$ in such charts, and the canonial 1-form $\theta = \sum_i p^i \wedge dq^i$. In particular, the resulting Hilbert space is isomorphic to $L^2(Q)$, as one would expect.\\
In contrast to geometric quantization on $\R^n$, however, for a general manifold $Q$ there might be different prequantizations, depending on the cohomology of $Q$.\footnote{$M = T^*Q$ has the same cohomology groups as $Q$ since they are homotopy equivalent.} A simple example where we can observe this effect is the cotangent bundle of the circle $M = T^*S^1$. Let us denote the coordinates on $M$ by $(\phi,p)$, here $\phi$ is an angular coordinate and $p$ the corresponding momentum. In particular, the 1-form $d\phi$ is closed but not exact, i.e. defines a non-zero class in $H^1(M,\R)$, and we can shift the symplectic potential by any nonzero multiple of $d\phi$, 
\begin{equation}
\theta_\lambda = \theta + \hbar\lambda d\phi. 
\end{equation}
The corresponding Hilbert space is still isomorphic to square-integrable functions on the circle, $L^2(S^1)$. However, every $\theta_\lambda$ gives rise to a unitarily inequivalent quantization. Using the argument of \cite{Blau1992}, consider the quantization of the function $p$, then 
\begin{equation}
Q^\lambda_p = -i\hbar\frac{\partial}{\partial \phi} -\hbar\lambda
\end{equation} 
and the spectrum of this operator is the spectrum of $-i\hbar\frac{\partial}{\partial \phi}$, shifted by $\hbar\lambda$: 
\begin{equation}
\mathrm{spec}(Q^\lambda_p) = \{\hbar(n-\lambda), n \in \mathbb{Z}\}
\end{equation}
which shows that the different quantization $Q^\lambda$ cannot be unitarily equivalent (since any such equivalence would preserve the spectrum). One can see that the spectra coincide for $\lambda$ an integer, an indeed we have found here the $S^1$ worth of prequantizations promised in section \ref{sec: prequantum choices}. Similarly, there is an $(S^1)^n$ worth of prequantizations of $T^*(S^1)^n$ - in the physics language, in this case there are $n$ vaccuum angles. 

\begin{figure}[!ht]
\centering
\def\svgwidth{0.5\columnwidth}
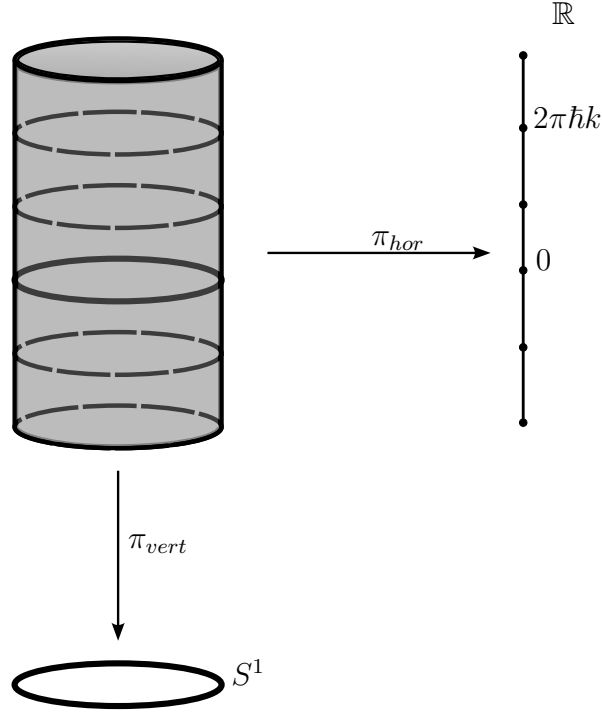
\caption{Horizontal and vertical polarization on $T^*S^1$ with their respective leaf spaces. For the horizontal polarization, the support of cohomological wave functions is concentrated at $2\pi\hbar\mathbb{Z}$.}
\end{figure}

\subsubsection{Horizontal polarizations}
Unlike the vertical polarization, which is defined for all contangent bundles, the horizontal polarization is defined only in special cases. The most prominent case is when the cotangent bundle is actually trivial, $T^*Q \cong Q \times \R^{\dim Q}$. This also happens in the case of the cylinder $M = T^*S^1 = S^1 \times \R$. Here, there is a globally defined horizontal polarization $P_{hor} = \mathrm{span}\left\lbrace\frac{\partial}{\partial \phi}\right\rbrace$. Here we can see that the leaves of this polarization are simply the horizontal circles in the cylinder, i.e. they are not simply connected! \\ 
Let us analyze the situation in more detail. The leaves of the horizontal polarization are precisely the horizontal circles $L_p = S^1 \times \{p\}$ and the holonomy of the prequantum connection $\nabla$ along the nontrivial loop $\gamma = L_p$ (which in this special case coincides with the leaf) is 
$$hol_\gamma\theta = \exp\left(\frac{ip}{\hbar}\right) = 1 \Leftrightarrow p = 2\pi\hbar k, k \in  \mathbb{Z}.$$
That is, the Bohr-Sommerfeld variety consists of the collection of circles at $2\pi\hbar$ times an integer value. Let us consider the cohomological wave functions: Since the prequantum line bundle is trivial, they are of the form $\tau_P = \tau \otimes \psi$ where $\psi$ is a half-form that we can take to be $\psi = \sqrt{dp}$. Notice that when restricting to $P$ the differential acts trivially on such half-forms. Therefore, to determine the cohomology of $\Omega^\bullet(M,P,L_P)$ in this case, it is enough to consider the complex of de Rham forms on the cylinder with twisted differential $d_\theta = d  - \frac{i}{\hbar}\theta \wedge$. \\
We claim that all cohomological wave functions have degree 1. We prove this statement degree by degree. In degree 0, we are simply considering functions $f(p,\phi)$. Such a function is $P$-closed if\footnote{There is no condition on the derivative with
respect to $p$ since we are restricting the differential form $d^\nabla f = \frac{\partial f}{\partial p}dp + \frac{\partial f}{\partial \phi }d\phi - \frac{i}{\hbar} f d\phi$ to the polarization $P$ spanned by $\frac{\partial}{\partial \phi}$}
\begin{equation}
\frac{\partial f}{\partial \phi} = \frac{i}{\hbar}p \cdot f \Leftrightarrow f(p,\phi) = \exp\left(\frac{i}{\hbar}p\cdot \phi \right).
\end{equation}
There are no such functions since $f$ is required to be a smooth function on the cylinder, which implies that $f(p,\phi) = f(p,\phi +1 )$. This in turn requires that $p \in 2\pi \hbar\mathbb{Z}$, which is impossible for a smooth function. Therefore there are no $P$-closed functions, and no cohomological wave functions in degree 0. In degree 2, all forms are closed (and in particular $P$-closed), but also, all forms are exact (and hence in particular $P$-exact). We can therefore concentrate on degree 1. Again, notice that all 1-forms are $P$-closed, since any 2-form on the cylinder vanishes when restricted to only vectors in $P$ (since it contains at least one $dp$ factor). A general $P$-closed 1-form is therefore of the form $\omega = \omega_\phi d\phi + \omega_pdp$. But again, since forms containing $dp$ vanish when restricted to $P$, they are in particular $P$-exact. Hence we can restrict ourselve to 1-forms of the form $\omega_kd\phi$. Such 1-forms are $P$-exact if there is a function $f(\phi,p)$ such that 
\begin{equation}
\omega_\phi(\phi, p) = \frac{\partial f}{\partial \phi} - \frac{i}{\hbar}p f(\phi,p).
\end{equation}
Expanding both $\omega_\phi = \sum_{k}e^{2\pi i k \phi}\omega_{\phi,k}(p)$ and $f(\phi,p) = \sum_k e^{2\pi i k \phi}f_k(p)$ in Fourier modes, we have the equation 
\begin{equation}
\left(2\pi i k - \frac{i}{\hbar}p\right)f_k(p) = \omega_k(p)
\end{equation}
Then we see that $\omega$ is $P$-exact if $\omega_k(2\pi k\hbar) = 0$ for all $k \in \mathbb{Z}$. In particular, we have a family of forms which are necessarily not $P$-exact, namely the 1-forms $\psi_k$ given by 
\begin{equation}
\psi_k = e^{2\pi ik\phi} \eta(2\pi k \hbar + p)d\varphi,
\end{equation}
where $\eta$ is a smooth function on $\mathbb{R}$ with total integral 1 which is non-zero only in the interval $(-\hbar/2,\hbar/2)$ and satisfies $\eta(0) >0$. With a little more work one can check that those indeed define a basis of $\mathcal{H}_P$ given by the cohomological wave functions. We can define an inner product on this space by declaring the $\psi_k$ to be an orthonormal family, this  make $\mathcal{H}_P$ into a Hilbert space. 
\subsubsection{K\"ahler polarization}
We briefly mention that one can also put a complex structure on the cylinder, which turns it into a K\"ahler manifold and work in the K\"ahler polarization. For instance, one can do this by identifying the cylinder with the punctured complex plane $\C^*$ via $(p,\phi) \mapsto \exp(p+i\phi)$. In the K\"ahler polarization, elements of the Hilbert space can be identified with holomorphic functions on the punctured plane.
\begin{exc}
Using the same steps as in Subsection \ref{sec: quant Rn complex}, show that the inner product of two sections represented by holomorphic functions $f,g$ on $\C^*$ is given by 
\begin{equation}
\langle f(z),g(z)\rangle = \frac{1}{2\pi \hbar}\int_\C f(z)\overline{g(z)}e^{\frac{1}{2\hbar}|z|^2}\frac{d\bar{z}dz}{|z|^2}. 
\end{equation}
\end{exc}
Holomorphic functions on the punctured plane can be decomposed into Laurent series, and therefore we have a basis $z^k, k \in  Z$, of the Hilbert space in the holomorphic polarization. Even though we will not discuss the details here, we observe that the Hilbert spaces arising from all three polarizations discussed (vertical, horizontal and holomorphic polarization) have a basis given by functions of the form $z \mapsto z^n$. With a little more work one can show that the various isomorphism are in fact unitary and intertwine the quantization maps. Even though those three polarizations all have some quite different features, the resulting quantizations turn out to be isomorphic! It is part of a pattern sometimes called ``invariance under polarization'' that different polarizations give rise to (unitarily) isomorphic quantizations if one can make sense of them, even though there is no general theorem telling us that it is so. \footnote{This should be compared with the fact that if we pick non-isomorphic line bundles with connection, we obtain manifestly non-isomorphic quantizations, as the ``vaccuum angles'' example shows.}

\subsection{The 2-sphere, quantization of angular momentum}
Finally, as an example of a geometric quantization of a compact manifold, let us consider the 2-sphere $S^2$. For prequantization, notice that the symplectic form on the unit sphere is necessarily not exact, since it has finite volume $\int_M \omega_0 = 1$, where $\omega_0$ is the volume form described in exercise \ref{exc:volume form sphere}. From the Weil integrality condition, we see that there exists a prequantization of $(S^2,\omega)$ if $\omega = 2\pi\hbar k\omega_0=:\omega_k$, and since $H^2(S^2,\mathbb{Z})$ is spanned by $\omega_0$, this is actually an if and only if. 
We now describe the associated prequantum line bundle .

Remember from Exercise \ref{ex: 2 sphere Kaehler} that on $S^2$ we have the two complex coordinates $z$ on $U_N = S^2 -\{N\}$ and $w$ on $U_S = S^2 -\{S\}$ given by 
\begin{equation*}
z = \frac{x^1 + i x^2}{1-x^3}, w = \frac{x^1 - i x^2}{1+x^3}
\end{equation*}
satisfying $z=w^{-1}$, and that in the $z$ coordinates we have $\omega_0 = \frac{1}{2 pi i} \frac{d\bar{z} \wedge dz }{(1+|z|^2)^2}$, and therefore $\omega_1 = -i\hbar  \frac{d\bar{z} \wedge dz }{(1+|z|^2)^2}$ and $\omega_k = k\cdot \omega_1$. 
\begin{exc}
\begin{enumerate}
\item Show that
\begin{equation}
\theta_N = -i\hbar\frac{\bar{z}dz}{(1+|z|^2}, \theta_S = -i\hbar\frac{\bar{w}dw}{(1+|w|^2)}
\end{equation} 
are primitives for $\omega$ on $U_N$ and $U_S$ respectively.
\item We define a line bundle $L$ on $S^2$ by declaring the transition function $g_{NS} = \frac{1}{z}$.  Show that the 1-forms $\theta_N, \theta_S$ define a connection $\nabla$ on this line bundle. Therefore $(L,\nabla)$ defines a prequantum line bundle for  $\omega_1$.
\item Show that the line bundle $L_k=(L^{\otimes k}, k\nabla)$ (defined by the transition function $g_{NS,k} = \frac{1}{z^k}$ and the connection $k\nabla$ defined by $k\theta_N, k\theta_S)$ is a prequantum line bundle for $\omega_k$.
\end{enumerate}
\end{exc}
Notice that the line bundle $L_k$ that we built above is \emph{holomorphic}. Polarized sections in the K\"ahler polarization are therefore given by holomorphic functions in both trivializing neighbourhoods $U_N$ and $U_S$ - since those are both $\C$, holomorphic functions are given by power series at zero: $\sigma_N(z) = \sum_{i=0}^\infty a_iz^i$ in $U_N$ and $\sigma_S = \sum_{i=0}^\infty b_iw^i$ in $U_S$. On the intersection $U_N \cap U_S$ we have $w = \frac{1}{z}$ and therefore $$\sigma_S(z) = \sum_{i=0}^\infty b_i z^{-i} = g_{NS,k}(z)\sigma_N(z) = \sum_{i=0}^\infty a_i z^{i-k}$$
from which we conclude that $a_i = b_i = 0$ for $i >k$ and $a_i = b_{k-i}$ for $i =0,\ldots, k$. In particular, the space of holomorphic sections of $L_k$ is finite dimensional and has dimension $k+1$! To describe the inner product we represent sections of $L_k$ by polynomials of degree at most $k$ in $U_N$, then we obtain 
\begin{equation}
\langle f_1(z),f_2(z) \rangle =\frac{1}{2\pi i} \int_\C\frac{f_1(z)\overline{f_2(z)}}{(1+|z|^2)^{k}} \frac{d\bar{z}\wedge dz}{(1+|z|^2)^2}
\end{equation}  
(here the denominator comes from a factor of $\exp(-k\cdot K/\hbar)$ where $K = \hbar\log(1+|z|^2)$ is the K\"ahler potential for $\omega_1$, that appears when identifying holomorphic functions with polarized sections).
\section{Chern-Simons theory}
In this final section, we will consider some elements of geometric quantization relevant for Chern-Simons theory, with the main goal of explaining some of the recent results in \cite{Cattaneo2022} and \cite{Cattaneo2022a}. We start by explaining some background on Chern-Simons theory.  
\subsection{Some basics of Chern-Simons theory} 
\subsubsection{Classical Chern-Simons theory}
We generally think of a field theory in $d$ spatial dimensions as an assignment that assigns to every $d+1$-dimensional manifold $X$ (the \emph{spacetime}) a space of fields $F_X$ and an action functional $S_X \colon F_X \to \mathbb{R}$. Chern-Simons theory is a field theory in 2 spatial dimensions. Here, we additionally fix a Lie group $G$ (usually $G=SU(2)$, but for our purposes it will not matter too much). Denote by $\mathfrak{g} = Lie(G)$ the Lie algebra of $G$ (for $SU(2)$, this is simply the vector space  $su(2)$ of traceless antihermitian matrices). The space of fields is then $F_X = \Omega^1(X,\mathfrak{g}) = \Omega^1(X)\otimes \mathfrak{g}$, which one should think of as the space of connections on a trivial principal $G$-bundle $P = M \times G \to M$ and the action functional is 
\begin{equation}
S_X[A] =\int_X \frac12 \langle A \wedge dA \rangle + \frac16 \langle A, [A,A]\rangle = \frac{1}{4\pi}\int_X \operatorname{tr} \left(A\wedge dA + \frac23 A\wedge A  \wedge A \right).  
\end{equation} 
In field theory the classical physics is described by the critical points of the action functional. For the Chern-Simons functional, those are precisely the flat connections 
\begin{equation}
\delta S_X[A] = 0 \Leftrightarrow F_A = dA + A \wedge A = 0. 
\end{equation}
The Chern-Simons theory is a gauge theory: The exponential of the action $e^{ikS_X[A]}$ is invariant under the transformation $A \mapsto gAg^{-1} + g^{-1}dg$, where $g \colon X \to G$ is a map from $X$ to the Lie group $G$, and so are the critical points. (Exercise?) We denote the space of all such maps by $\mathcal{G}_M = C^\infty(M,G)$ and call it the \emph{gauge group}. Usually in field theory one is interested in the value of observables - functionals $\mathcal{O}\colon F_M \to \R$. In the presence of gauge symmetries, such as in Chern-Simons theory, we are forced to consider only gauge-invariant observables, i.e. the ones that are invariant under this action of the gauge group. A class of gauge-invariant observables of interest in many gauge theories is formed by the Wilson loop observables. If $\gamma \colon S^1 \to X$ is a loop in $X$, and $R$ a representation of $G$, then we can define, for any connection $A \in \Omega^1(M, \mathfrak{g})$, the number 
\begin{equation}
W_\gamma(A) = \operatorname{tr}_R P\exp \oint_\gamma A.
\end{equation} 
Here $\operatorname{tr}_R$ is the trace of an element of $G$ in the representation $R$, and $P\exp \oint_\gamma A \in G$ is the holonomy of $A$ along $\gamma$. \footnote{That is, $P\exp \oint_\gamma A$ is the value at 1 of the solution of the initial value problem $\dot{g}(t) =g(t)\cdot A(\dot{\gamma}(t))$, where assume that $G$ is a matrix group for simplicity.} We also remark that this field theory is \emph{topological}: It does \emph{not} involve the metric or any other geometric structure on the 3-manifold $X$. For more information on the classical theory we refer e.g. to the works of Freed \cite{Freed1995}, \cite{Freed2002}. 
\subsubsection{Some aspects of Quantum Chern-Simons theory}
In quantum field theory one studies expectation values of observables, heuristically defined by integrating the observable over all allowed field configurations, weighted with the exponential of the action 
\begin{equation}
\langle \mathcal{O}\rangle \text{``}=\text{''} \frac{1}{Z}\int_{F_M}\mathcal{O}(\phi)e^{\frac{i}{\hbar}S[\phi]}d\phi
\end{equation}
where 
\begin{equation}
Z \text{``}=\text{''} \frac{1}{Z}\int_{F_M}\mathcal{O}(\phi)e^{\frac{i}{\hbar}S[\phi]}d\phi\label{eq: Z CS}
\end{equation}
is called the \emph{partition function}. The equalities here are in quotes because the integral on the right-hand side does not have a measure-theoretic definition in general. In physics, one usually tries to compute them by formally applying the principle of stationary phase, which yields an expression for $Z$ and $\mathcal{O}$ in terms of \emph{Feynman graphs}. For topological theories, the partition function, and expectation values of observables, are of interest in mathematics because they are expected to give rise to topological invariants. The enormous importance of Chern-Simons theory in mathematical physics comes from the seminal paper of Witten \cite{Witten1989} where he argued - actually using ideas from holomorphic quantization - that expectation values of Wilson loops $\mathcal{O}$ coincide with an invariant of knots known as the Jones polynomial \cite{Jones1985}. Shortly after, Reshetikhin and Turaev \cite{Reshetikhin1991} defined invariants $Z^{RT}_k[M]$ of 3-manifolds that can be understood as a mathematical model for the Chern-Simons partition function \eqref{eq: Z CS}. On the other hand, the perturbative approach to Chern-Simons theory has been developed by many authors - for an overview we refer to \cite{Mnev2019a}, \cite{Wernli2022}. 
\subsection{The phase space of Chern-Simons theory}
Let us consider the Chern-Simons theory on a manifold $X$ with boundary $\partial X = \Sigma$. Then, when trying to compute the variation of the action functional $S_X$, we get an additional term from integrating by parts: 
\begin{equation}
\delta S_X[A] = \int_X \operatorname{tr} F_A \wedge \delta A + \int_{\Sigma} \operatorname{tr} A \wedge \delta A 
\end{equation} 
The second term is a boundary term. Notice that the variation of the action functional is no longer zero when restricted to flat connections $F_A=0$: Instead, we also need to impose conditions on $A$ such that $\theta_\Sigma = \int_\Sigma \operatorname{tr} A \wedge \delta A$ is zero on the boundary - i.e., we need to impose \emph{boundary conditions}. To understand the structure of those boundary conditions, it is helpful to consider the space of boundary fields 
\begin{equation}
F_\Sigma = \Omega^1(\Sigma, \mathfrak{g}) 
\end{equation}
and think of $\delta$ as the de Rham differential on $F_\Sigma$ and $\theta_\Sigma$ as a 1-form on this (infinite-dimensional) space.\footnote{In this text we will not bother too much with the technicalities of infinite-dimensional vector spaces and manifolds, but here and in what follows, we are using the Fr\'echet topology on spaces of smooth functions and sections, such as differential forms.} We then obtain a symplectic form\footnote{To be precise, only a weak symplectic form. I.e. the map from vectors to covectors given by contracting with $\omega$ is only injective, but not surjective} on $F_\Sigma$, given by 
\begin{equation}
\omega_\Sigma = \int_\Sigma \delta_A \wedge \delta_A
\end{equation}
We call the symplectic vector space $(F_\Sigma,\omega_\Sigma)$ the \emph{phase space} of Chern-Simons theory. A \emph{boundary condition} for Chern-Simons theory is a Lagrangian $L \subset F_\Sigma$. It turns out that often it is natural to consider \emph{families} of boundary conditions: In the best possible case, we have a Lagrangian $L_x$ through every point $x \in F_\Sigma$, in general, such families correspond to polarizations of $F_\Sigma$. We refer to \cite{Cattaneo2011} and references therein for a detailed discussion of these matters, and a construction of $F_\Sigma$ for general Lagrangian field theories.  Picking a polarization $P$ of $F_\Sigma$ ensure that $\theta_\Sigma\big|_L = 0$, we can change the action by a boundary term 
\begin{equation}
S_{CS} \to S_{CS} + df, \theta \to \theta - \delta f
\end{equation}
where $f \colon F_\Sigma \to \R$ is a function on $F_\Sigma$, such that $\theta_\Sigma$ is adapted to $P$, i.e. vanishes on the fibers of $P$. From the point of view of geometric quantization, it is natural to think of this transformation as a gauge transformation on the trivial prequantum line bundle $F_\Sigma \times \C$.\footnote{This transformation is also sometimes called \emph{Weyl transformation} or simply \emph{$f$-transformation}, see \cite{Mnev2019}.} 

\subsection{Geometric quantization, Chern-Simons theory, and the quantization-commutes-with-reduction question}
The phase space of Chern-Simons theory that we described above is, in some sense, unphysical: the fields that can arise from solutions to equations of motion are only the flat connections on $\Sigma$, but in $F_\Sigma$ we have all the connections. Only the flat ones correspond to physical degrees of freedom. Also, on connections we have the action of the gauge group on $\Sigma$, $\mathcal{G}_\Sigma = C^\infty(\Sigma, G)$, and connections related to each other via gauge transformation describe the same physical configuration. Therefore, the ``physical'' (or sometimes also reduced) phase space is the quotient of the subspace of all connections by the action of the gauge group, 
\begin{equation}
F^{red}_\Sigma = \Omega^1(\Sigma,\mathfrak{g})_{flat} / \mathcal{G}_\Sigma =: MFC(\Sigma,G)
\end{equation}
the \emph{moduli space of flat connections on $\Sigma$}. This moduli space is actually finite-dimensional: Taking holonomies of the flat connection along generators of the fundamental group $\pi_1(\Sigma)$ of $\Sigma$, we obtain an isomorphism with the ``character variety'' of $\pi_1(\Sigma)$, 
\begin{equation}
MFC(\Sigma,G) \cong \operatorname{Hom}(\pi_1(\Sigma),G) / G
\end{equation}
where $\operatorname{Hom}$ denotes group homomorphisms and $G$ acts by conjugation.\footnote{This isomorphism is discussed in detail e.g. in the book of Taubes \cite{Taubes2011}. }  The character variety is finite-dimensional for finitely generated groups such as the fundamental group of a surface. It is, in fact, also a symplectic manifold, as one can see from the following short digression on symplectic reduction. 
\subsubsection{Symplectic reduction}
Consider the case of a compact group $G$ acting on a symplectic manifold $(M,\omega)$, preserving the symplectic structure. Differentiating the action of $G$, we obtain a map of Lie algebras $\rho\colon\mathfrak{g} \to \mathfrak{X}(M), \xi \mapsto \rho(\xi)=:\xi^\#$, and because the action is symplectic, it lands inside symplectic vector fields: 
\begin{equation}
L_{\xi^\#}\omega = 0, \qquad \text{ for all } \xi \in \mathfrak{g}.
\end{equation}
From Cartan's magic formula $L_{\xi^\#} = d\iota_{\xi^\#} + \iota_{\xi^\#} d$ and the fact that $\omega$ is closed we get that the 1-form $\iota_{\xi^\#}\omega$ is closed for all $\xi '\in \mathfrak{g}$. We say that the action is \emph{weakly hamiltonian} if all vector fields $\xi^\#$ are hamiltonian, i.e. the 1-forms $\iota_{\xi^\#}\omega = dH_\xi$ are all exact, and $\xi \mapsto H_{\xi}$ is linear.  The \emph{moment map} is the map $\mu \colon M \to \mathfrak{g}^*$ given by $x \mapsto (\xi \mapsto H_\xi(x))$. We say that the action is hamiltonian if the moment map $\mu$ is $G$-equivariant, where $G$ acts on $\mathfrak{g}^*$ by the coadjoint action. In this case, the map $\mu^* \colon \xi \mapsto H_\xi$ is a Lie algebra map $\mathfrak{g} \to C^\infty(M)$ with the Poisson bracket and is called the \emph{dual moment map}. The subset $\mu^{-1}(0)\subset M$ is invariant under the $G$-action and we call the quotient 
\begin{equation}
M // G := \mu^{-1}(0) / G
\end{equation}
the \emph{Marsden--Weinstein--Meyer symplectic reduction}, of $M$ by $G$. The theorem of Marsden and Weinstein then tells us that this is indeed a symplectic manifold if the $G$-action on $\mu^{-1}(0)$ is free: The symplectic structure is given by restricting $\omega$ to $\mu^{-1}(0)$ and evaluating it on $G$-orbits. We refer to \cite[Part IX]{Silva2008} for details. 
\subsubsection{Quantization commutes with reduction}
Symplectic reduction in the context of geometric quantization - where we assign Hilbert spaces to symplectic manifolds - give rise to the following natural question: \newline
What is the relation between the Hilbert space associated to $M$ and to its symplectic reduction $M // G$? \\ 
For instance, it makes sense to ask this question in K\"ahler quantization: One can show that the symplectic reduction of a K\"ahler manifold is again a K\"ahler manifold, therefore they both have the K\"ahler polarization and the Hilbert space associated to a prequantum line bunde. Then, the group $G$ acts on the Hilbert space $\mathcal{H}_M$ associated  to $M$ (since it acts on holomorphic sections of the prequantum line bundle) and one can show that the Hilbert space of the symplectic reduction is the $G$-invariant subspace of $\mathcal{H}_M$: 
\begin{equation}
\mathcal{H}_{M//G} = \mathcal{H}_M^G.
\end{equation}
One calls this phenomenon ``quantization commutes with reduction''. It was first proved by Guillemin and Sternberg \cite{Guillemin1982}.\footnote{Although they did not discuss \emph{unitarity} of the isomorphism - it was later realized that one needs to incorporate the metaplectic correction to obtain unitarity \cite{Hall2007}.} There is plenty of literature on the subject, we refer the interested reader e.g. to the recent survey \cite{Ma2021}.  
\subsubsection{Quantization and reduction in Chern-Simons theory} 
We return to the topic of Chern-Simons theory. Atiyah and Bott \cite{Atiyah1983} observed that the curvature 2-form of a connection gives a moment map for the action of the gauge group $\mathcal{G}_\Sigma$ on $F_\Sigma$. Namely, the Lie algebra of the gauge group is $Lie(G) = \Omega^0(M,\mathfrak{g})$, we can therefore identify $\Omega^2(M,\mathfrak{g} \subset Lie(\mathcal{G})^*$ via the pairing $\Omega^2(M,\g)) \times \Omega^0(M,\g) \to \mathbb{R}$, 
\begin{equation}
\langle \alpha, \beta \rangle = \int_{\Sigma}\mathrm{tr}(\alpha \wedge \beta)   \label{eq: AB symplectic}
\end{equation}
The zero set of the moment map coincides with the space of flat connections, and its quotient by the gauge group is the moduli space of flat connections.\footnote{This moduli space is in general singular because the action of the gauge group is not free, e.g. it fixes the zero connection. In this introductory text we gloss over those issues.} It therefore carries a natural symplectic structure, known as the Atiyah-Bott symplectic structure. In fact, this symplectic structure is K\"ahler: Picking a compatible complex structure $J$ on the surface $\Sigma$, we naturally obtain a complex structure $J^{up}$ on $\Omega^1(\Sigma,\g)$ such that the $\pm i$-Eigenspaces of $J^{up}$ are $\Omega^{1,0}(\Sigma)$ and $\Omega^{0,1}(\Sigma)$. Those naturally give us two polarizations on $F_\Sigma = \Omega^1(M,\g)$, denoted by 
\begin{align}
P_{ahol} &= {\rm span }\left(\frac{\delta}{\delta A^{0,1}}\right) \\
P_{hol} &= {\rm span }\left(\frac{\delta}{\delta A^{1,0}}\right)
\end{align} We also have a K\"ahler structure also on its symplectic quotient, the moduli space of flat connections. One can then ask if there is a  ``quantization commutes with reduction'' statement for Chern-Simons theory: On the one hand, one can try to geometrically quantize the moduli space of flat connections with a K\"ahler polarization coming from a complex structure on $\Sigma$, with prequantum line bundle given by the reduction of the trivial line bundle on $F_\Sigma$, called the \emph{Chern-Simons line bundle}. To construct the Hilbert space, one considers holomorphic sections of the Chern-Simons line bundle.  The main reference is \cite{Axelrod1991a}, see also \cite{Ramadas1989}. The dimension of those vector spaces is given by the famous Verlinde formula (\cite{Verlinde1988},\cite{Tsuchiya1989}), which for $G = SU(2)$ is 
\begin{equation}
\dim H^0(\Sigma_g,\mathcal{L}^k) = \left(\frac{k+2}{2}\right)^{g-1}\sum_{j=1}^{k+1}\left(\sin^2\frac{j\pi}{k+2}\right)^{1-g},
\end{equation} 
where $g$ is the genus of $\Sigma$ and $k$ an integer called the level, from the geometric quantization viewpoint it just means we are quantizing $k$ times an approriately normalized Atiyah-Bott symplectic form \eqref{eq: AB symplectic} by using the $k$ tensor power of the Chern-Simons line bundle. 
This formula is quite hard to prove, for a review see \cite{Schottenloher2008}. \\
On the other hand, we can try to geometrically quantize Chern-Simons theory before reducing it. That is, we decompose $$F_\Sigma \otimes \C = \Omega^1(M,\g ) \otimes \C = \Omega^{1,0}(\Sigma) \oplus \Omega^{0,1}(\Sigma) \ni (A^{1,0},A^{0,1}).$$
If we pick the holomorphic polarization, our states are represented by functionals the antiholomorphic part of the connection $\psi(A^{0,1})$. However, the \emph{physical states} are the ones satisfying the constraint equation 
\begin{equation}
F(A)\psi(A^{0,1}) = \left(\partial A^{0,1} + \frac{\pi}{k}\bar\partial \frac{\delta}{\delta A^{0,1}} - \frac{\pi}{k}\left[\frac{\delta}{\delta A^{0,1}}, A^{0,1}\right]\right)\psi(A^{0,1})=0 \label{eq:curvature constraint}
\end{equation} It is argued in \cite{Gawedzki1991} for genus 0,  \cite{Falceto1994} for genus 1 and \cite{Gawedzki1995} for higher genera that solutions to \eqref{eq:curvature constraint} coincide with solutions to chiral ward identities in the WZW model and that their dimension is also given by the Verlinde formula, thus establishing a quantization commutes with reduction statement in Chern-Simons theory. However, those argument rely on the WZW path integral, and to the best of my knowledge, the natural scalar product coming from Chern-Simons theory on the state spaces has not been fully understood yet. In the next and final section, to connect to recent work, we show via a Feynman diagram argument that the partition function of Chern-Simons theory (the vaccuum state in the state space) coincides with the WZW partition function.  
\subsection{Path integral quantization of Chern-Simons theory on cylinders and the CS-WZW correspondence}
Finally, we comment on the (perturbative) path integral quantization of Chern-Simons theory on cylinders, see \cite{Cattaneo2022}, \cite{Cattaneo2022a}. Since the Chern-Simons theory is a gauge theory, the natural setting for the perturbative path integral quantization is the BV-BFV formalism \cite{Cattaneo2014},\cite{Cattaneo2017}, but for simplicity here we will do the gauge-fixing ``by hand''. In the presence of boundary, we will use the holomorphic and antiholomorphic polarizations used above, i.e. our quantum states will be functionals of $B_\Sigma$, which is  either $\Omega^{1,0}_{\Sigma_i}$ or $\Omega^{0,1}_{\Sigma_i}$ on each connected component $\Sigma_i$ of $\Sigma$. The path integral is then formally given by integrating over the fibers of the map $p\colon F_M \otimes \C \to F_\Sigma \otimes \C \to B_\Sigma$, i.e. 
\begin{equation}
Z_M[b] = \int_{p^{-1}(b) \subset F_M} e^{\frac{i}{\hbar}S_M[\phi]}D\phi
\end{equation}  We consider Chern-Simons theory on a cylinder $I \times \Sigma$, and we will put holomorphic boundary conditions at $t = 0$ and anti-holomorphic boundary conditions at $t=1$, i.e. our quantum states will be functionals of $A^{0,1}$ on the in-boundary ($t = 0$) and $A^{1,0}$ at the out-boundary ($t=1$). In order for the symplectic potential coming from the action functional to be adapted to the polarization, we add to $S_{CS}$ the boundary term 
\begin{equation}
f = \frac{1}{4\pi}\int_{\{1\}\times \Sigma}\tr A^{1,0} \wedge A^{0,1} - \frac{1}{4\pi}\int_{\{0\} \times \Sigma} \tr A^{0,1} \wedge A^{1,0}
\end{equation} 
and denote $S_{M}^f = S_{M} + f$. 
Then the symplectic potential reads 
\begin{equation}
\theta = \frac{1}{2\pi}\int_{\{1\} \times \Sigma } \tr A^{1,0} \wedge \delta A^{0,1}  + \frac{1}{2\pi} \int_{\{0\} \times \Sigma} \tr A^{0,1} \wedge \delta A^{1,0}. 
\end{equation} We denote the coordinate on the interval by $t$, accordingly, we can split the field $A = dt \cdot A_I + A^{1,0} + A^{0,1}$ where $A_I$ is a function on $I \times \Sigma$ and $A^{1,0}$, $A^{0,1}$ are ($I$-dependent) $(1,0)$ and $(0,1)$ forms on $\Sigma$ respectively.  Next, decompose the Chern-Simons action into the free and interacting parts as
\begin{align*}
S^f_{I \times \Sigma}[A] &= S_{free}[A] + S_{int}[A] \\
S_{free} &= \frac{1}{4\pi}\int_{I\times\Sigma} \tr  A \wedge  dA = \frac{1}{2\pi}\int_{I\times \Sigma}\tr A^{1,0} \wedge d_IA^{0,1} + \frac{1}{2\pi}\int_{I \times \Sigma}\tr  dt\cdot A_I \wedge d_\Sigma ( A^{0,1} + A^{1,0})   \\
 &+ \frac{1}{4\pi} \int_{\{1\} \times \Sigma }\tr A^{1,0} \wedge A^{0,1} - \frac{1}{4\pi}\int_{\{0\} \times \Sigma }\tr A^{1,0} \wedge A^{0,1} \\
S_{int} &= \frac{1}{6\pi}\int_{I\times \Sigma} \tr  A \wedge A \wedge A = \frac{1}{\pi}\int_{I\times \Sigma
} A_I \wedge A^{0,1} \wedge A^{1,0}.
\end{align*}
Notice that the term in the second line has the same sign as $f$ and therefore adds up with it. We denote the two boundary components by $\Sigma_{in}  = \{ 0\} \times \Sigma $ and $\Sigma_{out}  = \{1\} \times \Sigma$ respectively. We are then imposing the boundary condition that $A^{0,1}\big|_{\Sigma_{out}} = A^{0,1}_{out}$ and $A^{1,0}\big|_{\Sigma_{in}}=A^{1,0}_{in}$, here $A^{0,1}_{out}$ and $A^{1,0}_{in}$ are fixed forms on $\Sigma$ (not $I$-dependent). Then, the Chern-Simons partition then formally function reads 
\begin{equation}
Z_{I \times \Sigma }[A^{1,0}_{out},A^{0,1}_{in}] = \int_{p^{-1}(A^{1,0}_{out},A^{0,1}_{in})}e^{\frac{i}{\hbar}S^f_{I \times \Sigma}[A_I,A^{1,0},A^{0,1}]}
\end{equation} 
We parametrize $p^{-1}(A^{1,0}_{out},A^{0,1}_{in})$ by choosing specific extensions $\tilde{A}^{1,0}_{out},\tilde{A}^{0,1}_{in}$ of $(A^{1,0}_{out},A^{0,1}_{in})$ to the bulk of the interval, then we have $A^{1,0} = \tilde{A}^{1,0}_{out} + a^{1,0}$, where $a^{1,0}\big|_{\Sigma_{out}} = 0$, and similary for $A^{0,1}$. The gauge-fixing condition that we impose\footnote{This condition has been used in Chern-Simons theory in various disguises at least since \cite{Froehlich1989}. It is used extensively in the works of Blau and Thompson \cite{Blau1993},\cite{Blau1995},\cite{Blau2006},\cite{Blau2013},\cite{Blau2016}, as well as the authors own work together with Mnev and Cattaneo, see \cite{Cattaneo2017a},\cite{Cattaneo2021},\cite{Wernli2019}.} is  
\begin{equation}
\partial_t A_I = 0.
\end{equation} In other words, we have $A_I = \sigma \in \Omega^0(\Sigma,\mathfrak{g})$. In other words, we are performing only a partial gauge fixing, and the partition function will be a function of $\sigma$ as well as the boundary conditions. 
One can show that the result of the Feynman diagram computation we are about to perform does not depend on the choice of the extensions $\tilde{A}^{1,0}_{out}, \tilde{A}^{0,1}_{in}$. \footnote{This argument has not been written down for this particular problem, but it is maybe not surprising, as it simply amounts to a reparametrization in the path integral.} Therefore we can let the support of $\tilde{A}^{1,0}_{out}$ approach the boundary, and similarly for $\tilde{A}^{0,1}_{in}$. In the limit, the gauge-fixed action functional becomes 
\begin{multline*}
S^f_{I\times M,gf}[A^{1,0}_{out},A^{0,1}_{in}; \sigma; a^{1,0}, a^{0,1}]=\\
\frac{1}{2\pi}\int_{ I\times\Sigma} \tr a^{1,0} d_I a^{0,1}  
+\frac{1}{2\pi} \int_{ I\times\Sigma} \tr dt\, (a^{1,0}+a^{0,1}) d_\Sigma \sigma+
\frac{1}{2\pi}\int_\Sigma \tr A^{1,0}_{out} a^{0,1}\big|_{t=1}  - \frac{1}{2\pi}\int_\Sigma A^{0,1}_{in} a^{1,0}\big|_{t=0} \\ 
- \frac{1}{\pi}\int_{I\times \Sigma}\tr  dt \tr a^{1,0} \wedge \sigma \wedge a^{0,1}.  
\end{multline*}
We now want to compute 
\begin{equation}
Z_{I \times \Sigma}[A^{1,0}_{out},A^{0,1}_{in}; \sigma] = \int_{a^{1,0},a^{0,1}} e^\frac{i}{\hbar}S^f_{I\times M,gf}[A^{1,0}_{out},A^{0,1}_{in}; \sigma; a^{1,0}, a^{0,1}]
\end{equation}
as a formal Fresnel integral through Feynman graphs and rules.\footnote{For an introduction to Feynamn graphs from a mathematical viewpoint, one can consult for instance on of the excellent texts \cite{Polyak2005},\cite{Reshetikhin2010}, \cite{Mnev2019a}. } We can invert the operator $d_I$ appearing in the first term on the gauge fixed action, the integral kernel of the inverse is called ``the propagator'' and is given by:
\begin{equation}
\langle a^{0,1}(t,z) a^{1,0}(t',z') \rangle  =-i\hbar\, \theta(t-t')\, \delta^{(2)}(z-z')\frac{i}{2}d\bar{z}\,dz' ,  \label{non-ab phys prop} 
\end{equation}
Then one uses Wick's theorem to define the integral: It means that all pairs of fluctuations  $a^{1,0}a^{0,1}$ coming from exponentials of the remaining terms are replaced by a propagator. The resulting terms are conveniently collected in diagrams called Feynman diagrams that are depicted below. 
The resulting Feynman diagrams are depicted in Figure \ref{fig: Feynman diagrams} below. 

\begin{figure}[h!]
\centering 
\def\svgwidth{\columnwidth}
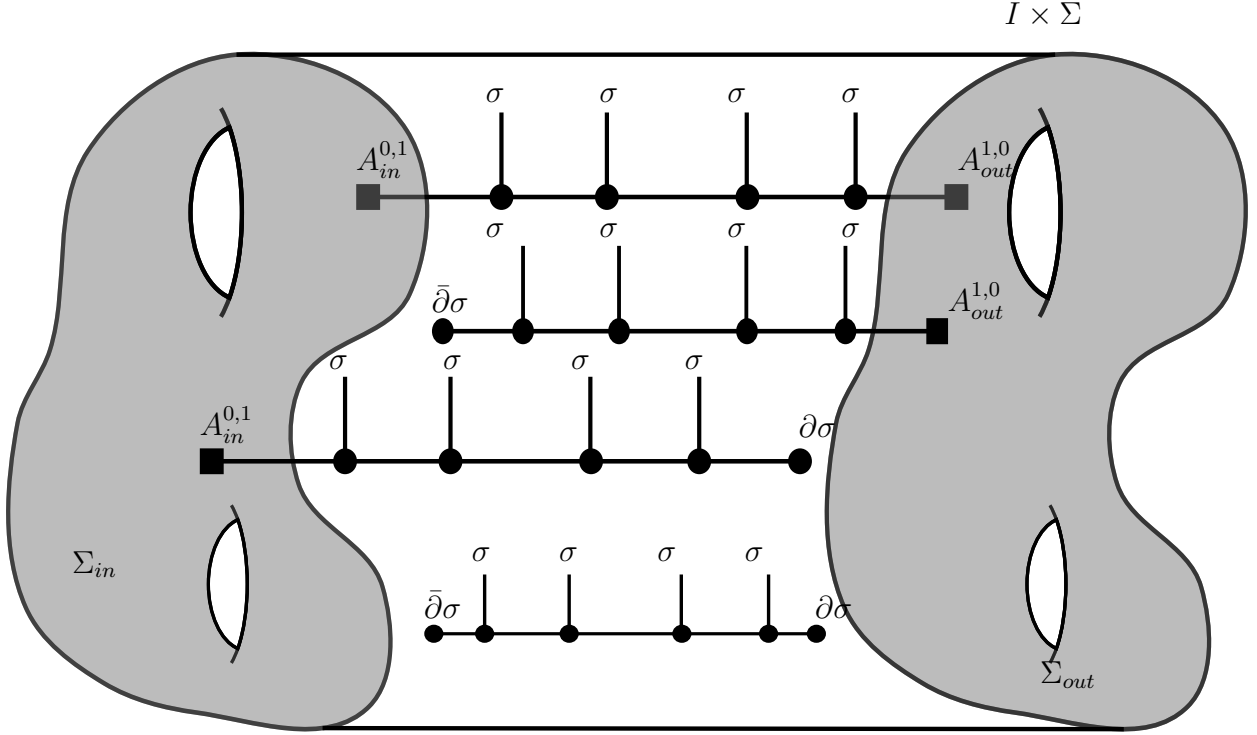\caption{Feynman diagrams appearing in the Feynman diagram computation of $Z$. Round vertices are integrated over the bulk $I\times M$ and square vertices over the respective boundary component.}\label{fig: Feynman diagrams}
\end{figure}

Resumming the Feynman diagrams, one can show the following result. 
\begin{prop}[\cite{Cattaneo2022a}]
Let $g = \exp( - \sigma)\colon \Sigma \to G$. Then we can write the Chern-Simons partition function as 
\begin{equation}
Z_{I\times M}[A^{1,0}_{out}, A^{0,1}_{in}; g]= e^{\frac{i}{\hbar}S^{eff}_{I\times M}[A^{1,0}_{out}, A^{0,1}_{in}; g]}
\end{equation}
where $S^{eff}_{I\times M}$, the \emph{effective action} is 
\begin{multline}\label{S eff non-abelian via g}
S^\mathrm{eff}_{I \times M}[A^{1,0}_{out}, A^{0,1}_{in}; g] =
\int_\Sigma \Big( \langle A^{1,0}_{out}, g\, A^{0,1}_{in} g^{-1} \rangle -
\langle A^{1,0}_{out},\bar{\partial} g\cdot g^{-1} \rangle 
-\langle  A^{0,1}_{in}, g^{-1}\,  \partial g  \rangle \\
+\mathrm{WZW}(g) = 
\end{multline}
\begin{equation}\label{WZW def}
\mathrm{WZW}(g)=- \frac12 \int_\Sigma  \langle \partial g\cdot g^{-1}, \bar\partial g\cdot  g^{-1} \rangle -\frac{1}{12}  \int_{I\times \Sigma}\langle d\tilde{g}\cdot \tilde{g}^{-1},[d\tilde{g}\cdot \tilde{g}^{-1},d\tilde{g}\cdot \tilde{g}^{-1}] \rangle .
\end{equation}
\end{prop} 
Here we have used the notation $\langle A,B \rangle = \frac{1}{2\pi}\tr A B$. 
That is, the effective action of Chern-Simons theory on a cylinder coincides with the action of a gauged WZW model, and thus the vaccuum state of Chern-Simons theory -- in the holomorphic polarization on phase space -- will coincide with the vaccuum state of the WZW model.\footnote{In fact, taking care of the ghost sector one also obtains the corresponding modification of the path integral measure. See \cite{Cattaneo2022a} (also \cite{Blau1993} for similar results in a different setup). } The gauged WZW action is known to satisfy the Polyakov-Wiegmann identity, namely 
\begin{equation}
S^{eff}_{I \times M }\left[{}^{h_{in}}(A^{0,1}_{in}),{}^{h_{out}}(A^{1,0}_{out});h_{out} gh_{in}^{-1}\right] = S^{eff}_{I \times M }(A^{0,1}_{in} , A^{1,0}_{out};g) - S^{eff}_{I \times M }(A^{0,1}_{in},0;h_{in}) +S^{eff}_{I \times M }(0, A^{1,0}_{out};h_{out}). \label{eq:PW}
\end{equation}
This implies the constraint \eqref{eq:curvature constraint}, see e.g. \cite{Gawedzki1991} or \cite{Cattaneo2022a}. 
\pagebreak
\printbibliography

\end{document}